\documentclass[floatfix,aps,twocolumn,showpacs,superscriptaddress,amsmath,amssymb,longbibliography]{revtex4-2}

\usepackage[autostyle=true]{csquotes}

\usepackage{times}

\usepackage{mathtools}
\usepackage{textcomp}
\usepackage{gensymb}
\usepackage{graphicx}
\usepackage{siunitx} 
\usepackage{ulem}
\usepackage{physics}
\usepackage{appendix}
\usepackage{xcolor}

\usepackage[unicode=true,
 bookmarks=true,bookmarksnumbered=true,bookmarksopen=true,bookmarksopenlevel=2,
 breaklinks=false,pdfborder={0 0 1},backref=false,colorlinks=true]
 {hyperref}

 \hypersetup{linkcolor=blue, citecolor=blue, urlcolor=blue, filecolor=blue, pdfpagelayout=OneColumn,
  pdfnewwindow=true, pdfstartview=XYZ, plainpages=false}

\usepackage[all]{hypcap}
\usepackage{color}

\makeatletter
\def\mathcolor#1#{\@mathcolor{#1}}%
\def\@mathcolor#1#2#3{%
  \protect\leavevmode%
  \begingroup\color#1{#2}#3\endgroup%
}%
\newcommand{\msout}[1]{\text{\sout{\ensuremath{#1}}}}%

\newcommand{\sam}[2]{%
\ifmmode%
  \msout{#1}\mathcolor{red}{#2}%
\else%
  \sout{#1}\textcolor{red}{#2}%
\fi}
\newcommand{\samO}[2]{%
\ifmmode%
  \msout{#1}\mathcolor{magenta}{#2}%
\else%
  \sout{#1}\textcolor{magenta}{#2}%
\fi}

\definecolor{orange}{rgb}{0.6,0.33,0.73}
\definecolor{bronze}{rgb}{0.8, 0.5, 0.2}

\newcommand{\gab}[2]{%
\ifmmode%
  \msout{#1}\mathcolor{green}{#2}%
\else%
  \sout{#1}\textcolor{green}{#2}%
\fi}

\newcommand{\newcomment}[1]{\color{black}}

\makeatother

\makeatletter

\newcommand\Autoref[1]{\@first@ref#1,@}
\def\@throw@dot#1.#2@{#1}
\def\@set@refname#1{
    \edef\@tmp{\getrefbykeydefault{#1}{anchor}{}}%
    \xdef\@tmp{\expandafter\@throw@dot\@tmp.@}%
    \ltx@IfUndefined{\@tmp autorefnameplural}%
         {\def\@refname{\@nameuse{\@tmp autorefname}s}}%
         {\def\@refname{\@nameuse{\@tmp autorefnameplural}}}%
}
\def\@first@ref#1,#2{%
  \ifx#2@\autoref{#1}\let\@nextref\@gobble
  \else%
    \@set@refname{#1}
    \@refname~\ref{#1}
    \let\@nextref\@next@ref
  \fi%
  \@nextref#2%
}
\def\@next@ref#1,#2{%
   \ifx#2@ and~\ref{#1}\let\@nextref\@gobble
   \else, \ref{#1}
   \fi%
   \@nextref#2%
}

\makeatother

\begin{document}

\def\be{\begin{equation}}
\def\ee{\end{equation}}

\def\myvec#1{{\bf #1}}
\def\Esw{\myvec{E}_{sw}}
\def\Hsw{\myvec{H}_{sw}}

\title{Taming a Maxwell's demon for experimental stochastic resetting}

\author{R\'emi Goerlich}
\affiliation{Raymond \& Beverly Sackler School of Chemistry, Tel Aviv University, Tel Aviv 6997801, Israel}
\affiliation{Universit\'e de Strasbourg, CNRS, Institut de Physique et Chimie des Mat\'eriaux de Strasbourg, UMR 7504, F-67000 Strasbourg, France}
\affiliation{Universit\'e de Strasbourg, CNRS, Centre Europ\'een de Sciences Quantiques \& Institut de Science et d'Ing\'enierie Supramol\'eculaires, UMR 7006, F-67000 Strasbourg, France}
\author{Minghao Li}
\affiliation{Department of Physics, University of Basel, Klingelbergstrasse 82, 4056 Basel, Switzerland}
\author{Lu\'is B. Pires}
\affiliation{Universit\'e de Strasbourg, CNRS, Centre Europ\'een de Sciences Quantiques \& Institut de Science et d'Ing\'enierie Supramol\'eculaires, UMR 7006, F-67000 Strasbourg, France}
\affiliation{Departamento de Física, Universidade Federal de Viçosa. Viçosa, Minas Gerais, Brazil}
\author{Paul-Antoine Hervieux}
\affiliation{Universit\'e de Strasbourg, CNRS, Institut de Physique et Chimie des Mat\'eriaux de Strasbourg, UMR 7504, F-67000 Strasbourg, France}
\author{Giovanni Manfredi}
\affiliation{Universit\'e de Strasbourg, CNRS, Institut de Physique et Chimie des Mat\'eriaux de Strasbourg, UMR 7504, F-67000 Strasbourg, France}
\author{Cyriaque Genet}
\affiliation{Universit\'e de Strasbourg, CNRS, Centre Europ\'een de Sciences Quantiques \& Institut de Science et d'Ing\'enierie Supramol\'eculaires, UMR 7006, F-67000 Strasbourg, France}

\date\today

\begin{abstract}

A diffusive process that is reset to its origin at random times, so-called stochastic resetting (SR), is an ubiquitous expedient in many natural systems \cite{Evans2011}.
Yet, beyond its ability to improve the efficiency of target searching, SR is a true non-equilibrium thermodynamic process that brings forward new and challenging questions \cite{Fuchs2016}.
Here, we show how the recent developments of experimental information thermodynamics renew the way to address SR and can lead, beyond a new understanding, to better control on the non-equilibrium nature of SR.
This thermodynamically controlled SR is experimentally implemented within a time-dependent optical trapping potential.
We show in particular that SR converts heat into work from a single bath continuously and without feedback.
This implements a Maxwell's demon that constantly erases information \cite{Roldan2014, Ciliberto2019}.
In our experiments, the erasure takes the form of a protocol that allows to evaluate the true energetic cost of SR.
We show that using an appropriate measure of the available information, this cost can be reduced to a reversible minimum while being bounded by the Landauer limit \cite{Lutz2015}.
We finally reveal that the individual trajectories generated by the demon all break ergodicity and thus demonstrate the non-ergodic nature of the demon's modus operandi. 
Our results offer new approaches to processes, such as SR, where the informational framework provides key experimental tools for their non-equilibrium thermodynamic control.

\end{abstract}

\maketitle

\noindent{\textbf{\large{Introduction}}}\\

In a stochastic resetting (SR) process, a Brownian object diffuses, either freely or in a potential, for a random time before being reset to the origin \cite{Manrubia1999, Evans2011, kusmierz2014first, Montero2017, Chechkin2018, Evans2020}.
This simple yet rich paradigm has drawn a lot of attention recently in various fields of research \cite{Magoni2020, Kumar2020, Perfetto2021}.
Because it minimizes first passage times in search processes \cite{Stanislavsky2021} SR is an efficient solution to numerous problems in nature \cite{Boyer2014, Roldan2016, Bressloff2020}, devices \cite{roberts2024ratchet, pal2024channel} and in algorithms used, for instance, in molecular dynamics \cite{Blumer2022, blumer2024combining}.
SR has been explored shown to be experimentally accessible using optical tweezers revealing the importance of finite-time dynamics and energetic costs \cite{TalFriedman2020, Besga2020, Besga2021}.
More recently, and closer to natural systems, experimental SR  has been realized using active self-propelled robots, in a memory-keeping environment \cite{Altshuler2023}.
Thermodynamically, resetting brings the system to a non-equilibrium steady state (NESS) \cite{Sokolov2023, evans2013optimal}. This implies that improving the efficiency in search process comes with an energetic cost, which has to be balanced with the gain in search time \cite{Fuchs2016, Sunil2023}.
Quantitative characterization of that cost is currently not only driving intense theoretical efforts \cite{Fuchs2016, Pal2017, Gupta2020_thermo, Pal2021, Gupta2022, Gupta2022Thermo, Mori2022} but is also challenging experimentalists to fill-in the gap between a theoretical, idealized SR process and its actual physical realization.

Here, we meet this challenge by implementing SR on a Brownian microsphere opticaly trapped in water and provide a radically new understanding of the thermodynamic significance of SR. 
We first measure all the appropriate thermodynamics quantities associated with idealized SR. This lead us to show that idealized SR drives the microsphere in a NESS where, surprisingly, heat is constantly converted into work despite the fact that the microsphere is only coupled to a single heat bath (surrounding water). We explain how this apparent breach of the Second Law of thermodynamics summons a Maxwell's demon who acts, behind the scene, on the trap potential as an external agent to maintain the microsphere in this NESS.
We then recognize that the nature of the demon is informational: each resetting action is an instantaneous ``teleportation'' event, which erases a finite amount of information available before resetting. The thermodynamics at play in the demon's work extraction scheme is therefore the one of an information machine \cite{toyabe_experimental_2010, admon_experimental_2018}.

Of course, an experimental implementation of SR necessarily involves continuous trajectories where resetting takes a finite time and in turns, consumes a larger amount of energy \cite{Gupta2020_NonInst, Bodrova2020, MercadoVasquez2020, Santra2021}. 
Indeed, looking at an idealized SR process or at its continuous implementation corresponds either to ignore or to include the demon's cost in the description.
More precisely, we show that the full energy consumption of the continuous, physical SR process, accounting for the work experimentally injected in the demon, is always larger than the work the information machine can extract. This injected work naturally restores the Second Law \cite{Fuchs2016, Esposito2011}.
The information machine description of resetting consisting of sequences of erasures leads us to the Landauer bound of SR, which sets the minimal amount of injected work necessary to operate the machine.
The information erased during each resetting event is explicitly measured as the distance between the SR NESS and a corresponding equilibrium state clearly defined.
In the spirit of the Szilard's engine \cite{szilard1929, Parrondo2015}, this access of information allows us to build a reversible protocol and to bring the power injected in the information machine to its minimal value.

Adopting this framework finally leads to another essential, yet seldom analyzed, aspect of SR. In the optical trap, SR distinguishes the region where the microsphere position is reset from the rest of the available space. While being necessary to process information, this distinction is however known to break ergodicity \cite{Parrondo2015}. By measuring the non-ergodic nature of SR from the ensemble of all reset trajectories, we clearly confirm experimentally this link between ergodicity breaking and information processing.\\

Our work gathers three main results:
\textit{(i)} an extensive experimental thermodynamic description of SR which captures its Maxwell's demon nature. Using this framework, we demonstrate that the protocol used to reset the Brownian object can be designed in order to reduce the cost of maintaining a steady-state based on SR;
\textit{(ii)} a characterization of the information processed by the demon by which, exploiting a Szilard-inspired driving, we bring the cost of the information machine to its reversible minimum;
\textit{(iii)} a connection between the information thermodynamics description of SR and its non-ergodic properties. This leads us to propose new quantitative links between both aspects.\\

\noindent{\textbf{\large{From physical resetting in an optical trap to idealized SR process}}}\\

In this work, we set up an experimental platform to implement SR. It consists of a Brownian microsphere subjected to a time-dependent optical potential with two states: (i) a weak potential state, allowing the stochastic position $x_t$ of the microsphere along the optical axis at time $t$ to diffuse during a time $\tau$ and (ii) a strongly confining state, quenching the particle very close to $x = 0$.
This quench constitutes what we call a single resetting event.
The time $\tau$ between two consecutive quenches is drawn from an exponential distribution $P_r(\tau) = \lambda e^{-\lambda\tau}$ with constant rate $\lambda$, hence implementing within our optical trap a Poissonian SR process \cite{Evans2011}.

Our experimental setup is sketched on Fig.~\ref{fig:TrajDist}~(a) and detailed in Supplemental Material Sec.~I \cite{Supplemental}. It uses a laser beam focused on a microsphere in water to induce a confining harmonic optical potential.
A secondary low-power laser beam acts as a passive probe to record in real-time the successive positions $x_t$ of the overdamped microsphere that compose a trajectory.
The stiffness $\kappa$ of the harmonic confinement is linearly related to the intensity of the trapping beam, controlled via an accousto-optic modulator programmed numerically (see details in Supplemental Material Sec.~I \cite{Supplemental}).
If the stiffness of the optical trap is low $\kappa = \kappa_{\rm min}$, the particle explores a shallow potential, with a diffusion coefficient $D = k_B T / \gamma$ given by Boltzmann constant $k_B$, temperature $T$, Stokes drag coefficient $\gamma$ and with a relaxation time $\tau_{\rm min}=  \gamma/ \kappa_{\rm min}$.
When the stiffness abruptly increases to a high value $\kappa_{\rm max} \gg \kappa_{\rm min}$, the particle relaxes exponentially fast towards the center of the potential $x = 0$, with a relaxation time $\tau_{\rm max} = \gamma / \kappa_{\rm max}$.
This stiffness quench acts as a resetting event and, to ensure that the particle is well reset, the potential is kept stiff for a waiting time $\tau_{\rm wait}$ several times larger than the relaxation time $\tau_{\rm max}$.

This experimental realization of SR cannot but departs from theoretical resetting processes \cite{Gupta2020_NonInst, Gupta2022Thermo,  Bodrova2020, MercadoVasquez2020} because of the non-instantaneous nature of the relaxation.
Despite this, however, an idealized SR process characterized by instantaneous "teleportation" events can be built from experimental trajectories if one removes the parts recorded during the waiting times $\tau_{\rm wait}$ over which the microsphere is relaxing inside the optical trap.
Having both decimated and full trajectories is absolutely crucial to connect the thermodynamics of the system to the cost of the operating demon.

\begin{figure}[htb!]
	\centerline{\includegraphics[width=0.8\linewidth]{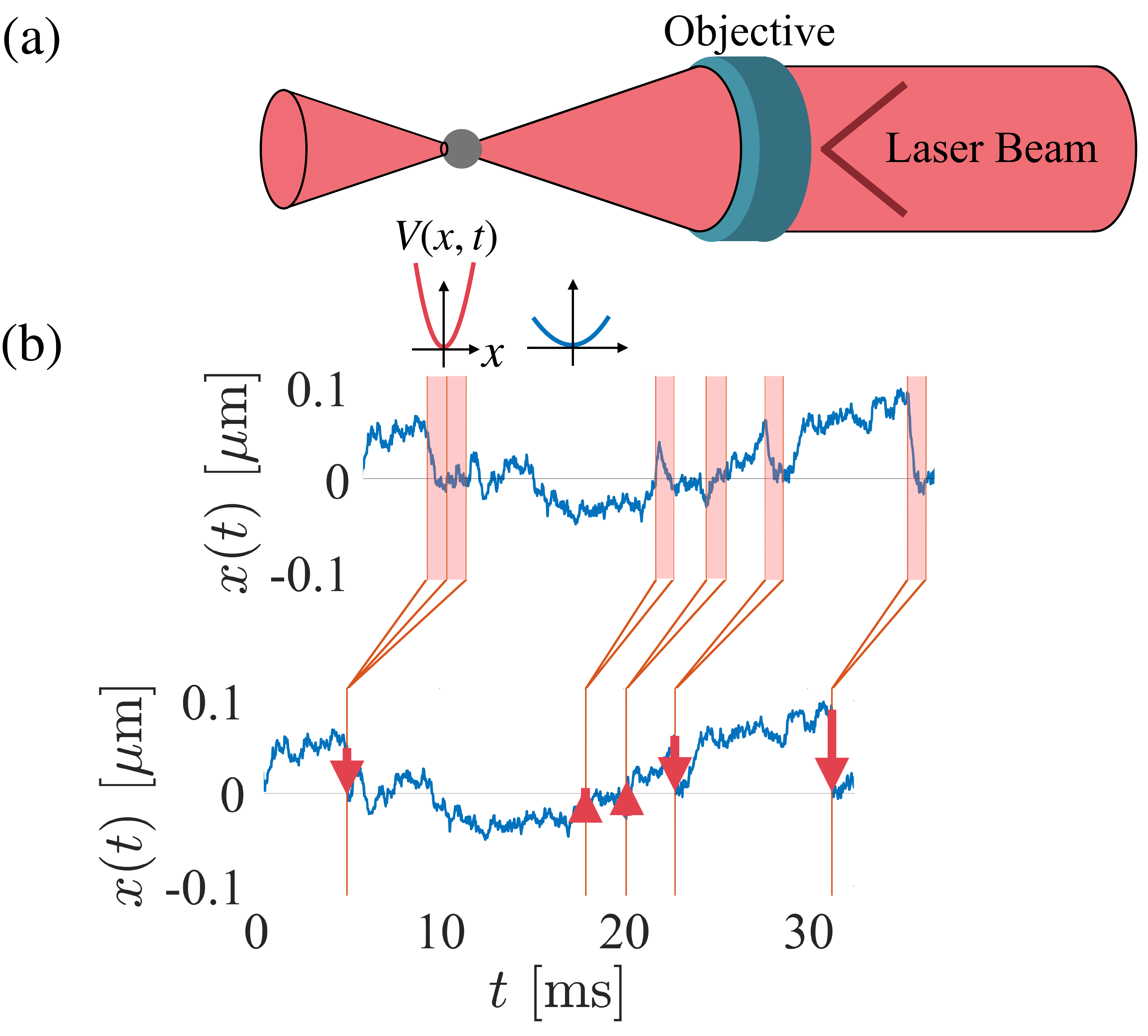}}
	\centerline{\includegraphics[width=0.85\linewidth]{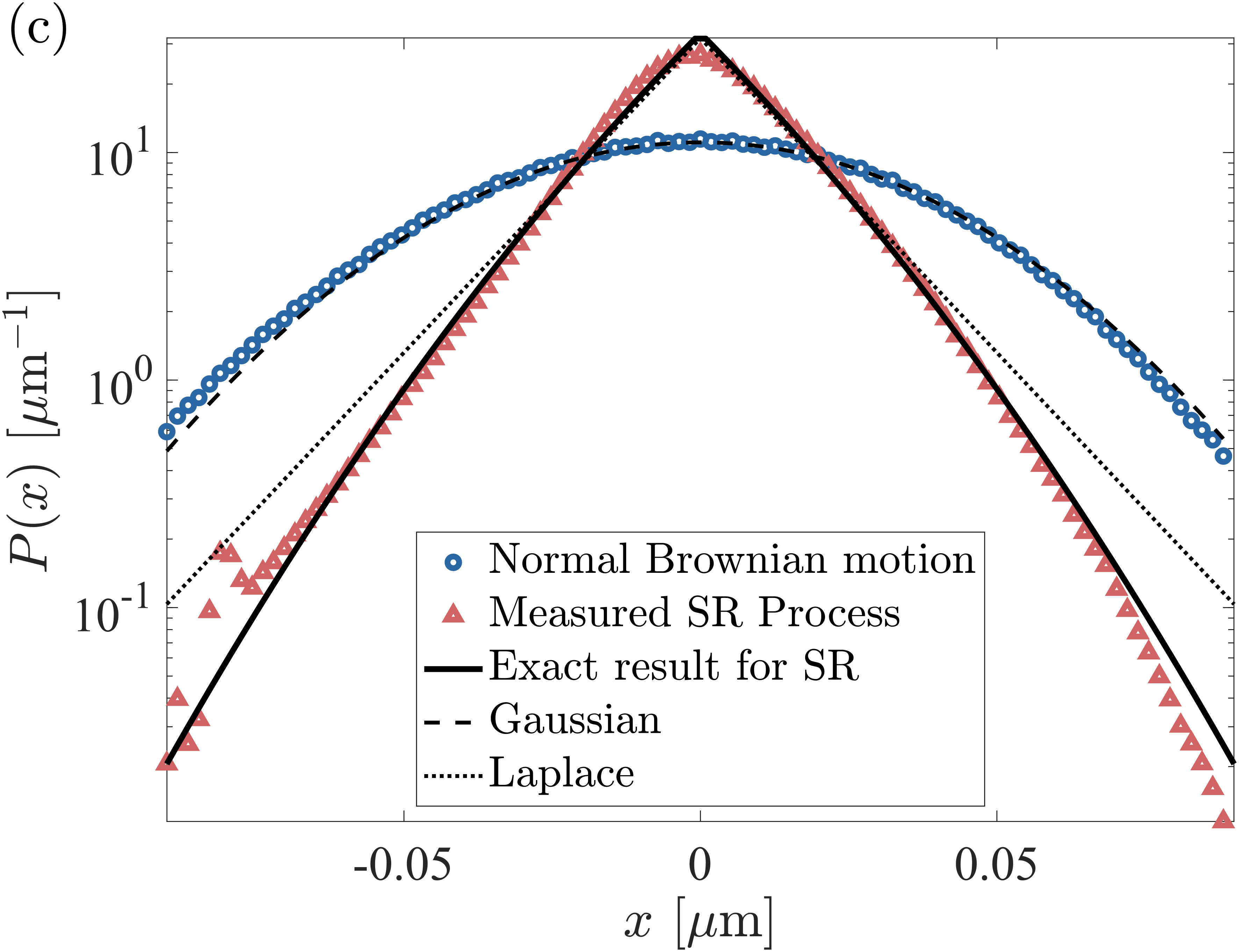}}
	\caption{(a) Simplified view of the optical trapping system: a 820 nm laser beam, tightly focused through a high numerical aperture objective is confining the overdamped motion of a 3 $\mu$m polysterene bead in water.
	(b) Experimental trajectory of the bead diffusing within the confining harmonic potential $V(x) = \kappa_{\rm min} x^2 / 2$ subjected to stochastic resetting at an inverse rate $\lambda^{-1} = 1.5 ~ \rm{ms}$. This corresponds to the sequence of potentials drawn above the trajectory, with the state of the potential $V(x)$ in blue for $\kappa_{\rm min}$ (leading to $\tau_{\rm min} = 7.9 ~\rm{ms}$) and in red for $\kappa_{\rm max}$ (leading to $\tau_{\rm max} = 0.27~ \rm ms$). The first trajectory shown is a full one, meaning that it contains the finite-time relaxation, visible inside the red patches, when $\kappa = \kappa_{\rm max}$. The bottom trajectory is a decimated one, for which all points during the transient have been removed. This procedure yields the idealized -hence shorter- instantaneous SR process (resetting events are marked with red arrows). See Supplemental Material Sec.~I \cite{Supplemental} for experimental details.
	(c) Probability distribution of position $P(x)$ of a $300$ seconds-long ideal, decimated SR process (red triangles) with rate $\lambda$ in potential $V(x)= \kappa_{\rm min} x^2 / 2$ together with the exact result derived in Supplemental Material Sec.~ II \cite{Supplemental}, using all experimental parameters (black solid line). The equilibrium distribution $P_{\rm eq}(x)$ of a normal Ornstein-Uhlenbeck process in the same potential $V(x)$ (blue circles) is a Gaussian (black dashed line) as expected for trapped Brownian object. The stready-sate SR probability distribution $P(x)$ significantly differs both from $P_{\rm eq}(x)$ and from the Laplace distribution of a free SR process \cite{Evans2011} (black dotted line). This clearly demonstrates the combined effect of a confining potential $V(x)$ and SR.}
	\label{fig:TrajDist}
\end{figure}

On Fig.~\ref{fig:TrajDist}~(b), we show an experimental SR process in a potential $V(x) =  \kappa_{\rm min} x^2 /2 $ with $\kappa_{\rm min} = 3.2 \pm 0.13$, $\kappa_{\rm max} = 97 \pm 4.2 ~\si{\pico\newton/\micro\meter}$ and with an inverse resetting rate $\lambda^{-1} = 1.5 ~\si{\milli\second}$.
The trajectory of the microsphere within the reset potential is recorded for $300$ seconds at a frequency $2^{15} = 32768 ~ \si{\hertz}$.
Points acquired during the waiting times $\tau_{\rm wait}$ are removed to form the idealized trajectory shown in Fig.~\ref{fig:TrajDist}~(b).
Importantly, this decimation erases the memory of the trajectory: there are no correlations between the position before and after each resetting event since $\tau_{\rm wait} > \tau_{\rm max}$.
The resetting events are marked by red arrows on Fig. ~\ref{fig:TrajDist}~(b).

For Poissonian SR in a harmonic potential, the process reaches a non-equilibrium steady-state distribution $P(x)$ that can be computed as $ P(x) = \lambda \int_0^\infty e^{-\lambda t} P(x | t, x=0) dt$ \cite{Pal2015, gradshteyn2007, Mori2022, Trajanovski2023}. Here, $P(x | t, x=0)=\exp (-x^2/2\sigma^2(t))/\sqrt{2\pi \sigma^2(t)}$ is the standard Ornstein-Uhlenbeck probability density of position for Brownian diffusion within the harmonic potential, giving the probability for the particle to diffuse from $0$ to $x$ in a time $t$ in the absence of resetting.
We derive in Supplemental Material Sec.~II \cite{Supplemental} the exact steady-state distribution that generalizes to all real values of the $\lambda / \omega_0$ ratio the known expression for SR in a harmonic potential \cite{Pal2015}.

On Fig.~\ref{fig:TrajDist}~(c), we plot the probability distribution built from this decimated trajectory.
The agreement between the experimental non-equilibrium steady-state (red triangles) and the analytical result confirms that the decimated experimental trajectories are very close to those of an idealized SR process.
The agreement also shows that the experimental error on the resetting position due to the finite value of $\kappa_{\rm max}$, can be neglected.
On the same graph, we plot in blue the equilibrium distribution of the normal Ornstein-Uhlenbeck process without resetting.
This diffusion in the harmonic potential $V(x)$ is characterized by a Gaussian distribution $P_{\rm eq}(x) =  \sqrt{\frac{\kappa_{\rm min}}{ 2 \pi k_B T}} e^{-\kappa_{\rm min} x^2 / 2 k_B T}$. It is clear that the SR process, confining the motion inside the trap, reaches a distribution that strongly differs from the equilibrium Gaussian probability density.\\

\noindent{\textbf{\large{SR in a harmonic potential emulates a Maxwell's demon}}}\\

\begin{figure*}
	\centerline{\includegraphics[width=0.9\linewidth]{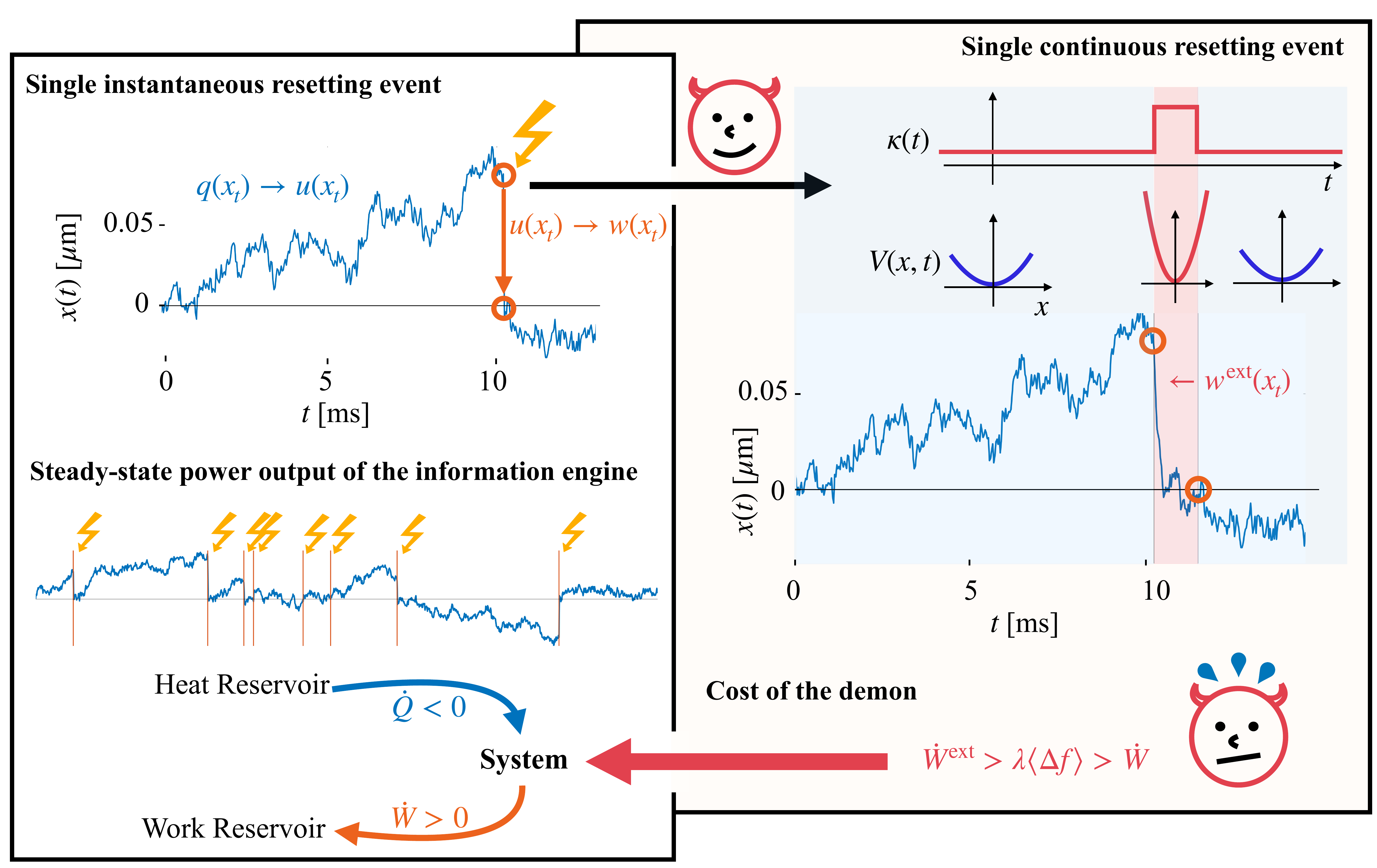}}
	\caption{Working principle of the SR as a Maxwell's demon. On the top left of the figure, a single instantaneous resetting event is shown : a single Brownian trajectory is diffusing from $x=0$ at $t=0$ until $x(\tau)$ at random time $t=\tau\approx 10$ ms, time at which it is instantaneously reset to $x = 0$. The explicit mechanism triggering the resetting event is for now unknown and is represented by a yellow spark. From a thermodynamic perspective, this trajectory is composed of two sequences: before resetting, the diffusing system absorbs a stochastic amount of heat from the surrounding heat bath and transforms it into internal energy as it explores the external potential $q(x_t) \rightarrow u(x_t) = \kappa_{\rm min} x_t^2/2$. At the resetting event, the particle experiences a net displacement $\Delta x = x_t$ induced by the same external potential. This event corresponds to some stochastic work $w(x_t) = -u(x_t)$ exerted against the potential. Combining both sequences, it clearly appears that the absorbed heat is converted into work.
	The accumulation of sequential resetting events in steady-state (bottom left) therefore constitutes a stationary Maxwell's demon where the system absorbs heat at a constant rate $\dot Q$ (blue arrow) from the single heat bath and fully converts it into work $\dot W$ (orange arrow).
	However, the agent triggering the resetting event can be explicitly described (top large black arrow) as an external \textit{demon}. As explained in the main text, each resetting event is achieved by a stiffness protocol $\kappa(t)$ under the form of a quench of the potential from a low $\kappa_{\rm min}$ to a high $\kappa_{\rm max}$ for a finite time. The stochastic work $w^{\rm ext}(x_t)$ needed for the quench can be evaluated with the standard tools of stochastic energetics. Again, in the steady-state (bottom right), this corresponds to a net power $\dot W^{\rm ext}$ engaged by the demon. This cost is bounded by the non-equilibrium Second Law and must be considered on the energetic balance of the whole system (red horizontal arrow) to recover thermodynamic consistency.
	Of course this external work injected in the system by the demon is fully dissipated as heat in the bath $\dot W^{\rm ext} = -\dot Q^{\rm ext}$. Therefore, the thermodynamics of the full trajectories is well described by the measure of the external work.
	}
	\label{fig:demon}
\end{figure*}

Both for decimated or full trajectories, SR brings the system into a NESS with continuous, exchanges energy with the heat bath.
As we now explain and describe in Fig.~\ref{fig:demon}, each resetting event can be seen as the action of a Maxwell's demon.

For an idealized process, SR yields trajectories by which stochastic internal energies $u(t)$ and system's stochastic entropies $s(t)$ can be measured \cite{Seifert2005}.
Both are then combined in non-equilibrium free energies $f(t)$. 
We use the center of the trap $x=0$ as a reference to build the free energy difference associated with one resetting event $\Delta f(t) = \Delta u(t) - T \Delta s(t)$ where $\Delta u(t) = V(x = 0) - V(x_t)$.
The abrupt change in stochastic internal energy results from a resetting event triggered by an external agent whose role is played by the Maxwell's demon sketched in Fig.~\ref{fig:demon}. The resetting event is also accompanied by a difference in the system's stochastic entropy stemming from the shift in position $\Delta s(t)= s(x=0)-s(x_t)$.

A single resetting event is detailed in the top left part of Fig.~\ref{fig:demon}.
First, during diffusion in-between resetting events, heat builds up, absorbed from the bath by the microsphere starting at $x=0$.
This occurs in the constant shallow potential ($\kappa_{\rm min}$) with no work exchanged. Heat therefore is fully converted into internal energy.
Then, at the resetting event, this stored internal energy is instantaneously converted into the work $\Delta w(t) = -\Delta u(t)$ \cite{Fuchs2016}.
No heat is involved during this instantaneous event and the free energy describing this sequence reads $\Delta f(t) = -\Delta w(t) - T \Delta s(t)$.
Importantly, at the center of the potential, $V(x = 0) = 0$ which implies that $\Delta w > 0$, \textit{i.e.} that stochastic work is extracted from the system.
Overall therefore, the heat absorbed during the diffusive sequence is fully converted into work at the resetting event.

Our platform gives access to the motion of the optically trapped microsphere over long times and hence to idealized trajectories that have undergone many such resetting events and have converged to their steady-state distribution (see Fig.~\ref{fig:TrajDist}~(c)). With such trajectories, we measure the average work production rate as $\dot{W} = \lambda \langle \Delta w \rangle=\lambda \langle V(x_t) \rangle$ where the brackets $\langle ... \rangle$ denotes the average taken in the steady-state distribution $P(x)$ as $\int ... P(x) dx$.

\begin{figure}[htb!]
	\begin{center}
		\centering{
			\includegraphics[width=0.23\textwidth]{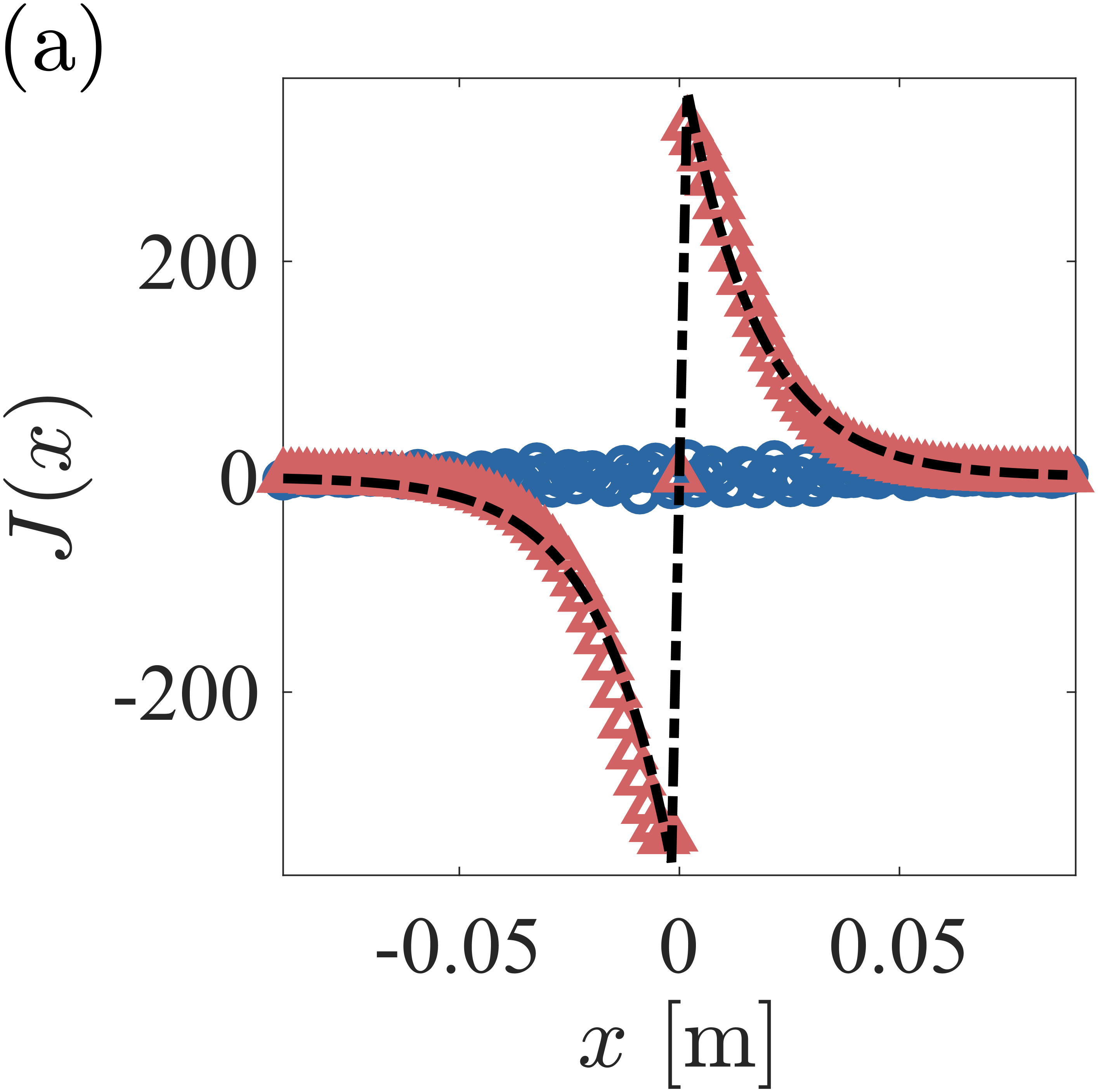}
			\includegraphics[width=0.23\textwidth]{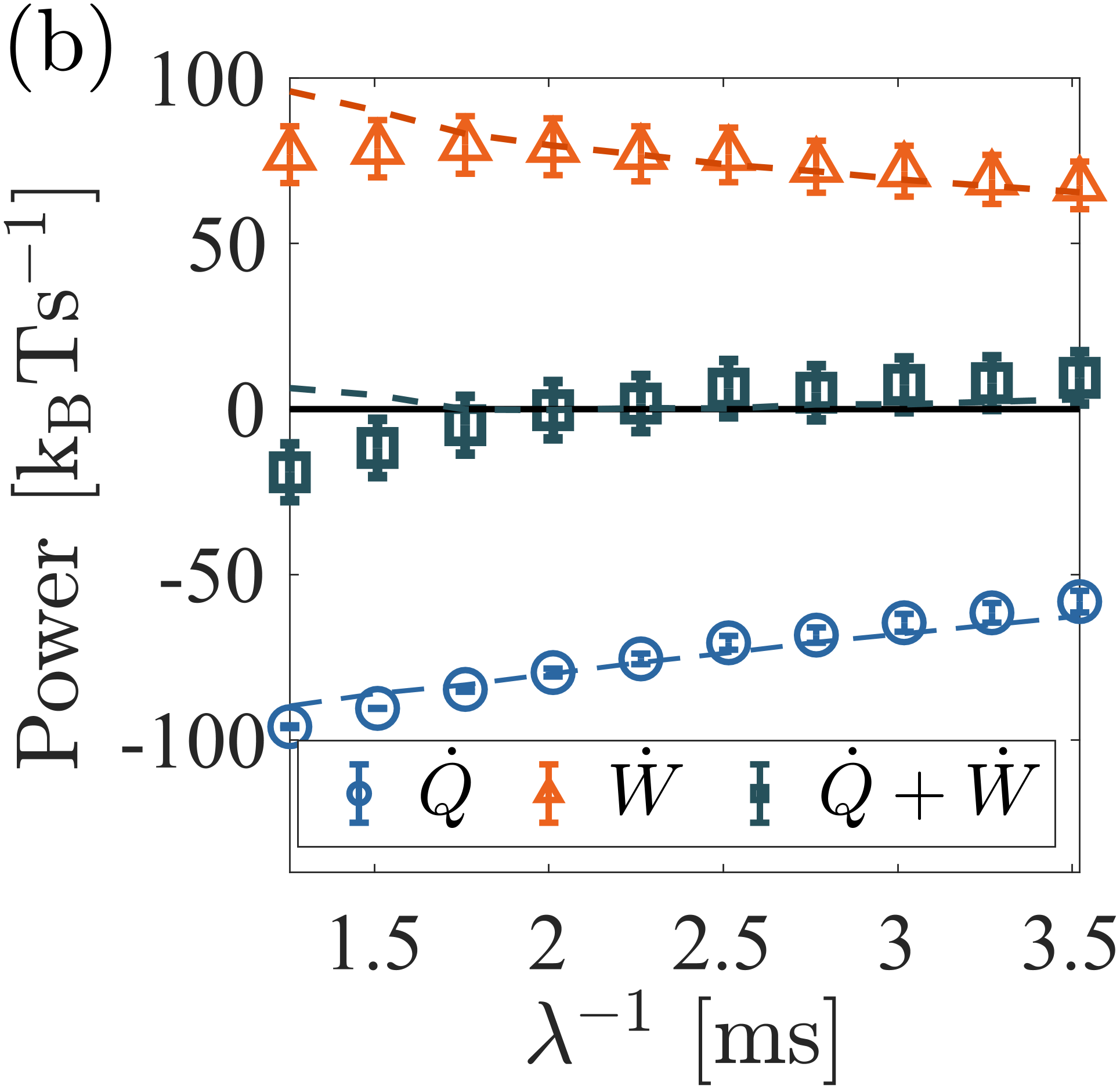}
			\includegraphics[width=0.46\textwidth]{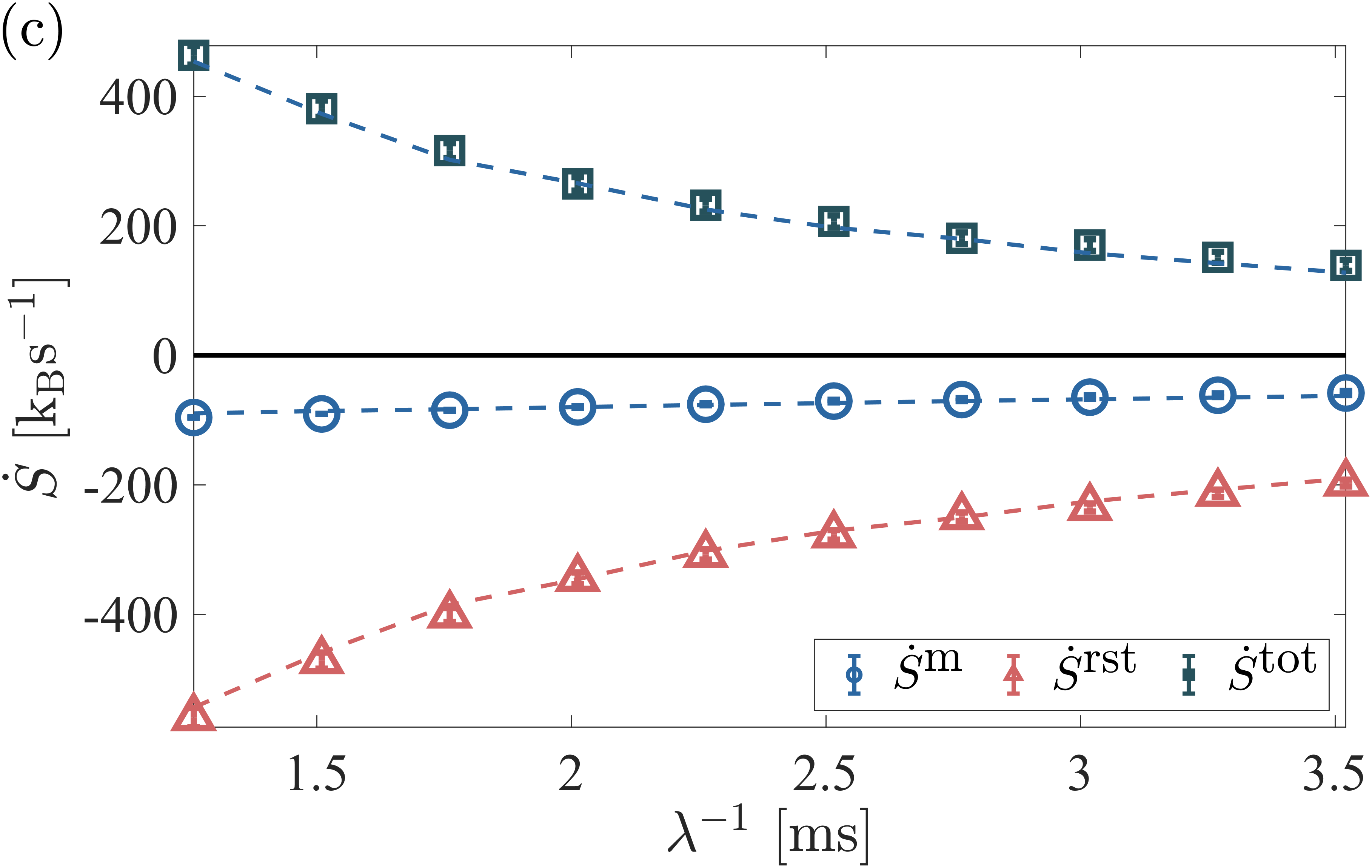}
			\label{fig:FirstPrinciple}}\\
		\caption{{ 
        (a) Probability currents $j(x)$ for the equilibrium trajectories in the optical trap (blue circles) and for a trajectory undergoing resetting with inverse rate $\lambda^{-1} = 1.26$ ms (red triangles). The exact result (black dashed line) derives from the analytical SR NESS probability density $P(x)$.
        (b) Work $\dot W$ and heat $\dot Q$ production rates experimentally measured for an SR process with inverse resetting rates $\lambda^{-1}$ ranging from $1.26$ ms to $3.52$ ms. The vanishing total rate $\dot W+\dot Q = 0$ measured for all $\lambda$ demonstrates the First Law of thermodynamics for resetting. Experimental data are compared with numerical simulations (dashed lines) performed in conditions similar to the experiments. 
        Comparing the motional variances of the measured trajectories to the numerical results leads to a correction of the experimental calibration, as detailed in Supplemental Material Sec.~I \cite{Supplemental}.
            (c) Experimentally measured medium entropy production rate $\dot{S}^{\rm m}  = \dot{Q}/T$ (blue circles), resetting entropy production rate $\dot{S}^{\rm rst}$ (red triangles) and total entropy production rate $\dot{S}^{\rm tot} $ (grey-blue squares), positive for all $\lambda$. All entropy productions decrease in the $\lambda^{-1} \rightarrow \infty$ limit of equilibrium.
		}}
	\end{center}
\end{figure}

This average work production rate is always positive, as seen on Fig.~\ref{fig:FirstPrinciple}~(b), where we display $\dot{W}$ experimentally extracted from our  idealized SR trajectories, as a function of the mean resetting time $\lambda^{-1}$, keeping the same $\kappa_{\rm min}$ and $\kappa_{\rm max}$.
The experimental results are complemented by numerical simulations (see Supplemental Material Sec.~I \cite{Supplemental} for details).

In the NESS, the average heat production rate is related to the non-vanishing probability current $j(x) = (-\frac{1}{\gamma}\frac{d V(x)}{dx}  - D \partial_x )P(x)$ maintaining the NESS distribution different from the equilibrium solution $P_{\rm eq}(x)$ in the external potential \cite{Fuchs2016, evans2013optimal}.
We plot the experimentally measured current on Fig.~\ref{fig:FirstPrinciple}~(a) both in equilibrium and in the SR-induced NESS.
In the latter, a net current is induced from both sides of the center $x=0$ of the optical potential, with opposite signs as the manifestation of the confining effect of SR -see Fig.~\ref{fig:TrajDist}~(c). This is the consequence of the breaking of the detailed-balance condition imposed by the resetting events at the level of each single trajectory.
Since the current can be evaluated using the experimental distribution $P(x)$ of the recorded trajectories, the average heat build-up rate exchanged with the bath given by $\dot{Q} = \int j(x) \frac{d V(x)}{dx} dx$ can be experimentally measured \cite{Fuchs2016}.
On Fig.~\ref{fig:FirstPrinciple}~(b), we plot the measured average heat production rate $\dot{Q} < 0$ for all probed $\lambda^{-1}$: heat is on average absorbed from the bath over the full resetting sequence and transformed into work $\dot{W} > 0$. 

This analysis clearly shows that the idealized SR in a confining potential emulates a true Maxwell's demon that converts heat into work from a single bath at constant temperature, as sketched in Fig.~\ref{fig:demon}
The steady-state thermodynamic balance is summarized in the bottom left part of the sketch.
Our results thus experimentally confirm the prediction of \cite{Fuchs2016}.
Importantly, in the idealized framework, the external potential is constant and there is no change in internal energy in the steady-state.
As a direct thermodynamic consequence, the First Law for resetting reads $\dot{Q} + \dot{W} = 0$, as verified on Fig.~\ref{fig:FirstPrinciple}~(b).
The magnitudes of both heat and work production rates decrease when $\lambda^{-1}$ increases \textit{i.e.} when the trajectory undergoes fewer resetting events.
In the limit of infinite $\lambda^{-1}$, the system is back to equilibrium, without resetting and no exchange of energy.\\

\noindent{\textbf{\large{Protocol-dependent erasure of information and Landauer's principle for real SR}}}\\

We now take a different look at the action of the demon by noting that each resetting event also implies a change in information \cite{Fuchs2016}. This is seen in the expression taken by $\Delta s = s(x=0) - s(x_t) = k_B\ln[P(x_t)/P(0)]$ that writes as a difference in stochastic Shannon entropy $s(x_t) = -k_B\ln[P(x_t)]$.
Since $P(0)$ is the maximum of the distribution function (see Fig~\ref{fig:TrajDist}~(c)), $\langle \Delta  s \rangle < 0$, meaning that the entropy of the SR trajectory is reduced. In a steady-state SR process with rate $\lambda$, the average difference gives the \textit{resetting} entropy production rate $\dot{S}^{\rm rst} = \lambda \langle \Delta  s \rangle$.

The heat build-up rate $\dot{Q}$ discussed above gives a second entropic contribution associated with the \textit{medium} entropy $\dot{S}^{\rm m}  = \dot{Q}/T$.
In Fig.~\ref{fig:FirstPrinciple}~(c) we plot  the evolution of experimentally measured entropy production rates as a function of $\lambda^{-1}$.
As heat and work in the First Law (Fig.~\ref{fig:FirstPrinciple}~(b)), both entropic rates decrease as the system approaches thermodynamic equilibrium for large $\lambda^{-1}$.
In the steady-state, $\dot{S}^{\rm rst}$ and $\dot{S}^{\rm m}$ are combined into the total entropy production rate which is shown as grey-blue squares on Fig.~\ref{fig:FirstPrinciple}~(c).
As expected by the second Law, this total entropy production rate $\dot{S}^{\rm tot} = \dot{S}^{\rm m} - \dot{S}^{\rm rst}$ \cite{Fuchs2016} is always positive as detailed in Supplemental Material Sec.~ III \cite{Supplemental}.

The average rate of free energy, explicitly measured as $\lambda \langle \Delta f\rangle = -\dot{W} - T \dot{S}^{\rm rst}$, allows us to compare the minimal informative cost of the Maxwell's demon with the thermodynamics of its physical implementation. 
On the one hand, the informative cost associated with idealized SR, is composed of instantaneous events, with no additional energetic footprint other than $\Delta f$.
On the other hand, the full trajectories recorded in the experimental implementation of SR are generated by a series of optical potential quenches. Such quenches, where the stiffness increases from $\kappa_{\rm min}$ to $\kappa_{\rm max}$ following a given protocol have their own thermodynamic cost that must be evaluated.
This constitutes the actual energy injected in the system by the Maxwell's demon for the information machine to operate (right half of Fig.~\ref{fig:demon}), its contribution to the thermodynamic balance is shown with the red horizontal arrow in the bottom.

In order to measure the real energetic cost of the demon, the full trajectories, including the waiting times $\tau_{\rm wait}$ are analyzed.
They are Langevin trajectories experiencing a time-dependent potential. In our experiment, the choice of the individual protocol $\kappa(t)$ connecting $\kappa_{\rm min}$ and $\kappa_{\rm max}$ to implement resetting will induce a work that can be computed within the framework of stochastic energetics as \cite{Sekimoto1998, SekimotoBook, CilibertoPRX2017, Bechhoefer2020, Cabara2020, Gupta2022Thermo}:
\be
	w^{\rm ext}(t) = \frac{1}{2} \int_0^t \dot{\kappa}(t') x_{t'}^2 dt'.
	\label{Eq:ExtWork}
\ee
This expression provides an explicit measure of the \textit{external work} needed to reset the system. According to the non-equilibrium Second Law, this work is bounded from below by the average non-equilibrium free energy with $w^{\rm ext}(t) > \Delta f(t)$ \cite{Fuchs2016}.

The time-averaged work production rate over a long trajectory of duration $t_{\rm tot} \gg \lambda^{-1}$ (containing many resetting events) is simply $\dot{W}^{\rm ext} = w^{\rm ext}(t_{\rm tot}) / t_{\rm tot}$.
This external applied work obviously depends on the choice of $\kappa(t)$ connecting the same initial $\kappa_{\rm min}$ and final $\kappa_{\rm max}$ stiffnesses at the same rate $\lambda$.
Therefore, the non-equilibrium free-energy bound can be tested by seeking for protocols $\kappa(t)$ that minimize $\dot{W}^{\rm ext}$.
Doing so, we test the dependence on the choice of $\kappa(t)$ and measure the real thermodynamic cost of maintaining the SR system in a NESS.

\begin{figure}[htb!]
	\begin{center}
		\centering{
			\includegraphics[width=0.42\textwidth]{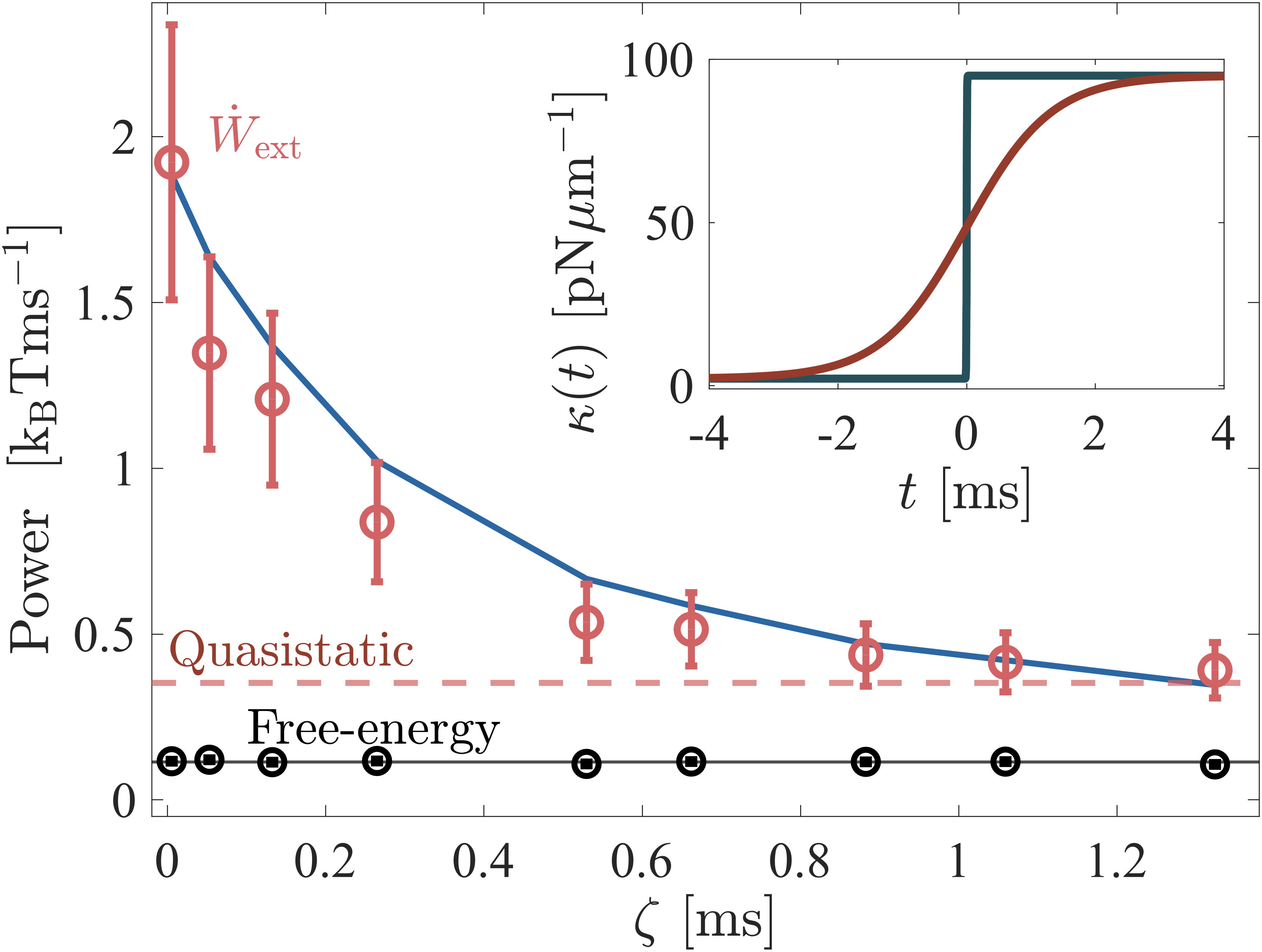}
			\label{fig:ExpBound}}
		\caption{ (inset) Time-evolutions of two different protocols $\kappa(t)$ for a single resetting event connecting the same $\kappa_{\rm min} = 2.18 ~\rm{pN/\mu m}$ and $\kappa_{\rm max} = 95 ~\rm{pN/\mu m}$. The parameter $\zeta$ governs the shape of the protocol, ranging from a very abrupt step-like protocol with $\zeta = 0.03 ~ \rm{ms}$ (grey-blue line), to a very slow close-to-quasi-static protocol for a large $\zeta \approx 1 ~ \rm{ms}$ (red line), compared to the relaxation time in the stiff resetting potential is $\tau_{\rm max} = \gamma / \kappa_{\rm max} = 0.26 ~ \rm{ms}$. The relaxation time of the sphere optically trapped in the external potential is $\tau_{\rm min} = \gamma/\kappa_{\rm min} = 11.5 ~ \rm{ms}$. Each class of protocol is then implemented for every individual resetting event forming a long SR process. Hence, in each case, the recorded trajectory experiences SR with the exact same parameters $\kappa_{\rm min}$, $\kappa_{\rm max}$ and $\lambda$, the only difference being the abruptness of every single potential quench.
		(main plot) Associated stochastic work rate $\dot{W}^{\rm ext}$ (red circles) experimentally measured with Eq.~(\ref{Eq:ExtWork}), applying the standard tools of stochastic thermodynamics to the full trajectory with the time-dependent protocol.	 As imposed by the Second Law, this external work rate $\dot{W}^{\rm ext}$ is always larger than the free energy rate $-\dot W - T \dot S^{\rm rst}$, measured on the idealized instantaneous resetting process (black circles and black solid line). As clearly seen, $\dot{W}^{\rm ext}$ decreases for large $\zeta$, when each resetting event is closer to the quasi-static limit but does not reach the free-energy bound. Instead, it asymptotically reaches a quasi-static limit (dashed red line) derived with the exact experimental parameters in the Supplementary File. The experimental data are in good agreement with numerical simulations (blue dashed line) performed in conditions similar to the experiments.
		}
	\end{center}
\end{figure}

 We design smooth protocols for the same resetting sequences at a given rate $\lambda$, but where each increase in stiffness follows $\kappa(t) = \frac{1}{2}(\kappa_{\rm max} - \kappa_{\rm min}) \textrm{tanh}( \frac{t}{\zeta}) +  \frac{1}{2}(\kappa_{\rm max} + \kappa_{\rm min})$.
This allows us to perform SR with protocols ranging from abrupt step-like changes for small $\zeta$ to slow drivings for large $\zeta \gg \tau_{\rm max} $, thus approaching the quasi-static limit. Two example protocols connecting the same $\kappa_{\rm min}$ and $\kappa_{\rm max}$ are displayed in the inset of Fig.~\ref{fig:ExpBound}.
Each protocol is applied in the optical trap and the corresponding long-time series of positions $x_t$ are recorded with their associated cost $\dot{W}^{\rm ext}$ measured.
The protocol $\kappa(t)$ determines the shape of each single resetting event, but it is independent of the sequence of resettings, entirely determined by the rate $\lambda$.
In other words, the same sequence of SR can be applied on the system with different local protocols $\kappa(t)$.

In parallel to $\dot{W}^{\rm ext}$, we measure the average rate of the resetting free energy $\lambda \langle\Delta f\rangle = -\dot{W} - T \dot{S}^{\rm rst}$ for each time series. As clearly seen on Fig.~\ref{fig:ExpBound} $\lambda \langle\Delta f\rangle$ is, as expected, independent on the choice of $\kappa(t)$ while the external work rate $\dot{W}^{\rm ext}$ strongly depends of $\zeta$.
Strikingly, we observe that the external work rate is always larger than this constant non-equilibrium free energy with $\dot{W}^{\rm ext} > \lambda \langle \Delta f \rangle = 0.12 \pm 7\times10^{-3} ~ \rm{[k_B T / ms]}$, confirming in the SR steady-state the non-equilibrium Second Law. We also calculate that the average free energy is larger than the work $\dot W = 0.032 \pm 3\times10^{-3} ~ \rm{[k_B T / ms]}$ the demon can extract. This last result is easy to understand: First and Second Law combined imply $-T\dot{S}^{\rm rst} \geq \dot{W}$ which, in turns, implies $\lambda \langle \Delta f \rangle \geq \Delta W$.
In a steady-state such as the one considered here with constant internal energy, the free energy $\lambda \langle\Delta f\rangle$ is proportional to the total entropy production rate.

The free-energy bound $\dot{W}^{\rm ext} > \lambda \langle \Delta f \rangle$ can be interpreted as the ultimate Landauer's limit for an idealized SR NESS \cite{Fuchs2016}.
In idealized SR indeed, we remind that each resetting event of the position $x_t$ is an instantaneous erasure of information. According to Landauer, this erasure must necessarily release a quantity of heat in the bath equal to the cost of energy needed to reset \cite{Berut2012, Jun2014, Lutz2015, Ciliberto2019, Bechhoefer2020}.
This minimal cost of energy is given by $\Delta f$. Averaged over the entire SR process, $\lambda \langle \Delta f \rangle$ thus corresponds to the Landauer's limit.

As seen Fig.~\ref{fig:ExpBound}, $\dot{W}^{\rm ext}$ decreases for slower protocols with larger $\zeta$.
This is a first central result of this work: the protocol used to implement resetting in an experimental set-up can be designed in order to reduce the cost of maintaining the SR system in a NESS. We however stress that, while this cost $\dot{W}^{\rm ext}$ decreases, it does not reach the free-energy bound $\lambda \langle \Delta f \rangle$. Within our experimental realization, this bound is therefore not tight \cite{Mori2022}. At large $\zeta$, the external work rate only reaches asymptotically a quasi-static limit which is carefully derived in the Supplemental Material Sec.~ IV \cite{Supplemental}.

Pushing down further $\dot{W}^{\rm ext}$ is only possible if the thermodynamic transformation corresponding to the erasure process becomes reversible. But in the context of SR, the duration between the repeated interruptions of the diffusion performed at random times is shorter than equilibration. Each resetting event is thus triggered on a system out of equilibrium. Whether it is possible or not to perform in a reversible way the series of interruptions that makes up SR is a key thermodynamical question. As we now show, this question is positively answered by handling more precisely the quantity of average information available at each resetting.\\

\noindent{\textbf{\large{Handling information allows to operate the machine reversibly}}}\\

We focus on a single resetting event along a continuous trajectory with time $t = 0$ taken just after one last resetting and waiting time sequence. At $t = 0$, the trajectory starts diffusing from  $x \approx 0$ in the (shallow) potential.
In our case of harmonic confinement, the state of the system at time $t$ is described by the Gaussian probability density $P(x | t, x=0)$ of the position of the diffusing sphere --see above-- with a variance that grows deterministically, starting with a very small value $\sigma^2_i = k_B T / \kappa_{\rm max}$.
If the system is let to diffuse for infinite time, the variance will eventually reach $\sigma^2_f = k_B T / \kappa_{\rm min}$, obeying equipartition of energy in the shallow potential : the system would have reached equilibrium.
But when a resetting event happens at a random finite time $\tau >0$, the variance still possesses an intermediate value $\sigma^2(\tau)$ which does not corresponds to its equilibrium value $\sigma^2_f$.

Therefore, if the demon triggers the next resetting event at time $\tau$ by merely increasing the stiffness starting from $\kappa_{\rm min}$, even quasistatically, the process would not be reversible.
To be reversible, the transformation used to reset must ensure that the instantaneous state of the system described by $P(x_t)$ is at equilibrium at every instant of time. 

\begin{figure}[htb!]
	\begin{center}
		\centering{
			\includegraphics[width=0.41\textwidth]{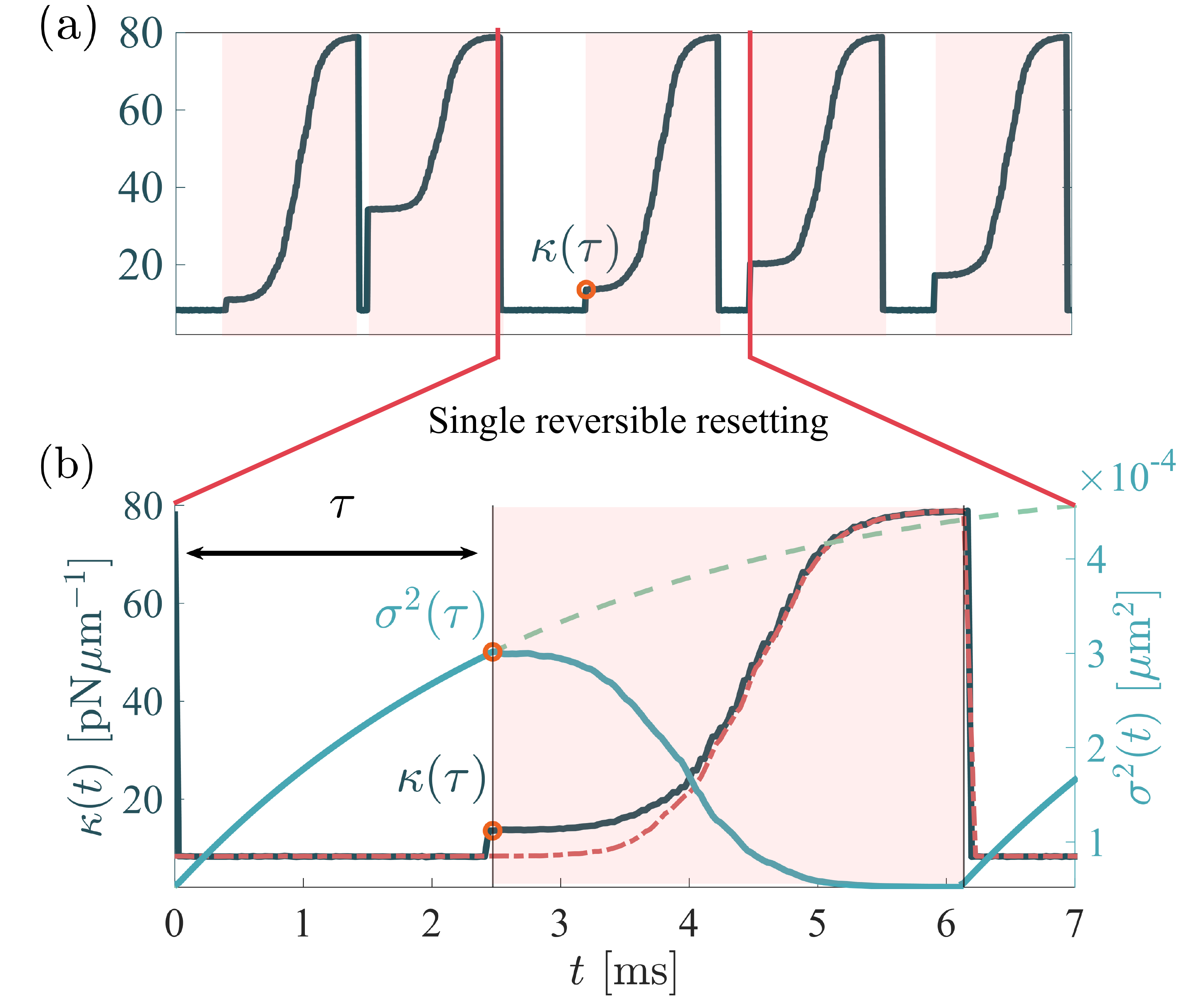}
			\label{fig:ProtocolTraj}}\\
		\centering{
			\includegraphics[width=0.39\textwidth]{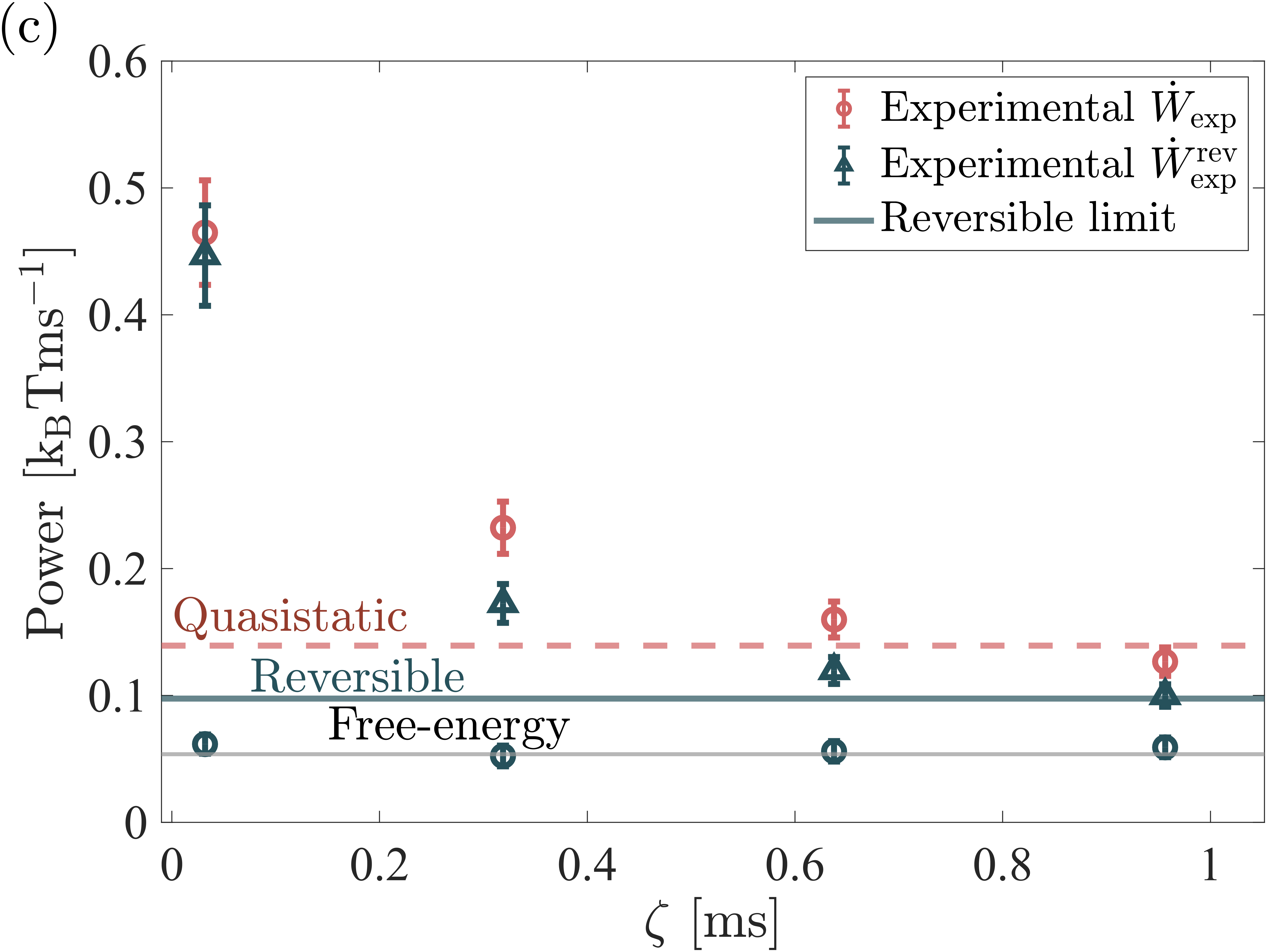}
			\label{fig:OptimalWork}}\\
		\caption{{(a) A succession of reversible quenches from $\kappa_{\rm min} = 8.18 ~\rm{pN/\mu m}$ to $\kappa_{\rm max} = 78.90~\rm{pN/\mu m}$. In contrast with the simple quasi-static drive used in Fig.~\ref{fig:ExpBound}, here the value of $\kappa$ is abruptly increased at the beginning of each resetting protocol. The amplitude of the discontinuity is a random variable, function of the time elapsed since last resetting.
		(b) Detailed explanation of a single reversible protocol. From time $t = 0$ to the random time $t = \tau$ the particle diffuses in the harmonic potential according to a Gaussian probability density $P(x|t,x=0)$ of increasing variance $\sigma^2(t)$ (blue line). At time $\tau$ the variance $\sigma^2(\tau)$ has not yet reached its equilibrium value $k_BT/\kappa_{\rm min}$ (its slow relaxation is continued as blue dashed line). In order to reversibly bring back the system into a potential of stiffness $\kappa_{\rm max}$, the first step consists in instantaneously adapting the potential to the variance $\sigma^2(\tau)$. This is done by increasing the stiffness to $\kappa(\tau) = k_B T / \sigma^2(\tau)$. The stiffness is then slowly (quasistatically) increased until $\kappa_{\rm max}$, keeping the system close to equilibrium for all times $t>\tau$.
		(c) Experimentally measured external power as a function of the protocol smoothness parameter $\zeta$ while keeping a constant $\lambda^{-1} = 5 ~\rm{ms}$ with $\tau_{\rm min} = 3.1~\rm{ms}$ and $\tau_{\rm max} = 0.23~\rm{ms}$.
		 The external power is shown for protocols approaching quasistaticity for large $\zeta$ ($\dot W^{\rm ext}$, red circles) but also for protocols approaching reversibility for large $\zeta$, as described above (denoted $\dot W^{\rm ext}_{\rm rev}$, blue-grey triangles).
		 Both analytical values of the quasi-static limit (dashed red line) and the reversible limit (blue-grey line) are shown as well as the measured average rate of free-energy difference (black circles and line).
		The measured work with instantaneous equilibration reaches the reversible value for large $\zeta$ while the simple slow protocol agrees with the quasi-static limit.
		}}
	\end{center}
\end{figure}

In order to build a reversible transformation starting from a non-equilibrium state, one can start by adjusting the external constrains to the state in which the system lies so that it becomes an equilibrium state with the new
constrains. On our platform, we adjust the stiffness of the potential (our external constrain) as to enforce equipartition at the time $\tau$ of resetting with $\kappa(\tau) = k_BT / \sigma^2(\tau) > \kappa_{\rm min}$ so that the state characterized $\sigma^2(\tau)$ now stands for an equilibrium state in the new potential.
The transformation used to reset now starts at equilibrium and the stiffness can then increase quasistatically towards $\kappa_{\rm max}$. Doing so, the system remains at equilibrium at every time along the protocol.
This protocol is schematized on Fig.~\ref{fig:OptimalWork}~(a) and (b) where the instantaneous adjustment of $\kappa$ at $\tau = 2.5$ s is clearly visible.
For all times $t > \tau$ the system is at equilibrium and the variance obeys equipartition $\sigma^2(t>\tau) = k_B T / \kappa(t>\tau)$ until the next resetting sequence.
Being reversible, this transformation does not dissipate any entropy.

Considering now the whole sequence starting from $t = \tau^-$ bringing the system from a non-equilibrium state to an equilibrium state, the reversible work is composed of two contributions \cite{Esposito2011, Parrondo2015}.
First, of the equilibrium free-energy difference $\Delta F_{\rm eq} = \frac{k_B T}{2} \ln\left( \frac{\kappa_{\rm max}}{\kappa_{\rm min}} \right)$ between both equilibrium states characterized by $\sigma^2_i$ and $\sigma^2_f$.
Second, of the excess of free energy between the non-equilibrium and the equilibrium states. As detailed in the Suppl. Mat. file, this excess is given by the relative entropy $k_B I(\tau)$ between the state described by $P(x|\tau,x=0)$ and the equilibrium state whose probability density is approached for long times as $P(x|\tau \rightarrow \infty, x=0) = P_{\rm eq}(x)$, and multiplied by $T$. As such therefore, this quantity quantitatively measures the distance of the system from equilibrium.

This quantity truly embodies the average information available at the instant $t=\tau$ of resetting and is evaluated as a Kullback-Leibler divergence between $P(x|\tau,x=0)$ and $P_{\rm eq}(x)$ \cite{Esposito2011}.
In the case of Gaussian densities, this divergence simplifies into a relation between variances (as detailed in Supplemental Material  Sec.~V \cite{Supplemental})
\be
	k_B T I(\tau) = \frac{-k_B T}{2} \left( \ln\left[ \theta \right] + 1 - \theta \right) \geq 0
\ee
where $\theta = \frac{\sigma^2(\tau)}{\sigma^2_f} = \frac{\sigma^2(\tau)\kappa_{\rm min}}{k_B T}$.
Importantly $I(\tau)$ is still a random variable of the random resetting time $\tau$. Its average contribution is evaluated by integrating it over the resetting times probability distribution $ \langle I(\tau) \rangle_\tau = \int I(\tau) P(\tau) d\tau$.
A positive $\langle I(\tau) \rangle_\tau$ allows to engage a work smaller than equilibrium free-energy difference.
In our experimental realization, each resetting event ends by the abrupt decrease of the stiffness back to $\kappa_{\rm min}$, initiating the next diffusing sequence.
This re-expansion of the trapping volume also produces a work $ \frac{k_B T }{2} \left[ \frac{\kappa_{\rm min}}{\kappa_{\rm max}} - 1 \right]$ which must be added to the contribution of the reversible compression in the total average reversible power
\be
	\dot W^{\rm ext}_{\rm rev} = \lambda \bigg( \underbrace{\Delta F_{\rm eq} - k_B T \langle I(\tau) \rangle_\tau}_{\rm reversible ~ compression} + \underbrace{\frac{k_B T }{2} \left[ \frac{\kappa_{\rm min}}{\kappa_{\rm max}} - 1 \right]}_{\rm re-expansion} \bigg)
	\label{eq:RevWork}
\ee
injected in the system by our Maxwell's demon.\\

\noindent{\textbf{\large{Szilard-like protocols perform SR at minimal reversible costs}}}\\

Interestingly, this reversible transformation is analogous to operating a Szilard engine made of a box containing a single ideal gas particle \cite{szilard1929}.
In the Szilard thought-experiment, the probability density of the position of the particle is uniform across the volume of the box at thermal equilibrium.
The first informational step consists in measuring on which side of the box the particle is. This input of information evidently breaks the uniformity of the probability density and thereby brings the system into a non-equilibrium state. This first step corresponds in our experiments to the resetting event when information on the system (namely, its distance from equilibrium) can be extracted through the knowledge of the motional variance $\sigma^2(\tau)$ at time $\tau$ of the trigger.
The second step in the Szilard experiment consists in instantaneously inserting a partition isolating both halves of the box.
This corresponds to our instantaneous increase of stiffness from $\kappa_{\rm min}$ to $\kappa(\tau)$ matching the external constrains to bring the state of the system at equilibrium.
Finally the partition is quasistatically moved until the full volume is retrieved, as in our quasistatic increase of $\kappa(t>\tau)$.
This protocol is ensuring minimal cost since it prevents dissipation \cite{Esposito2011, Parrondo2015, Ciliberto2019}.

We implemented experimentally this reversible protocol using the fact that $\sigma^2(\tau)$ is uniquely related to the time elapsed since the last resetting. This implies that the random values of $\kappa(\tau)$ are known for a given time-series of resetting times $\tau$.
We measure the Brownian trajectories under a Poissonian  sequence of reversible quenches and compare on Fig.~\ref{fig:OptimalWork}~(c) the injected power $\dot W^{\rm ext}_{\rm rev}$ to the case of quasi-static drive.
Remarkably, the reversible protocols bring the demon's cost below the quasi-static limit, reaching the minimal cost of reversibility.
This constitutes the second central result of our work and shows the strong experimental relevance of the informational framework: a correct Szilard-like control allows to perform resetting at the minimal reversible cost.

We however note again that the Landauer bound $\Delta f$ \cite{Fuchs2016} is not reached, even in the reversible limit.
This discrepancy probably stems from two main reasons.
First, the reversible work $ W_{\rm rev}$ is derived within experimental parameters, notably, as a transition between two finite stiffnesses. This contrasts with $\Delta f$ that assumes instantaneous resetting to the exact, non-fluctuating position $x=0$.
Second, as underlined above, even if the relative entropy $k_{\rm B}I(\tau)$ is a random variable, it remains an average quantity with respect to the stochastic variable $x_t$ since it only depends on the variance $\sigma^2(\tau)$.
There necessarily exist rare fluctuations, where the microsphere diffusion during this finite time strongly differs from its expected behavior characterized by $\sigma^2(\tau)$.
We postulate that taking into account the microscopic stochastic trajectory $x_t$ would reduce further the injected power, bringing it closer to $\Delta f$.
We propose that a microscopic instantaneous equilibration can be implemented by using --instead of the Gaussian probability density-- the variance of the empirical density $p(x, \tau) = \frac{1}{\tau} \int_0^{\tau} \delta(x - x_s) ds$ to chose the stiffness discontinuity at time $\tau$.
The quantity $p(x, \tau)$ is known to obey large deviation principle \cite{touchette2009large} and would allow the reversible limit to include the fluctuations of the microscopic trajectories. 
From an experimental perspective however, such protocol relies on feedback loops using the instantaneous $x_\tau$ to build the protocol $\kappa(\tau)$. This feedback strategy however goes beyond the scope of the present work\\

\noindent{\textbf{\large{Stochastic resetting, information, and ergodicity breaking}}\\

Finally, we experimentally verify the non-ergodic nature of SR. A breaking of ergodicity is expected by the mere presence of instantaneous jumps towards $x = 0$ that distinguish the center of the potential with respect to other spatial regions within the potential extension. 
Ergodicity breaking can be understood directly from a single trajectory perspective. Indeed, a long but finite SR trajectory contains large but rare jumps, that necessarily make it different from another trajectory drawn from the same process.
This is observed in our experiments.
in Fig.~\ref{fig:FirstPrinciple}~(a), where $x = 0$ is the converging point of the probability current, breaking detailed-balance as discussed above. 
For that reason, ergodicity and detailed-balance are broken together by the same mechanism: resetting events at the level of single trajectories \cite{Stojkoski2022, Barkai2023}.

Since the distinction of the origine with respect to all other position is the necessary condition to process information, the Landauer's bound must therefore be closely connected to the non-ergodic nature of the stochastic trajectories.
In this section we will use a strong ergodic criterium to demonstrate the non-ergodic nature of the SR process recorded on our platform while verifying the ergodicity of a normal Brownian diffusion without resetting.

For stationary processes, ergodicity is defined as the equality of time average and ensemble average in the limit of infinite time and infinitely large ensemble.
This definition becomes operational by studying Mean-Squared-Displacement (MSD) of the Brownian trajectory.
The discrepancy between its time-average and ensemble-average is a measure of the non-ergodic nature of the process and can be evaluated with finite time and ensemble \cite{METZLER2000, Barkai2009, Metzler2015}.

\begin{figure}[htb!]
	\begin{center}
		\centering{
			\includegraphics[width=0.42\textwidth]{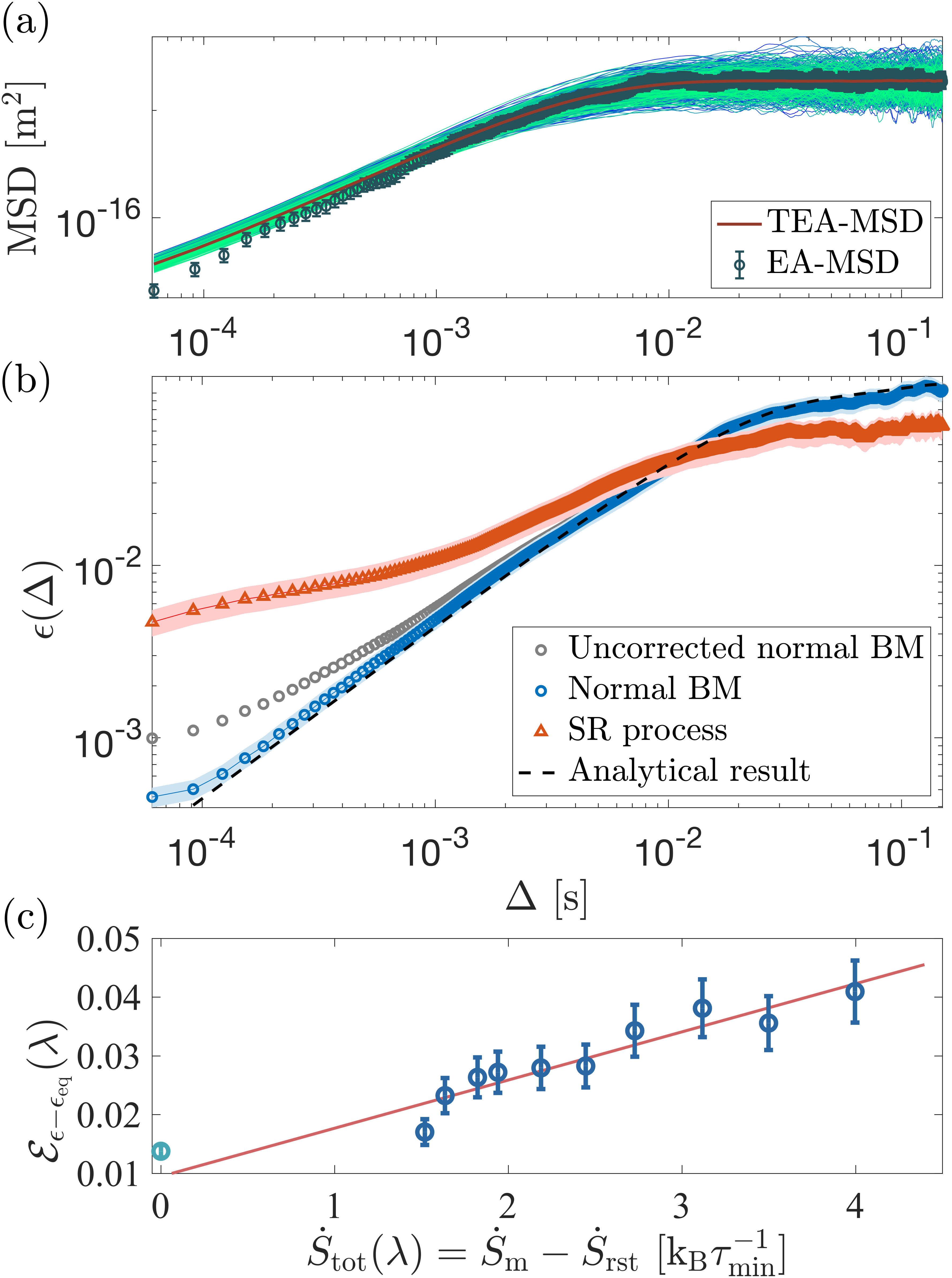}
			\label{fig:ErgodicParam}}\\
		\caption{{
		(a) Time-averaged-MSDs (TA-MSD) of an ensemble of $800$ individual sub-trajectories (light lines from blue to green), together with the instantaneous ensemble-averaged MSD (EA-MSD, grey-blue circles). 
		Averaging again all the TA-MSDs over the ensemble leads to the time-ensemble-averaged TEA-MSD (red solid line). Importantly, the dispersion of TA-MSDs is due to the intrinsic variability of individual sub-trajectories (as explained in the main text), revealing the non-ergodic nature of the process.
		(b) Ergodic criteria $\epsilon(\Delta)$ both for an equilibrium trajectory in the harmonic potential without resetting (normal BM, blue circles) and for an SR process in the same potential (red triangles). The analytical expression for $\epsilon(\Delta)$ (black dashed line) coincides with the experimental result for normal BM while the measured values for an SR process are significantly different.
		Long time drift in the experimental setup had to be corrected with a method detailed in the Supplemental Material  Sec.~ VI \cite{Supplemental}.
		(c) Root-mean-square error $\mathcal{E}(\lambda)$ between the measured ergodic criteria  $\epsilon(\Delta)$ and its equilibrium counterpart as a quantitative measure of ergodicity breaking (deep blue circles) as a function of the total entropy production in units of $k_B T / \tau_{\rm min}$ for the same resetting rates $\lambda$ as the thermodynamics of Fig.~\ref{fig:FirstPrinciple}~(b) and (c). We also represent the same error measured in the same experiment in the absence of resetting (turquoise circle in the bottom left) and a linear fit of all measured deviation points $\mathcal{E}_{\epsilon - \epsilon_{\rm eq}}(\lambda)$ (red line).
		}}
	\end{center}
\end{figure}

In our case where the SR process reaches a steady state, such an ensemble can be built by cutting the long recorded trajectory into shorter sub-trajectories.
Here we study SR process with $\kappa_{\rm min} = 2.9 \pm 0.15$ and $\kappa_{\rm max} = 83 \pm 2.1 ~ \si{\pico\newton / \micro\meter}$ at a rate $\lambda^{-1} \approx 6.1 ~ \si{\milli\second}$.
We also verify ergodicity of a normal Brownian motion in exactly the same $V(x)$ external harmonic potential.
After removing the waiting times, we obtain a $239$ seconds-long SR trajectory that is cut into an ensemble of $800$ individual trajectories of a total time $\mathcal{T} = 0.3$ seconds so that each individual sub-trajectory contains many resetting events.
On the one hand, a time-averaged MSD (denoted as TA-MSD) can be computed for each individual $i^{\rm th}$ trajectory $x_t^i$ from the ensemble as
\be
	\textrm{TA-MSD}_i \equiv  \overline{\delta{x^i}^2(\Delta)} \equiv \frac{1}{\mathcal{T}-\Delta} \int_{0}^{\mathcal{T}-\Delta} \left( x^i_{t+\Delta} - x^i_t\right)^2 dt
\ee
On the other hand, an instantaneous ensemble-averaged MSD $\langle \overline{\delta{x^i}^2(\Delta)}\rangle$ (denoted as EA-MSD)  can be computed on the whole ensemble, by summing over trajectories instead of integrating over time.
Finally, both averaging can be combined and the set of individual $\textrm{TA-MSD}_i$ can be averaged on the ensemble to build the time-ensemble averaged MSD (denoted as TEA-MSD).

As discussed in details in previous works \cite{metzler2014anomalous, Li2019, Ergodicity2021}, a first necessary condition for ergodicity is the convergence of TEA-MSD to the EA-MSD in the limit of large $\mathcal{T}/\Delta$.
This is verified on Fig.~\ref{fig:ErgodicParam}~(a) where we see that the TEA-MSD agrees with the EA-MSD apart from small deviation for short $\Delta$.
However, this is only a necessary condition for ergodicity since it can hide a strong dispersion of individual $\textrm{TA-MSD}_i$ (which is also visible on Fig.~\ref{fig:ErgodicParam}~(a) with individual curves from blue to green), implying that individual sub-trajectories strongly differ from each other, violating the ergodic condition.
Therefore, a strong ergodicity criterion (both necessary and sufficient condition) is the vanishing of the dispersion of $\textrm{TA-MSD}_i$ in the large $\mathcal{T}/\Delta$ limit.

This dispersion of individual TA-MSD is well captured by the estimator $\epsilon(\Delta) = \bigg\langle \overline{\delta{x^i}^2(\Delta)}^2 \bigg\rangle / \langle \overline{\delta{x^i}^2(\Delta)}\rangle ^2 - 1$, which can be analytically computed for standard Brownian motion in a harmonic potential \cite{Metzler2014, Ergodicity2021}.
The estimator evaluated in Fig.~\ref{fig:ErgodicParam}~(b) for the SR process (red triangles) clearly reveals  the non-ergodic nature of SR over all probed time-scales:  $\epsilon(\Delta)$ does not go to zero for short $\Delta$,  in agreement with previous numerical works \cite{Stojkoski2022}.
In contrast, the evolution of $\epsilon(\Delta)$ for normal Brownian motion in the same external potential (blue circles) in Fig.~\ref{fig:ErgodicParam}~(b) ensures the ergodicity of the experimental setup when no resetting is performed. These two results show that resetting is the sole mechanism breaking ergodicity.

We finally ask if the qualitative link between ergodicity breaking and the ability to process information can be made quantitative \cite{Stojkoski2021, Wang2021, Stojkoski2022}.
As a tentative answer in that direction, we measure the mean deviation between $\epsilon(\Delta)$ (triangles for NESS and circles for equilibrium, on Fig.~\ref{fig:ErgodicParam}~(b)) from the expected trend for equilibrium $\epsilon_{\rm eq}(\Delta)$ (black dashed line on Fig.~\ref{fig:ErgodicParam}~(b)).
The root-mean-square error $\mathcal{E}_{\epsilon - \epsilon_{\rm eq}}(\lambda) = \sqrt{ \frac{1}{\mathcal{T}} \int_0^\mathcal{T} |\epsilon(\Delta) - \epsilon_{\rm eq}(\Delta)|^2 d\Delta} $ gives a simple quantification of ergodicity breaking.
For equilibrium, this deviation should be zero, and only accounts for experimental errors (single turquoise circle on the left of Fig.~\ref{fig:ErgodicParam}~(c)).
When resetting is applied, the deviation (blue circles) increases with the resetting rate $\lambda$, as the system is pushed further away from equilibrium.
As seen on Fig.~\ref{fig:ErgodicParam}~(c), $\mathcal{E}_{\epsilon - \epsilon_{\rm eq}}(\lambda)$ scales linearly with total entropy production $\dot{S}_{\rm tot}(\lambda)$, meaning that both share the same dependency over the resetting rate $\lambda$.
We recall here that in our case, the total entropy production is also proportional to the average rate of free-energy difference.
This correlation draws a direct and quantitative link between ergodicity-breaking and the performance of the information engine and constitutes the third and last main result of this work.}\\

\noindent{\textbf{\large{Conclusion}}}\\

In summary, we studied SR induced on a microsphere diffusing in a confining optical potential by recording real-time stochastic trajectories.
A simple manipulation of these trajectories gave us access to the ideal, instantaneous limit of SR which reveals its Maxwell's demon nature, extracting work from a single temperature heat bath.
We then confronted the idealized thermodynamics of the demon to the actual cost involved in making SR real, showing that the cost of operating the demon always exceeds the work it extracts.
The nature of the continuous trajectories of real SR, namely a succession of transitions from non-equilibrium to equilibrium states, allowed us to extract the average information available at each resetting event.
We recovered the non-equilibrium Second Law where the SR free-energy bounds, from below, the external work needed to reset the system. This informational approach allows to interpret this bound as the Landauer's limit of idealized SR. In addition, the capacity to handle the available information led us to reach the reversible limit of operating SR. In this limit of reversibility, we showed how SR can be described as a series of Szilard-like protocols \cite{Esposito2011, szilard1929}.

This assessment of quantitative bounds put on this stochastic process as an information machine has obviously consequences in understanding natural phenomena, given the ubiquity of SR \cite{Cherstvy2008, Boyer2014, Roldan2016, Bressloff2020}.
Finally, our experiment gives a clear illustration of the fundamental connection between information processing and the breaking of ergodicity that manifests itself in each irreversible resetting event.
We anticipate that our results will provide new directions for the application of information thermodynamics in non-equilibrium experiments but will also offer new analysis tools in terms of ergodicity breaking \cite{toyabe_experimental_2010, Roldan2014, Parrondo2015}.
In addition, by demonstrating fundamental bounds on the operation of search processes such as stochastic resetting, our work directly impacts their frugal implementations in devices and their biological functioning \cite{Roldan2016, Bressloff2020}.
Such concerns on the thermodynamic costs of SR are currently fueling a intense theoretical research effort \cite{olsen2023thermodynamic, olsen2024thermodynamic}. The drive to getting closer to natural systems opens numerous challenges at the crossroad between resetting and active matter \cite{Kumar2020, Santra2020, Abdoli2021, Altshuler2023}.

\section*{Acknowledgments}

We sincerely thank Yael Roichman and Shlomi Reuveni for inspiring discussions. This work is part of the Interdisciplinary Thematic Institute QMat of the University of Strasbourg, CNRS and Inserm. It was supported by the following programs: IdEx Unistra (ANR-10-IDEX- 0002), SFRI STRATUS project (ANR-20-SFRI-0012) and USIAS (ANR-10-IDEX- 0002-02) under the framework of the French Investments for the Future Program. R.G. acknowledges support from the Mark Ratner Institute for Single Molecule Chemistry at Tel Aviv University.

\section{Experimental and numerical methods}
\label{App:Methods}

We detail here the realization of SR within an optical trap, the measurement technique and the method of calibration used. We also explicitly show how the trajectories are manipulated when going from their full, continuous form, to their ideal, decimated version. All the informational thermodynamic analysis of the main manuscript is based on this distinction. Finally, we provide a description of the numerical simulations.

\subsection{Experimental realization of SR in an optical trap}

\begin{figure}[h!]
	\centerline{\includegraphics[width=0.9\linewidth]{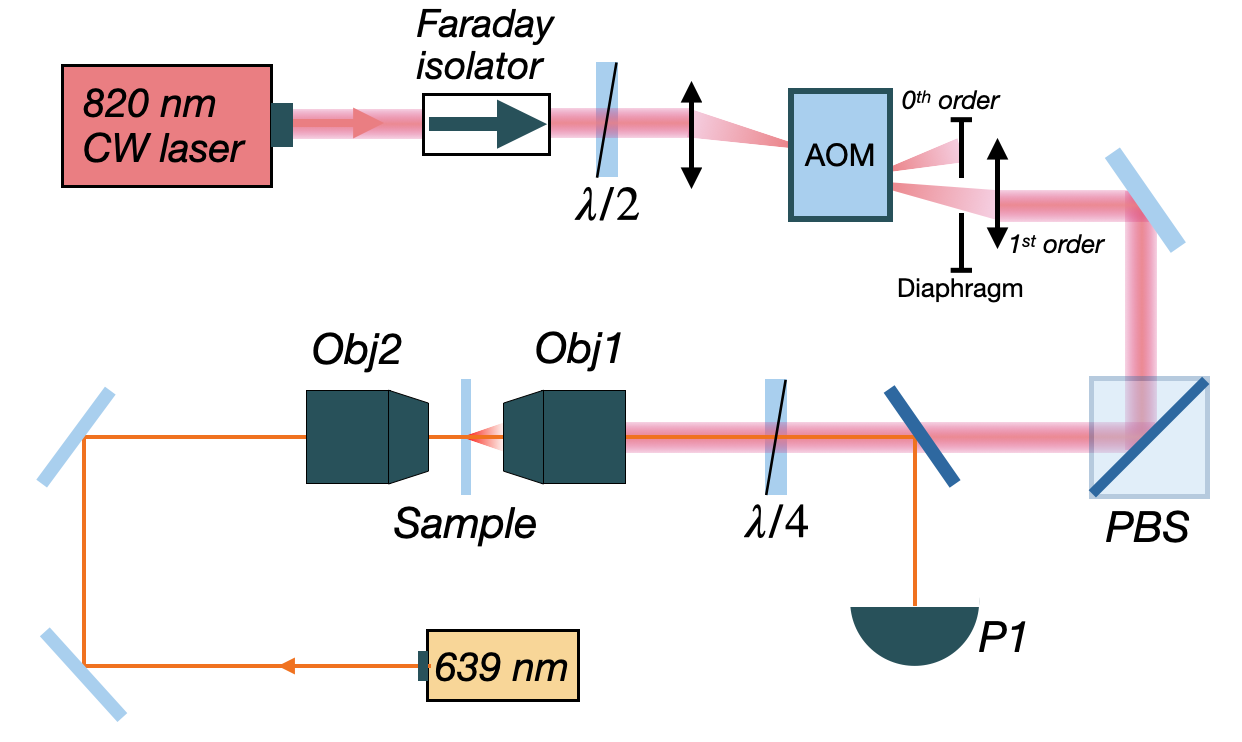}}
	\caption{Simplified view of the optical trapping setup. The sphere is suspended in water inside the \textit{Sample} cell inserted between the two objectives Obj1 and Obj2. The $820~\si{\nano\meter}$ trapping beam is drawn in pink. The intensity of this beam controlled by the acousto-optic modulator (AOM). The instantaneous position of the trapped bead is probed using the auxiliary $639~\si{\nano\meter}$ laser beam, drawn in orange, whose scattered signal is sent to a high-frequency photodiode.
		}
	\label{fig:schema}
\end{figure}

Our experimental setup consists in optically trapping, in a harmonic potential, a single dielectric bead ($3~\si{\micro \meter}$ polystyrene sphere) in a fluidic cell filled with dionized water at room temperature $T = 296 ~ \si{\kelvin}$.
The harmonic potential is induced by focusing inside the cell a linearly polarized Gaussian beam ($800~\si{\nano\meter}$, CW $5$ W Ti:Sa laser, Spectra Physics 3900S) through a high numerical aperture objective (Nikon Plan Apo VC, $60\times$, NA$=1.20$ water immersion, Obj1 on Fig.~\ref{fig:schema}).
The intensity of this trapping beam is controlled by an acousto-optic modulator (Gooch and Housego 3200s, AOM on Fig.~\ref{fig:schema}) using a digital-to-analogue card (NI PXIe 6361) and a \textsc{python} code.\\

The instantaneous position $x(t)$ of the sphere along the optical axis is measured by recording the light scattered off the sphere of a low-power $639~\si{\nano\meter}$ laser (CW $30$ mW laser diode, Thorlabs HL6323MG), sent on the bead via a second objective (Nikon Plan Fluor Extra Large Working Distance, $60\times$, NA$=0.7$, Obj2 on the figure). The scattered light is collected by Obj1 and recorded by a photodiode ($100$ MHz, Thorlabs Det10A).
The recorded signal (in V/s) is amplified using a low noise amplifier (SR560, Stanford Research) and then acquired by an analog-to-digital card (NI PCI-6251). The signal is filtered through a $0.3$ Hz high-pass filter at 6 dB/oct to remove the DC component and through a $100$ kHz low-pas filter at 6 dB/oct to prevent from aliasing. The scattered intensity varies linearly with the position of the trapped bead $x(t)$ for small enough displacements and we make sure to work in the linear response regime of the photodiode so that the recorded signal is linear with the intensity, resulting in a voltage trace well linear with $x(t)$.\\

\begin{figure}[htb!]
	\begin{center}
		\centering{
			\includegraphics[width=0.4\textwidth]{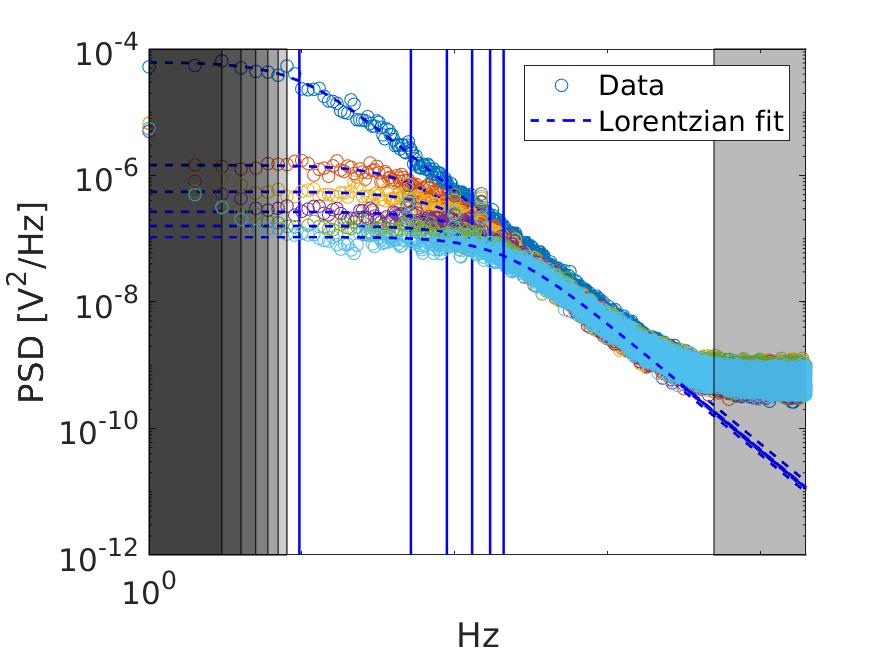}
			\label{fig:PsdCalib}}\\
		\centering{
			\includegraphics[width=0.35\textwidth]{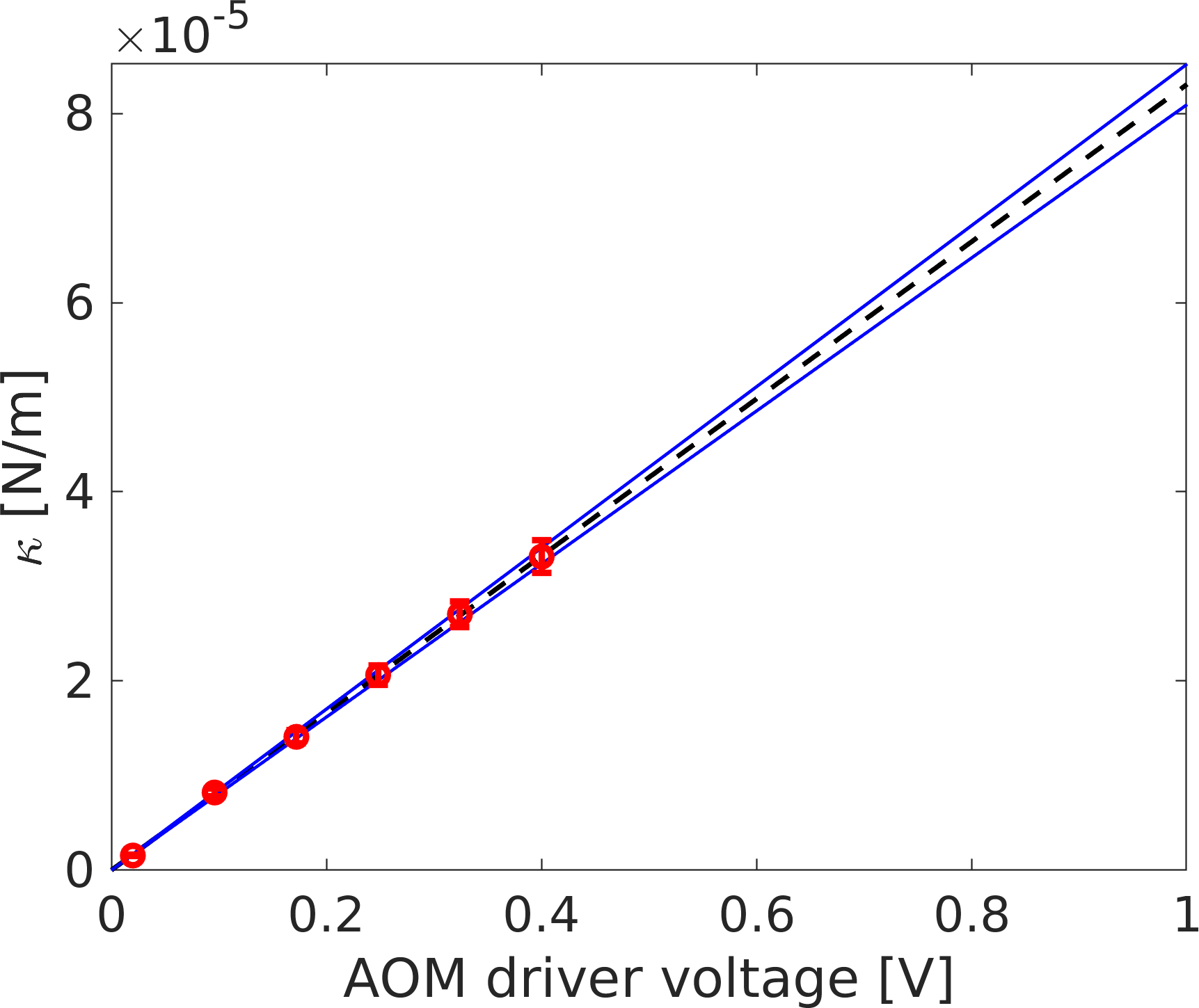}
			\label{fig:KappaCalib}}
		\caption{{ (Upper panel) Power spectral density of the microsphere motion for various trapping laser insensity. The circles are experimentally measured PSD, the dashed lines are Lorentzian fits, from which the cutoff frequency is extracted (blue vertical lines). The shaded patches represent the limits of the frequencies used for the Lorentzian fit, we gradually restrain the bandwidth as we go to high stiffnesses where the measured signal decreases.
		(Lower panel) Stiffnesses extracted from the fit of PSDs at different driving voltage of AOM (hence different trapping laser intensity) with a linear fit giving the relation between driving voltage and in situ stiffness. This allow us to know the maximal usable stiffness given this working power ($400 ~ \si{\milli\watt}$ in the input of AOM) of Ti:Sap laser at $\kappa_{\rm max} = 83.1 \pm 2.1 ~ \si{\pico\newton/\micro\meter}$.
		}}
	\end{center}
\end{figure}

To build an SR process with the expected characteristics, we rely on the knowledge of the stiffness of the optical trap in which the particle diffuses.
We therefore calibrate the relation between the voltage send to the AOM driver and the stiffness $\kappa$ of the optical potential.
On Fig.~\ref{fig:PsdCalib}~(a) we show the power spectral density (PSD) of the recorded trajectories for 6 values of voltages, spanning the beginning of the dynamical bandwidth of the AOM.
An issue here is that, as the stiffness increases, the motional variance of the microsphere decreases, making it harder to probe.
For stiffnesses larger than circa. $40 ~ \si{\pico\newton/\micro\meter}$ we can hardly obtain a good PSD.
We can however assume that the linear increase of $\kappa$ with driving voltage is unaffected by this probing issue.
We therefore probe the beginning of this linearity, (driving voltage from $0.01~\si{\volt}$ to $0.4~\si{\volt}$) and extrapolate the linear relation on the whole dynamical bandwidth of the AOM, as shown Fig.~\ref{fig:KappaCalib}~(b).
This gives a maximal stiffness $\kappa_{\rm max} = 83.1 \pm 2.1 ~ \si{\pico\newton/\micro\meter}$ for the resetting event.

From the same PSD fit, we can extract a calibration factor $\beta$ by the ration between the amplitude of the measured PSD in $\si{\square\volt/\hertz}$ and the expected value in $\si{\square\meter/\hertz}$ depending on the diffusion coefficient of the microsphere in water $D = k_BT/\gamma \approx 0.16 ~ \si{\micro\meter/\second}$.
This calibration factor allows to obtain trajectories in meter out of the recorded time-series of voltages.\\

\begin{figure}[h!]
		\centering{
			\includegraphics[width=0.4\textwidth]{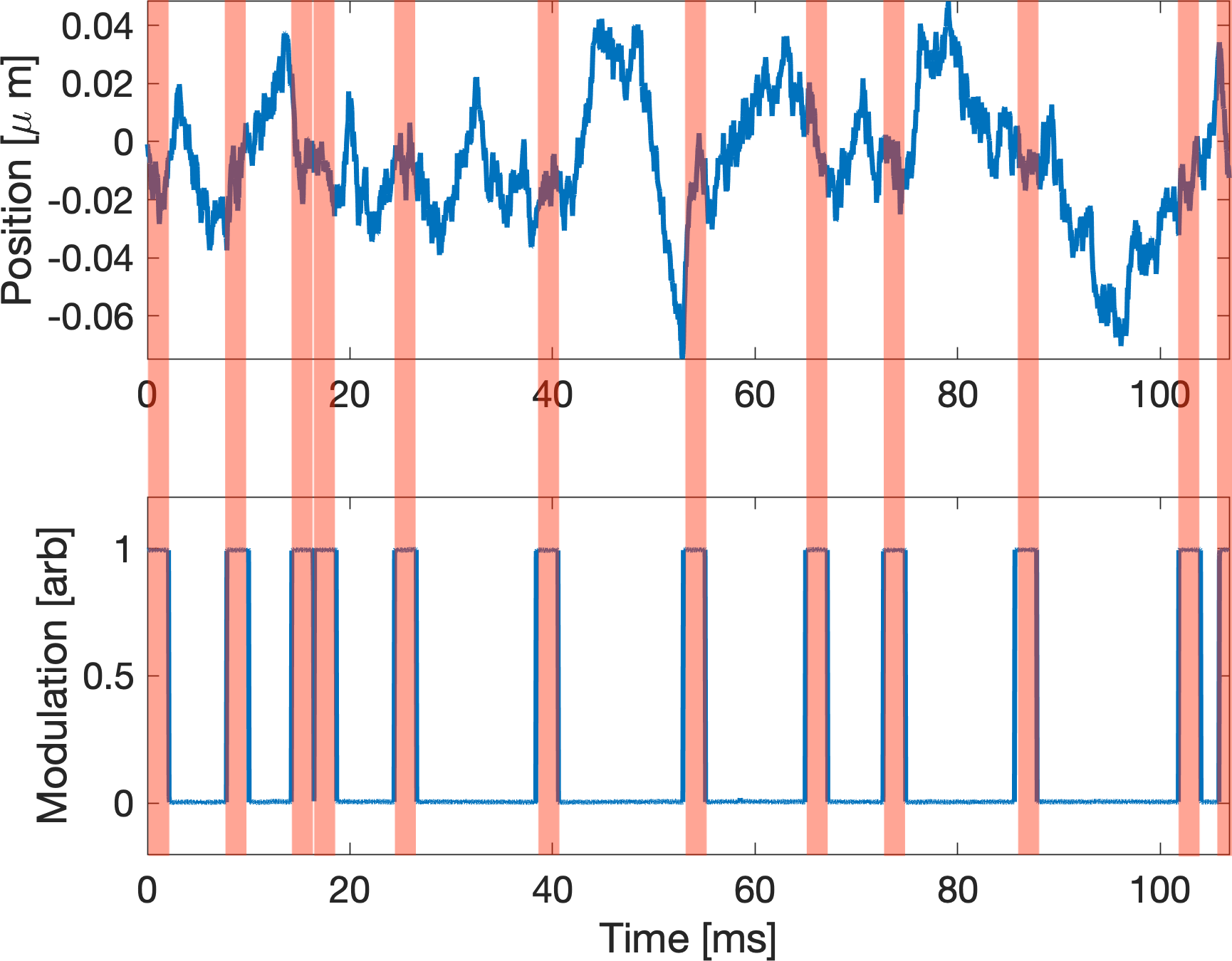} 
			\label{fig:ShemaMethods}}\\
		\centering{
			\includegraphics[width=0.35\textwidth]{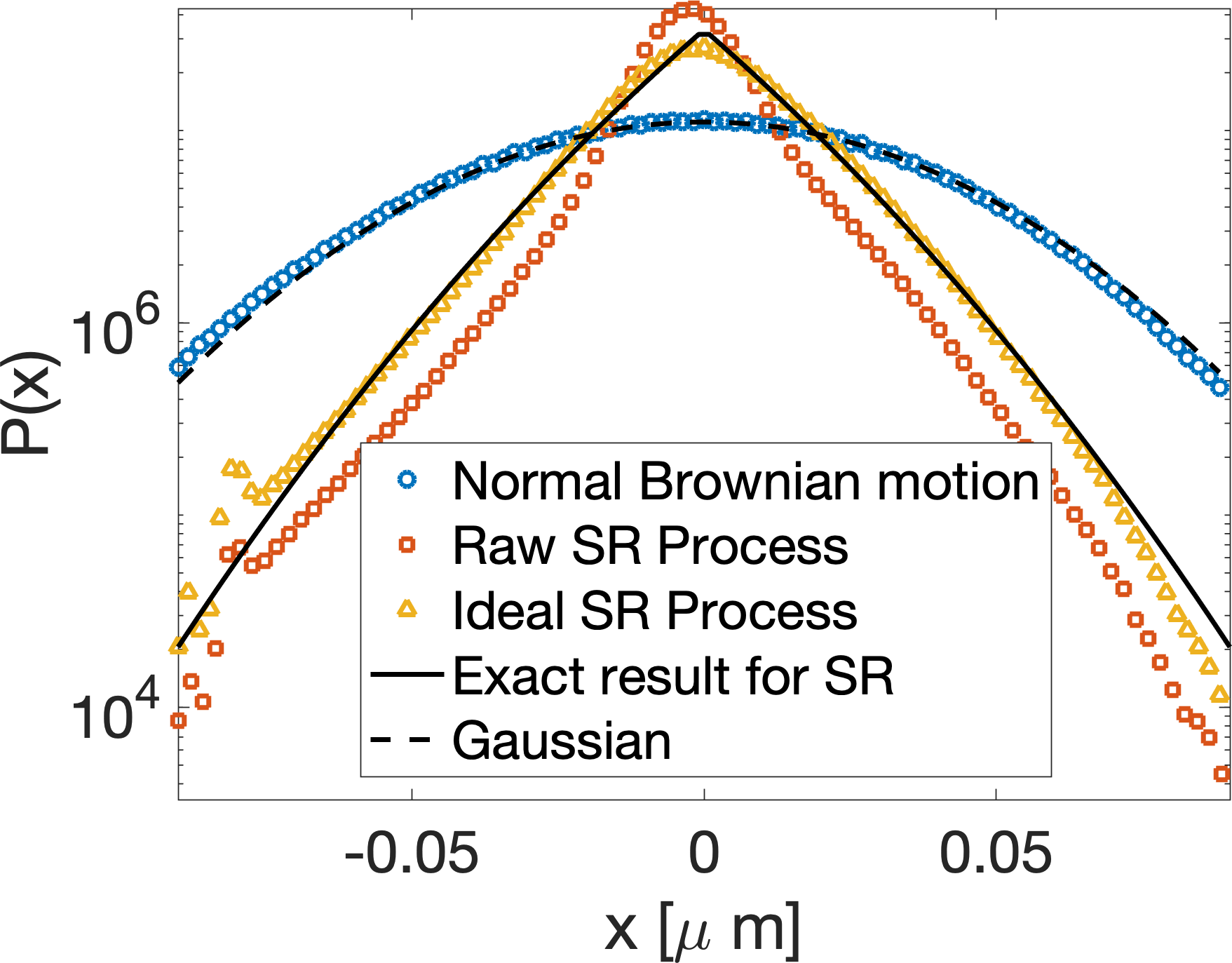}
			\label{fig:DistWithWait}}			
	\caption{(Upper panel) Small sub-part of an experimental realization of stochastic resetting. On this figure we show the full time series of both the modulation potential and the recorded time-series. We highlight with red stripes the times when the potential is kept stiff, which corresponds to the resetting events. By removing the points during these times, one obtains an idealized instantaneous resetting process. (Lower panel) Probability distribution of the raw trajectory measured in the optical trap (red squares) together with the distribution of the decimated idealized trajectory (yellow triangles) in agreement wit the analytical distribution as described in the main text (black line). We also show the equilibrium distribution (blue circles) together with the analytical Gaussian profile (black dashed line).
		}
	\label{fig:schemaSR}
\end{figure}

On Fig.~\ref{fig:schemaSR}, we show a full time-series of recorded points (for a few milliseconds). 
The synchronized recording of the trajectory $x(t)$ (Fig.~\ref{fig:schemaSR} top panel) and the time-series of stiffness $\kappa(t)$ (Fig.~\ref{fig:schemaSR} middle panel, normalized) allows to build the \textit{ideal} instantaneous resetting.
By removing points during the waiting times where $\kappa(t) = \kappa_{\rm max}$ (marked as vertical red stripes) we obtain instantaneous resetting events.
The confrontation of the full and decimated trajectories is the backbone of our analysis in the main text.
On Fig.~\ref{fig:schemaSR} (lower panel) we show the position's probability distribution for the full trajectories, for the decimated trajectory (also shown in the main text), together with the equilibrium reset-free distribution and both analytical results.
As seen, the presence of the waiting time in the trajectories affects the center of the distribution with, as expected, an accumulation of points near $x=0$. The tails are not affected.
Even in the decimated case (yellow triangles) the head of the distribution is slightly rounded with respect to the analytical cusp, due to the $ \sqrt{k_B T / \kappa_{\rm max}}$ intrinsic dispersion on the resetting position.
As seen in the main text, this slight deviation does not affects the thermodynamic result, which still fit the analytical result, derived for error-free resetting.

\subsection{Numerical simulation}

The experimental results presented in this paper are complemented by numerical simulations.
In order to be able to reproduce numerically both the idealized instantaneous resetting as well as the physical realization as we do with experimental data, we simulate the complete experimental scheme.
Numerical SR process are, like in the experiment, solutions of Langevin equations in time-dependent external potential, which stiffness varies from $\kappa_{\rm min}$ to $\kappa_{\rm max}$, both fed with experimental values.
The Langevin equation is solved using a standard Euler scheme \cite{Volpe2013}.
Both the random white noise simulating the thermal bath and the random distributions of times $\tau$ used to devise the resetting sequence are generated using built-in \textsc{python} pseudo-random number generators.\\

The trajectories obtained from these simulations are fully comparable to the experimental time-series and can be analyzed the same way.
They are used to correct small calibration offset in the recorded trajectory, obtained by comparing the motional variance of simulated and recorded trajectories in a steady-state.
This is a constant correction on calibration factor $\tilde \beta = 0.822 ~ \beta$ that does not change the dynamics of any observed effect. \\

\section{Stationary distributions}
\label{App:Dist}

Trapped Brownian trajectories undergoing an SR process are distributed according to a steady-state probability distribution function that depends both on the confining potential and the resetting parameters.
In our case, the Brownian microsphere is optically trapped in a locally harmonic potential $V(x) = \frac{1}{2} \kappa_{\rm min} x^2$, where $\kappa_{\rm min} \sim 10^{-6} ~ \si{\newton / \meter}$ is the stiffness of the confining potential (in contrast with $\kappa_{\rm max} \sim 10^{-4} ~ \si{\newton / \meter}$ which corresponds to the potential quench associated with each resetting event).
This confining potential, together with the viscous drag $\gamma$ experienced by the microsphere in the fluid, induces a characteristic pulsation $\omega_0 = \kappa_{\rm min} / \gamma \sim 10^{-3} ~ \si{\second}$ in the dynamics.

The steady state solution for the Fokker Planck equation with stochastic resetting is  \cite{Mori2022}
\begin{equation}
	P(x) = \lambda\int_0^{\infty} d\tau e^{-\lambda\tau}G(x |\tau,0),
\end{equation}
where $\lambda$ is the rate of the Poissonian stochastic resetting and $G(x| \tau,0)$ is the propagator of a given stochastic process that is reset to.
For a Wiener process, the propagator is 
\begin{equation}
	G(x| \tau,0) = \frac{1}{\sqrt{4\pi D\tau}}\exp \left( -\frac{x^2}{4D\tau} \right).
\end{equation}
Therefore, the corresponding integral to solve is
\begin{equation}
	P(x) = \frac{\lambda}{\sqrt{4\pi D}}\int_0^{\infty} \frac{d\tau}{\sqrt{\tau}} \exp \left[ -\left( \lambda\tau +\frac{x^2}{4D\tau} \right) \right]
\end{equation}
now, let $u = \sqrt{\tau}$, the integral becomes then
\begin{equation}
	\begin{aligned}
		P(x) &=  \frac{\lambda}{\sqrt{\pi D}}\int_0^{\infty} du \exp\left( -\lambda u^2 - \frac{x^2}{4Du^2} \right)\\
		&= \frac{1}{2}\sqrt{\frac{\lambda}{D}}\exp\left( -\sqrt{\frac{\lambda}{D}}|x| \right).
	\end{aligned}
\end{equation}
This corresponds to an Laplace (bi-exponential) distribution, known steady-state for free Brownian motion undegoing Poissonian resetting. This Laplace-distribution is also represented on the first figure of the main text.

For a Ornstein-Uhlenbeck process (Brownian motion in a harmonic potential), the propagator is given by
\begin{equation}
	G(x| \tau,0) = \sqrt{\frac{\omega_0}{2\pi D (1-e^{-2\omega_0\tau})}}\exp\left( -\frac{\omega_0}{2D}\cdot\frac{x^2}{1-e^{-2\omega_0\tau}} \right),
\end{equation}
where the parameter $\omega_0 = \kappa/\gamma$. For simplicity, we define $a = \omega_0/2D$ and $b = 2\omega_0$.
Thus, the steady state PDF can be written as
\begin{equation}
	P(x) = \sqrt{\frac{a}{\pi}}\int_0^{\infty} d\tau re^{-\lambda \tau} \frac{1}{\sqrt{1- e^{-b\tau}}}\exp\left( -\frac{a x^2}{1-e^{-b\tau}} \right).
\end{equation}
Applying the change of variable $u = e^{-\lambda \tau}$ leads to $e^{-b \tau} = u^p$ where $p = b/\lambda$. The integral thus becomes
\begin{equation}
	P(x) = \sqrt{\frac{a}{\pi}}\int_0^1 du \frac{1}{\sqrt{1-u^p}}\exp\left( -\frac{a x^2}{1-u^p} \right).
\end{equation}
A second change of variable defined by $w = 1 - u^p$, where $du = -\frac{1}{p}(1-w)^{\frac{1}{p}-1}dw$, gives the expression for the probability distribution
\begin{equation}
	P(x) = \sqrt{\frac{a}{\pi}}\int_0^1 dw \frac{1}{p}(1-w)^{\frac{1}{p} - 1} w^{-1/2} \exp\left( -\frac{a x^2}{w} \right).
\end{equation}
According to Gradshteyn \& Ryzhik 3.471.2 \cite{gradshteyn2007}, our integral leads to the final result:
\begin{equation}
	P(x) = \frac{1}{p} \sqrt{\frac{a}{\pi}} \Gamma\left(\frac{1}{p}\right) e^{-\frac{a x^2}{2}} \left( a x^2 \right)^{-\frac{1}{4}} W_{\frac{1}{4}-\frac{1}{p},\frac{1}{4}}(a x^2),
\label{eq:DistExact}
\end{equation}
where $W_{k,m}(x)$ is the Whittaker function. 

\begin{figure}[h!]
	\centerline{\includegraphics[width=0.9\linewidth]{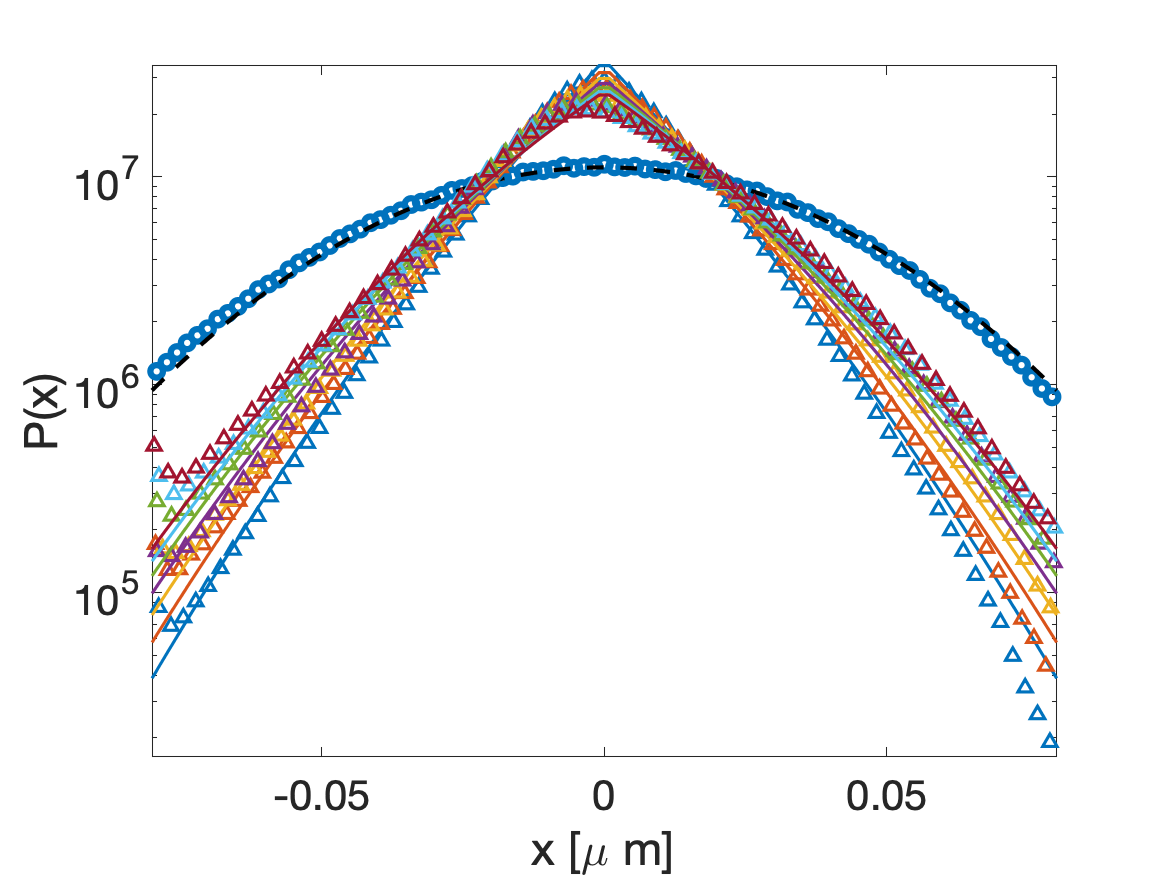}}
	\caption{Experimental histograms (triangles) together with the exact distribution Eq.~(\ref{eq:DistExact}) (solid lines) for a large range of resetting rates from $\lambda = 283$ Hz to $\lambda = 795$ Hz while $\omega_0 = 159$ Hz.
		 As a comparison, the Gaussian probability distribution of a Brownian trajectory in the same potential $V(x)$ with the characteristic pulsation $\omega_0$ is shown (black dashed line) together with the associated experimentally measured histogram (blue circles).
			}
	\label{fig:AllDistSR}
\end{figure}

The result Eq.~(\ref{eq:DistExact}) is used in the main text and agrees with the experimentally measured distribution for an SR process in a harmonic potential. This agreement validates our experimental realization of SR.
On Fig. \ref{fig:AllDistSR}, we show the agreement between experimentally measured steady-state distributions and the result Eq.~(\ref{eq:DistExact}) for various resetting rates, ranging from 283 to 795 Hz.
The analytical expression captures the experimental behavior on a large spacial range as well as on an extended parameter variation.
One can note that the steady-states share a sharp central peak, similar to the exponential distribution of \textit{free} SR process, while the tails tends to follow the (Ornstein-Uhlenbeck) Gaussian tails.
For small $\lambda / \omega_0$, the distribution is closer to the Gaussian corresponding to a trajectory without resetting while the distribution is progressively more sharply peaked around the resetting position $x = 0$ when the resetting rate increases, departing more strongly from the Gaussian distribution.

Importantly, this result generalizes the expression known in the literature for SR in a harmonic potential. This expression is expressed in terms of negative order Hermite's polynomials \cite{Pal2015, Trajanovski2023}.

Indeed, if the two results coincides exactly for integer values of $\lambda / \omega_0$, our result, based on the Whittaker function is not restricted to integer ratio between the resetting rate and the characteristic pulsation, since the parameter $k$ and $m$ of $W_{k,m}(x)$ function can take arbitrary real values.

\section{Entropy and Second law of thermodynamics for resetting} 
\label{App:Entropy}

The second term in the non-equilibrium free energy is proportional to the average rate of stochastic Shannon entropy change $\Delta s$.
The mean resetting entropy production rate is obtained by averaging  $\Delta s$ over resetting events:
\be
	\dot{S}^{\rm rst} = \lambda \langle \Delta s \rangle = k_B \lambda  \int\ln\left[\frac{P(x)}{P(0)}\right] P(x) dx.
\ee
This quantity does not account for the contribution of the diffusive trajectory between resetting.
This contribution enters in the heat production rate, to which is associated a dissipation of entropy in the medium, $\dot{S}^{\rm m} = \dot{Q}/T$.
Finally, the total entropy production rate of a Brownian trajectory experiencing SR reads $\dot{S}^{\rm tot} = \dot{S}^{\rm m} + \dot{S}^{\rm sys} - \dot{S}^{\rm rst} \geq 0$,  \cite{Fuchs2016, Gupta2022Thermo}.
The system entropy $\dot{S}^{\rm sys} = k_B \frac{d}{dt} \int P(x) \ln[P(x)]dx$ vanishes in the steady-state \cite{Busiello2020, Pal2021}.
Therefore, here, the total entropy production rate
\be
\dot{S}^{\rm tot} = \dot{S}^{\rm m} - \dot{S}^{\rm rst} \geq 0
\ee
is a positive quantity, only reaching zero in the limit of equilibrium. As such, it is interpreted as the Second law of thermodynamics for an SR process \cite{Esposito2011, Fuchs2016, Mori2022}.

\section{Quasistatic limit on the external work}
\label{App:QS}

Here we derive the external work rate expended in the limit of quastatic resetting events.
As detailed in the main text, the external work reads
\be
	\langle w^{\rm ext}(t) \rangle = \frac{1}{2} \int_0^t \dot{\kappa}(t') \langle x_{t'}^2 \rangle dt'.
	\label{Eq:ExtWork}
\ee
where $\dot \kappa(t)$ is the time-derivative of the protocol imposed and $\langle x^2(t) \rangle$ is the motional variance of the microsphere, characterizing the response of the system to this specific protocol $\kappa(t)$.
The motional variance obeys a simple differential equation reading
\be
	\frac{d \langle x^2(t) \rangle}{dt} = \frac{-2 \kappa(t)}{\gamma} \langle x^2(t) \rangle + 2 D.
	\label{Eq:Var}
\ee
We begin by imposing a protocol $\kappa(t) = \frac{\kappa_{\rm max} - \kappa_{\rm min}}{\tau_{\rm w}} (t - t_m) +  \kappa_{\rm min}$ connecting $\kappa_{\rm min}$ to  $\kappa_{\rm max}$ in a linear way with characteristic time $\tau_{\rm w}$ and starting at time $t_m$.
The response of the variance to this protocol can be obtained by solving numerically Eq.~(\ref{Eq:Var}) for this protocol.
In the limit of very large $\tau_{\rm w}$ (the quasistatic limit) we expect no difference with the hyperbolic tangent protocol used in the experiment.

\begin{figure}[htb]
	\begin{center}
		\centering{
			\includegraphics[width=0.36\textwidth]{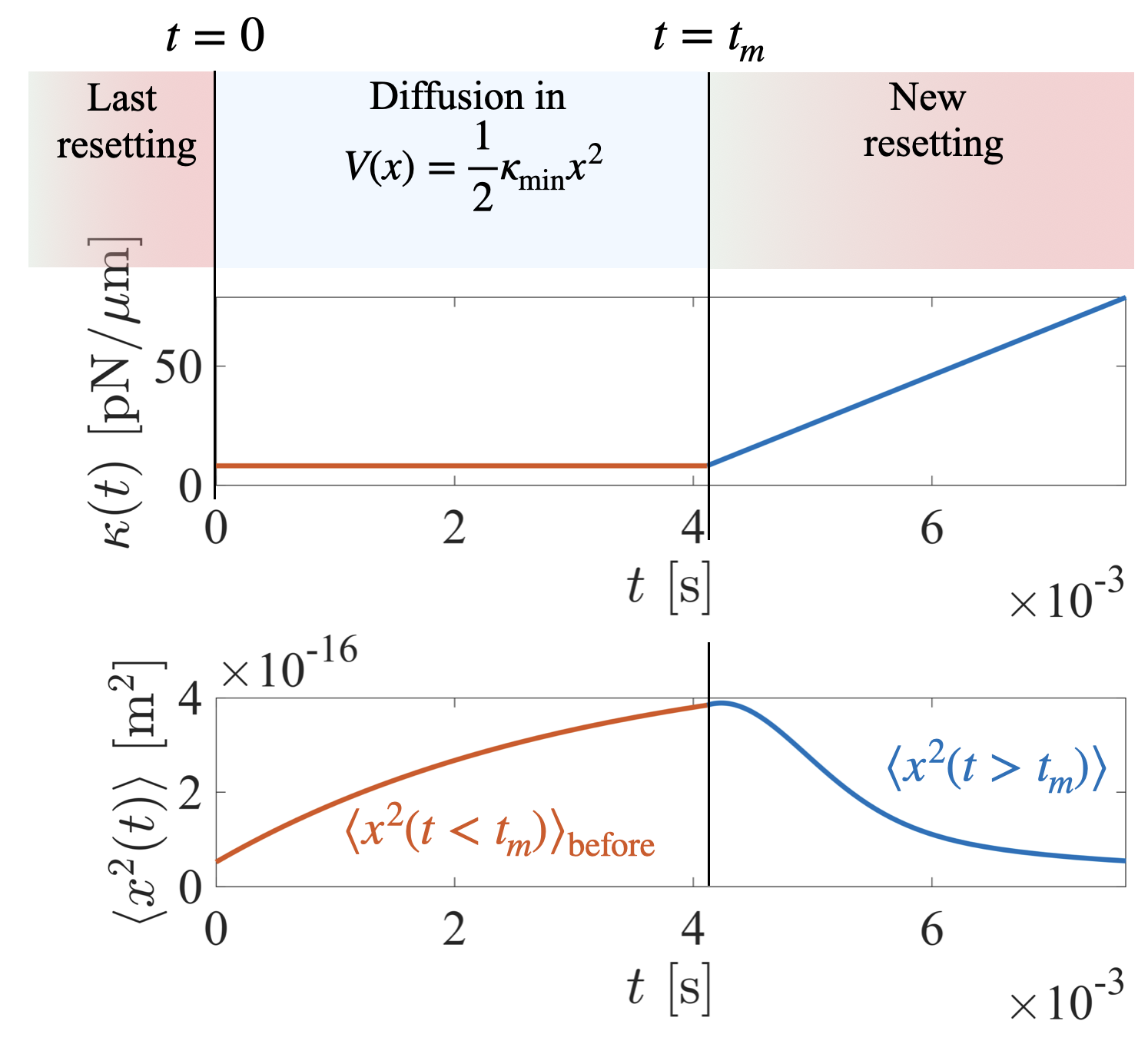}
			\label{Fig:VarSol1}}
		\caption{{Schematic explanation of a sequence of diffusion and quasistatic resetting. For $t \in [0, t_m]$ the particle diffuses and its variance $\langle x^2(t) \rangle_{\rm before}$ increases as $\kappa = \kappa_{\rm min}$. For $t \in [t_m, t_m + \tau_{\rm w}]$ the stiffness gradually increases and the variance $\langle x^2(t) \rangle$ evolves non-monotonously.
		}}
	\end{center}
\end{figure}

The solution takes a the form
\be
	\langle x^2(t) \rangle = \mathcal{C} \times \rm{exp}(f(t)) + \rm{exp}(g(t)) \times \rm{erf}(h(t))
	\label{Eq:VarSol}
\ee
where $f$, $g$ and $h$ are simple functions of time obtained in the numerical solution of Eq.~(\ref{Eq:Var}).
The evolution of $\langle x^2(t) \rangle$ describe the variance during the resetting event.
Before the beginning of the resetting event, for times $t \in [0, t_m]$, the variance was evolving according to
\be
	\frac{d \langle x^2(t) \rangle_{\rm before}}{dt} = \frac{-2\kappa_{\rm min}}{\gamma} \sigma^2(t) + 2 D
	\label{Eq:Var2}
\ee
with solution
\be
	\langle x^2(t_m) \rangle_{\rm before} = \left( \frac{k_B T}{\kappa_{\rm min}} - \frac{k_B T}{\kappa_{\rm max}} \right) e^{-2\kappa_{\rm max} t_m / \gamma} + \frac{k_B T}{\kappa_{\rm max}}.
	\label{Eq:Var2Sol}
\ee
This corresponds to the spontaneous relaxation of the variance in the shallow potential of stiffness $\kappa_{\rm min}$ after the last resetting event, at time $t=0$ evaluated at time $t_m$, with initial condition $\langle x^2(0) \rangle_{\rm before}) = k_B T / \kappa_{\rm max}$.
The evolution of the variance before and after the resetting starts is schematized on Fig.~\ref{Fig:VarSol1}.
The constant $\mathcal{C}$ in Eq.~(\ref{Eq:VarSol}) can be expressed by matching the value of the initial value of $\langle x_{t_m}^2 \rangle$ to the final value of $\langle x^2(t_m) \rangle_{\rm before}$ at the time $t_m$

This, combined with the solution Eq.~(\ref{Eq:VarSol}), allows to express the variance $\langle x^2(t) \rangle$ which is not only a functional of the protocol $\kappa(t>t_m)$ but also, through its initial condition set by $\sigma^2(t_m)$, a function of the time elapsed since the previous resetting event.

\begin{figure}[htb]
	\begin{center}
		\centering{
			\includegraphics[width=0.36\textwidth]{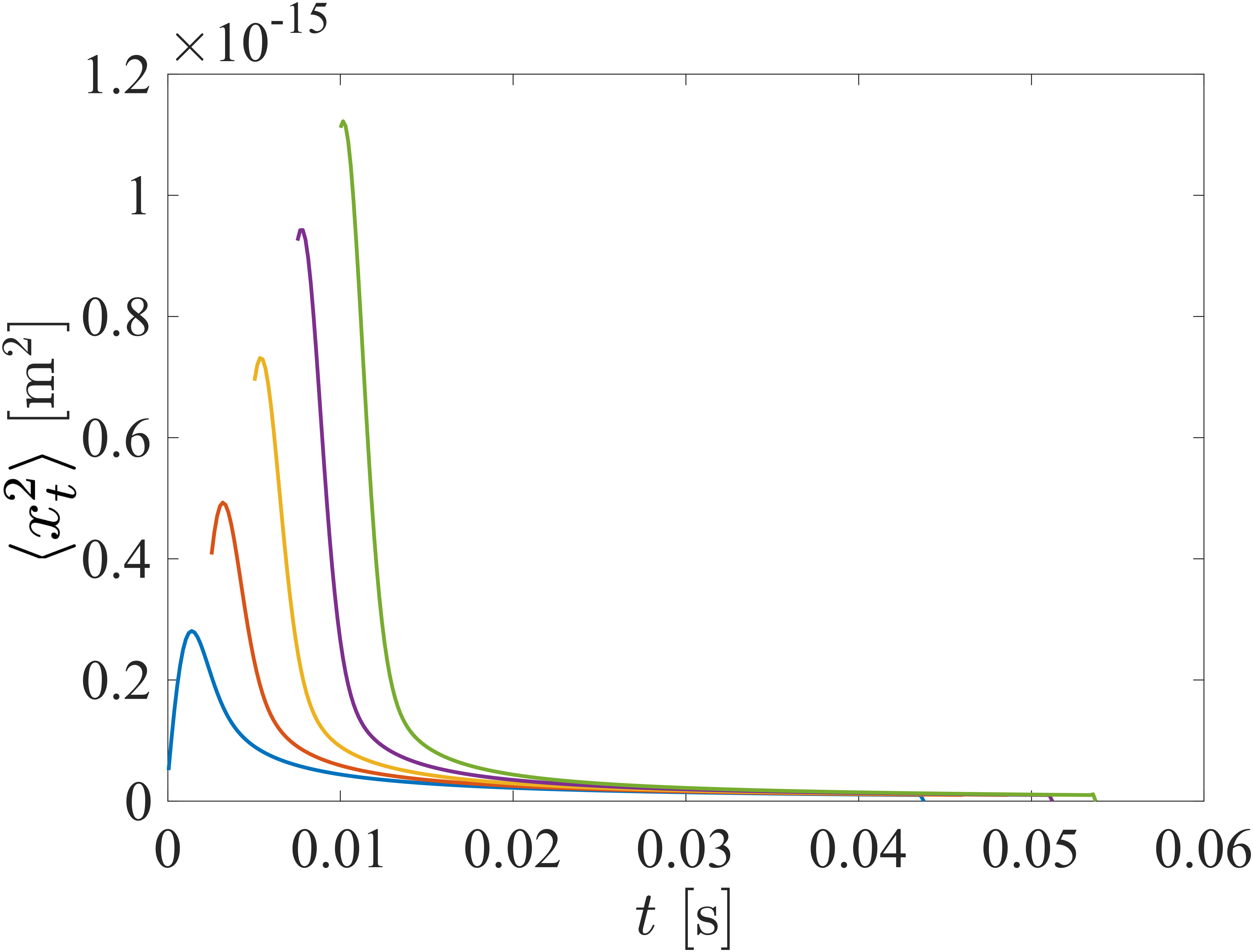}}\\
		\centering{
			\includegraphics[width=0.36\textwidth]{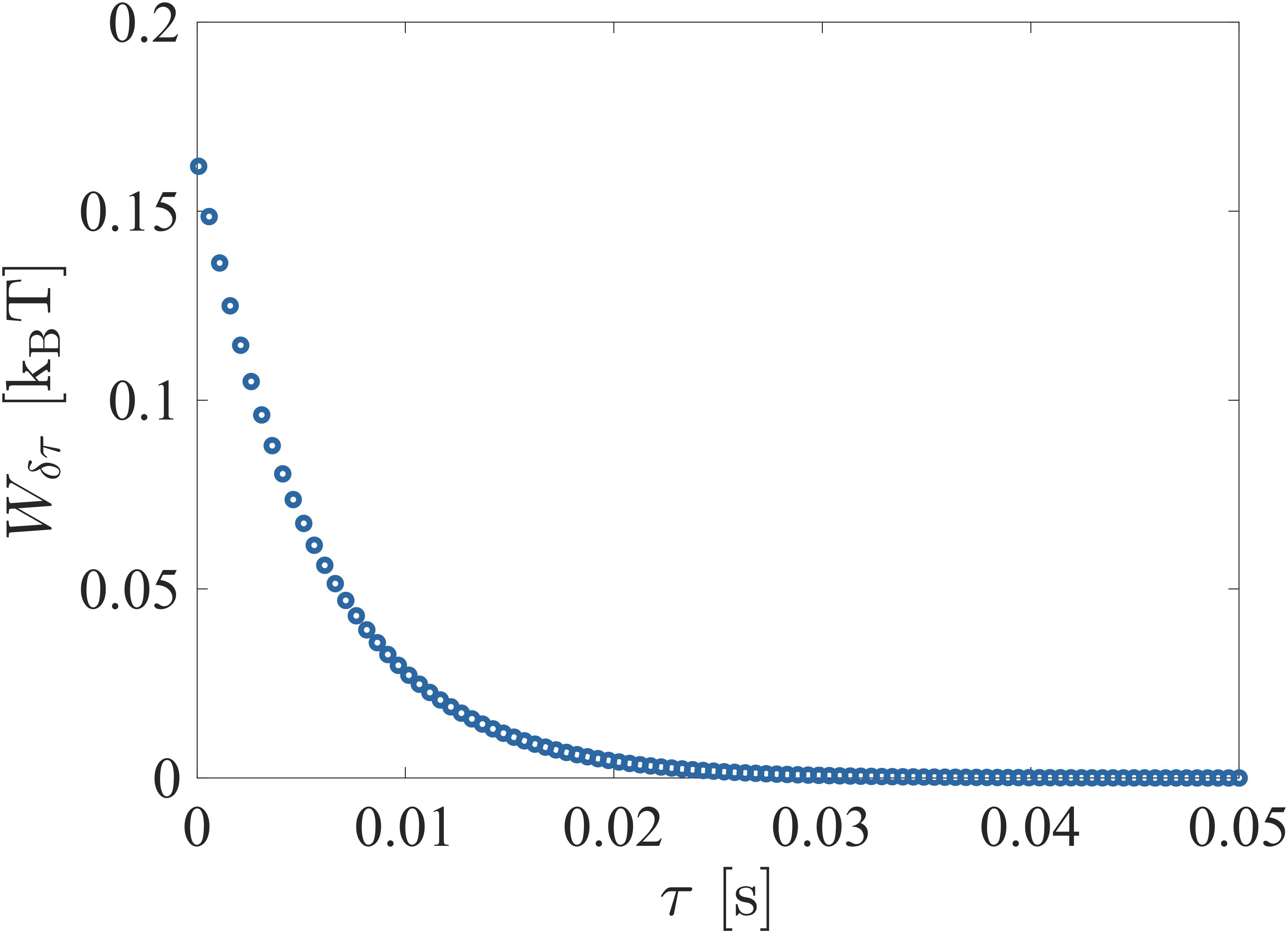}
			\label{Fig:VarSol}}
		\caption{{(Upper panel) Variance decay $\langle x^2(t) \rangle$ as a function of time for different initial conditions, set by the time since last resetting. For clarity we set the last resetting event to have ended at time $t=0$ and plot the variance evolution starting from the different values of $t_m$. Therefore, the relaxation of the variance between both resetting events is visible through the enveloppe of the initial value of $\langle x^2(t) \rangle$ which draws the exponential decay Eq.~(\ref{Eq:Var2Sol}).
		(Lower panel) Average work contribution for random times $t_m$ since last resetting $ t \in [t_m, t_m + \delta t_m]$.
		}}
	\end{center}
\end{figure}

The resulting expression is to long to be reasonably written but takes an intuitive shape which we represent on Fig.~\ref{Fig:VarSol1} in details for one set of parameters and on Fig.~\ref{Fig:VarSol} for different initial condition, \textit{i.e.} different times $t_m$.
In the following, we explain this behavior.
For very small $t_m$ (blue line on Fig.~\ref{Fig:VarSol}) the variance is strongly non-monotonic.
Indeed, after the previous resetting, the state is quenched and equilibrated in the stiff potential, with variance $ \frac{k_B T}{\kappa_{\rm max}}$.
The variance then starts to increase according to Eq.~(\ref{Eq:Var2Sol}) until the next resetting.
If it occurs after a short time $t_m$, the variance is still very close to its initial value, hence, when the stiffness $\kappa(t)$ starts quasistatically to increase starting from $\kappa_{\rm min}$, the variance transiently still increase, until later decreasing again, following equipartition $\langle x^2(t) \rangle = \frac{k_B T}{\kappa(t)}$ for long times $t \gg t_m$.
In the opposite limit, if the time since last resetting is very large (green line on Fig.~\ref{Fig:VarSol}), then the variance approaches $\frac{k_B T}{\kappa_{\rm min}}$ and when the stiffness starts to increase, equipartition is immediately reached and the variance evolves monotonically.
For all cases in between (red, yellow and purple), the variance keeps a transient non-monotonous behavior.

Importantly, the non-monotonous evolution of variance for short diffusive times $t_m$ reveals the non-equilibrium initial condition (breaking equipartition) and demonstrate why a quasistatic driving is not reversible in that case.
It cannot correspond to the minimal expanded work.
The instantaneous increase of $\kappa$ described in the main text as well as in the next section of the SI precisely serves at cancelling this transient evolution, ensuring equipartition at all times.
In the following we derive the work associated with such non-optimal quasistatic process, while the work cost of a reversible protocol will be derived in the Sec~\ref{App:Rev}.

The variance $\langle x^2(t) \rangle$ is an average quantity with respect to the stochastic position $x(t)$ but remains a random variable due to the randomness of the time since last resetting $t_m$ which is drawn from the distribution $P(t_m) = \lambda e^{-\lambda t_m}$.
We can therefore compute the average work contribution arising from trajectories undergoing resetting at times $ t \in [t_m, t_m + \delta t_m]$ as
\be
	W_{\delta t_m}(t_m) = \frac{1}{2} \lambda e^{-\lambda t_m} \delta t_m \int_{t_m}^{\tau_{\rm w}} \dot \kappa(t) \langle x^2(t) \rangle dt.
\ee
where we recall that $\tau_{\rm w}$ is the length of the potential.
This quantity is plotted on Fig.~\ref{Fig:VarSol} (lower panel) as a function of the time since last resetting $t_m$.
We can note its characteristic decaying behavior, consequence of the Poissonian nature of the resetting.

The average power injected in the system in the quasistatic limit of SR is finally evaluated as
\be
	\dot W_{\rm QS} = \lambda \sum_{t_m =0}^{\infty} W_{\delta t_m}(t_m)
\ee
taking into account the weighted work of each resetting event on average.
This is the quantity used as quasistatic limit for work in the main text.

\section{Reversible limit and Szilard-like protocols}
\label{App:Rev}

As seen in the previous section, when a resetting event occurs, the system is in a dynamical non-equilibrium state: its motional variance is in a transient between $\frac{k_B T}{\kappa_{\rm max}}$ and $\frac{k_B T}{\kappa_{\rm min}}$.
Therefore, a quasistatinc increase of the stiffness cannot correspond to a reversible driving of the system (as is revealed by the non-monotonous evolution of the variance after the new resetting started as seen Fig.~\ref{Fig:VarSol}).
To cancel this effect and obtain a reversible transformation, we rely on a protocol similar to the one proposed by Szilard \cite{szilard1929}.
At the new resetting time we instantaneously change the stiffness to $\kappa(t_m) = k_B T / \langle x^2(t_m) \rangle$.
Doing so ensure that the system is at equilibrium at time $t_m$ and, by quasistatically increasing $\kappa$ that it will stay at equilibrium until $\kappa = \kappa_{\rm max}$.
Importantly, the accumulation of these reversible equilibrium-to-equilibrium transformation in the long steady-state (composed of many of such event) remains a non-equilibrium steady-state.
Indeed, even if individual transformation are reversible, it still constitute a constant driving, maintaining the PDF different from the stationary equilibrium Gaussian in the potential $\kappa_{\rm min} x^2 / 2$.

We now look at the thermodynamic cost of such transformation.
It is composed of 4 steps:
\begin{itemize}
	\item initially, the system is out of equilibrium with $\langle x^2(t_m) \rangle \neq k_B T / \kappa(t_m)$
	\item the stiffness is increased to ensure equipartition $\kappa(t_m) = k_B T / \langle x^2(t_m) \rangle$, the system is now at equilibrium
	\item the stiffness is increased quasistatically until $\kappa_{\rm max}$ the system remains at equilibrium, with final variance $\langle x^2(t \gg t_m) \rangle = k_B T / \kappa_{\rm max}$. This terminates the resetting operation
	\item the stiffness is abruptly decreased to $\kappa_{\rm min}$ again, initiating the next diffusing sequence.
\end{itemize}
	
Considering the 3 first steps, it corresponds to a transformation from a non-equilibrium state to an equilibrium.

The total entropy produced $\Sigma$ to carry such a transformation is well described in the literature, via the non-equilibrium second law \cite{Esposito2011, Parrondo2015, Ciliberto2019} and reads
\be
	T \Sigma = W - \Delta F_{\rm eq} - k_B T \mathcal{D}(P_f || P_{\rm eq}) + k_B T \mathcal{D}(P_i || P_{\rm eq})
\ee
where $\mathcal{D}(P || P_{\rm eq}) = \int P(x) \ln (P(x) / P_{\rm eq}(x) ) dx$ is the Kullback-Leibler divergence (or relative entropy) between the state $P(x)$ and the corresponding equilibrium. The equilibrium free energy derives from the equilibrium partition functions and reads $\Delta F_{\rm eq} = \frac{k_B T}{2} \ln \left( \frac{\kappa_{\rm max}}{\kappa_{\rm min}} \right)$, it relates both equilibria in harmonic potentials of stiffness $\kappa_{\rm min}$ and $\kappa_{\rm max}$.

$P_i(x)$ corresponds to the state of the system when the transformation begins, and $P_f(x)$ to the state of the system when the transformation ends.
In our case, for all times $t \in [t_m, t_m + \tau_{w}]$ the corresponding (target) equilibrium state $P_{\rm eq}(x)$ is a Gaussian of variance $\sigma_{\rm eq} = k_B T / \kappa_{\rm max}$
\be
	P_{\rm eq}(x) = \frac{1}{\sqrt{2 \pi \sigma^2_{\rm eq}}} e^{-x^2 / 2 \sigma^2_{\rm eq}}
\ee
Since equilibrium is reached at the end of the transformation, hence $\mathcal{D}(P_f || P_{\rm eq}) = 0$, but at the beginning of the resetting sequence, where $t = t_m$ the system still lies in the a non-equilibrium state characterized by
\be
	P_i(x) = \frac{1}{\sqrt{2 \pi \sigma^2_i}} e^{-x^2 / 2 \sigma^2_i}.
\ee
The initial Kullback-Leibler divergence can therefore be evaluated using
\be
	\ln \left( \frac{P_i(x)}{P_{\rm eq}(x)} \right) = \frac{1}{2} \left( -x^2 \left[ \frac{1}{\sigma^2_i} - \frac{1}{\sigma^2_{\rm eq}} \right] + \ln \left[ \frac{\sigma^2_{\rm eq}}{\sigma^2_{i}} \right] \right),
\ee
leading to
\begin{widetext}
\begin{align*}
	\mathcal{D}(P_i || P_{\rm eq}) &= \frac{-1}{2 \sqrt{2 \pi \sigma^2_i}} \int_{-\infty}^{+\infty} e^{-x^2 / 2 \sigma^2_i} \left( -x^2 \left[ \frac{1}{\sigma^2_i} - \frac{1}{\sigma^2_{\rm eq}} \right] + \ln \left[ \frac{\sigma^2_{\rm eq}}{\sigma^2_{i}} \right] \right) dx,\\
		&= \frac{-1}{2} \ln \left( \frac{\sigma^2_i}{\sigma^2_{\rm eq}} \right)  -  \frac{-1}{2 \sqrt{2 \pi \sigma^2_i}} \left[ \frac{1}{\sigma^2_i} - \frac{1}{\sigma^2_{\rm eq}} \right]  \int_{-\infty}^{+\infty} x^2 e^{-x^2 / 2 \sigma^2_i} dx
\end{align*}
\end{widetext}
where we recognize the motional variance $\sigma^2_i \equiv \int x^2 P_i(x) dx$ in the last term. Using the notation $\theta = \sigma^2_i/\sigma^2_{\rm eq}$ for simplicity, the Kullback-Leibler divergence between two Gaussian distributions reads
\be
	\mathcal{D}(P_i || P_{\rm eq}) = \frac{-1}{2} \left( \ln \theta - \left[ \theta - 1 \right] \right)
\ee
which is a positive quantity since $\ln \theta < \left[ \theta - 1 \right]$.
This is the expression used in the main text.

From a thermodynamic perspective, a reversible driving where the system is at thermal equilibrium for all times corresponds to a transformation with non entropy production $\Sigma = 0$ in the expression above.
This lead to the definition of the minimal reversible work
\be
	w_{\rm rev}^{\rm rst} = \Delta F_{eq} - \frac{-k_B T}{2} \left( \ln \theta - \left[ \theta - 1 \right] \right)
\ee
which is smaller than the equilibrium free-energy difference.
Perfectly aligned with the non-equilibrium second law \cite{Esposito2011}, we show that using a correct evaluation of the initial distance to equilibrium allows to reduce the thermodynamic cost of the transformation below the standard equilibrium limit.
The work $w_{\rm rev}^{\rm rst}$ is still a random variable of the time since last resetting $t_m$, defining the value of the variance $\sigma^2_i$ entering in $\theta$.

The total work exchanged during all 4 steps described above should however also take into account the work exchanged when the stiffness is abruptly decreased at the end of the resetting sequence.
The cost of such step-like transformation of the potential can be evaluated \cite{Cabara2020} and reads $\frac{k_B T}{2} \left( \frac{\kappa_{\rm min}}{\kappa_{\rm max}} - 1 \right)$.

The average minimal power, needed to maintain the system in an SR-NESS using reversible increase of stiffness for each resetting event reads
\begin{align*}
	\dot W_{\rm ext}^{\rm rev} &= \lambda \bigg( \Delta F_{\rm eq}\\
	& - k_B T\int_0^\infty \mathcal{D}(P_i(x, \tau) || P_{\rm eq}(x)) \lambda e^{-\lambda \tau} d\tau\\
	& + \frac{k_B T}{2} \left[ \frac{\kappa_{\rm min}}{\kappa_{\rm max}} - 1 \right] \bigg)
\end{align*}
which is the expression used in the main text with the identification $\int_0^\infty \mathcal{D}(P_i(x, \tau) || P_{\rm eq}(x)) \lambda e^{-\lambda \tau} d\tau = \langle \mathcal{D}(P_i(x, \tau) || P_{\rm eq}(x)) \rangle_\tau = \langle I(\tau) \rangle_\tau$ as the average information erased by the successive resetting events.

\section{Ergodicity and drift correction}
\label{App:Ergo}

Ergodicity of the stochastic processes at play is evaluated with a statistical ensemble of individual sub-trajectories, drawn out of a long time-series of position $x(t)$.

\begin{figure}[htb]
	\begin{center}
		\centering{
			\includegraphics[width=0.40\textwidth]{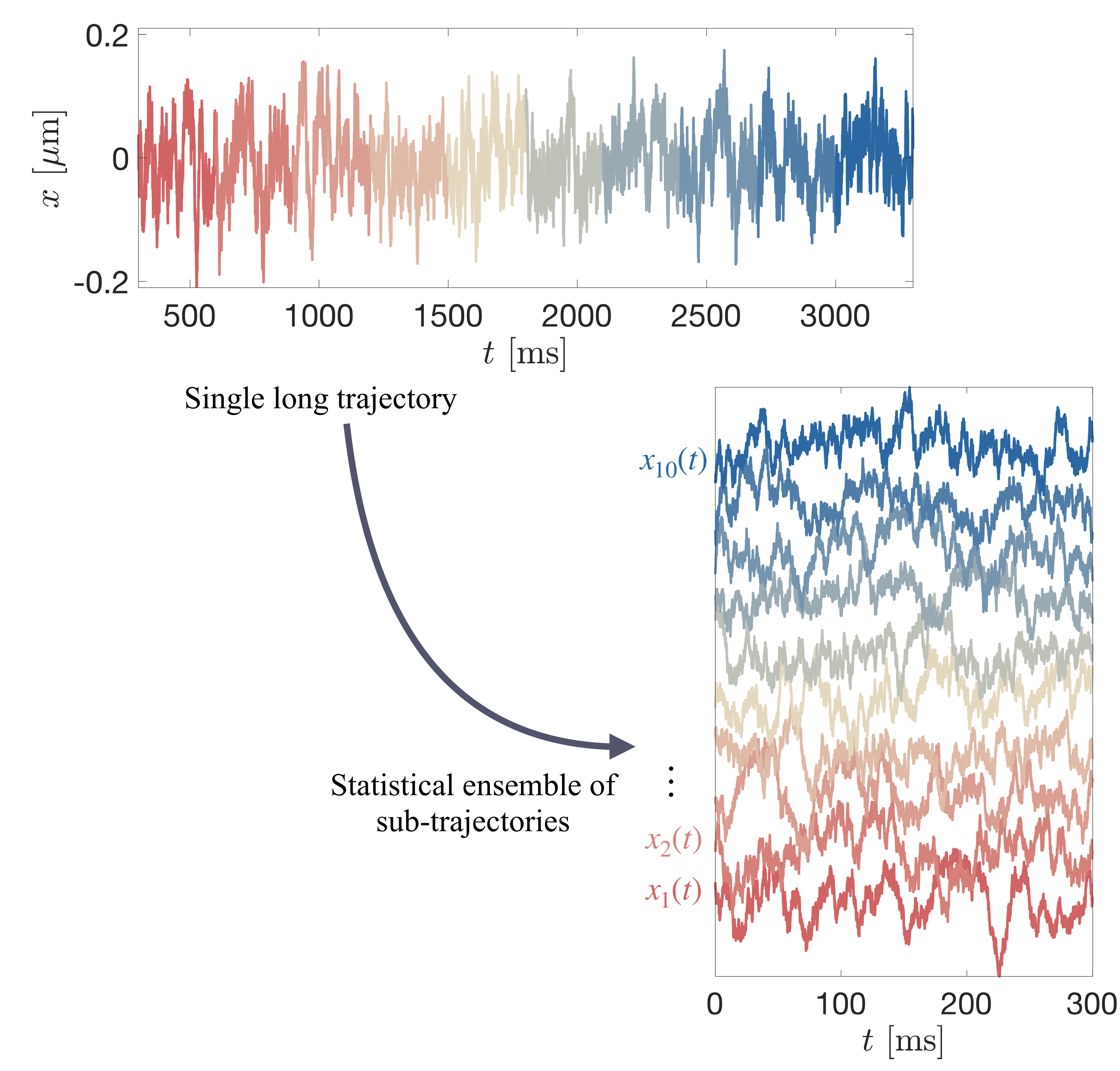}
			\label{fig:ExplanationEnsemble}}\\
		\centering{
			\includegraphics[width=0.35\textwidth]{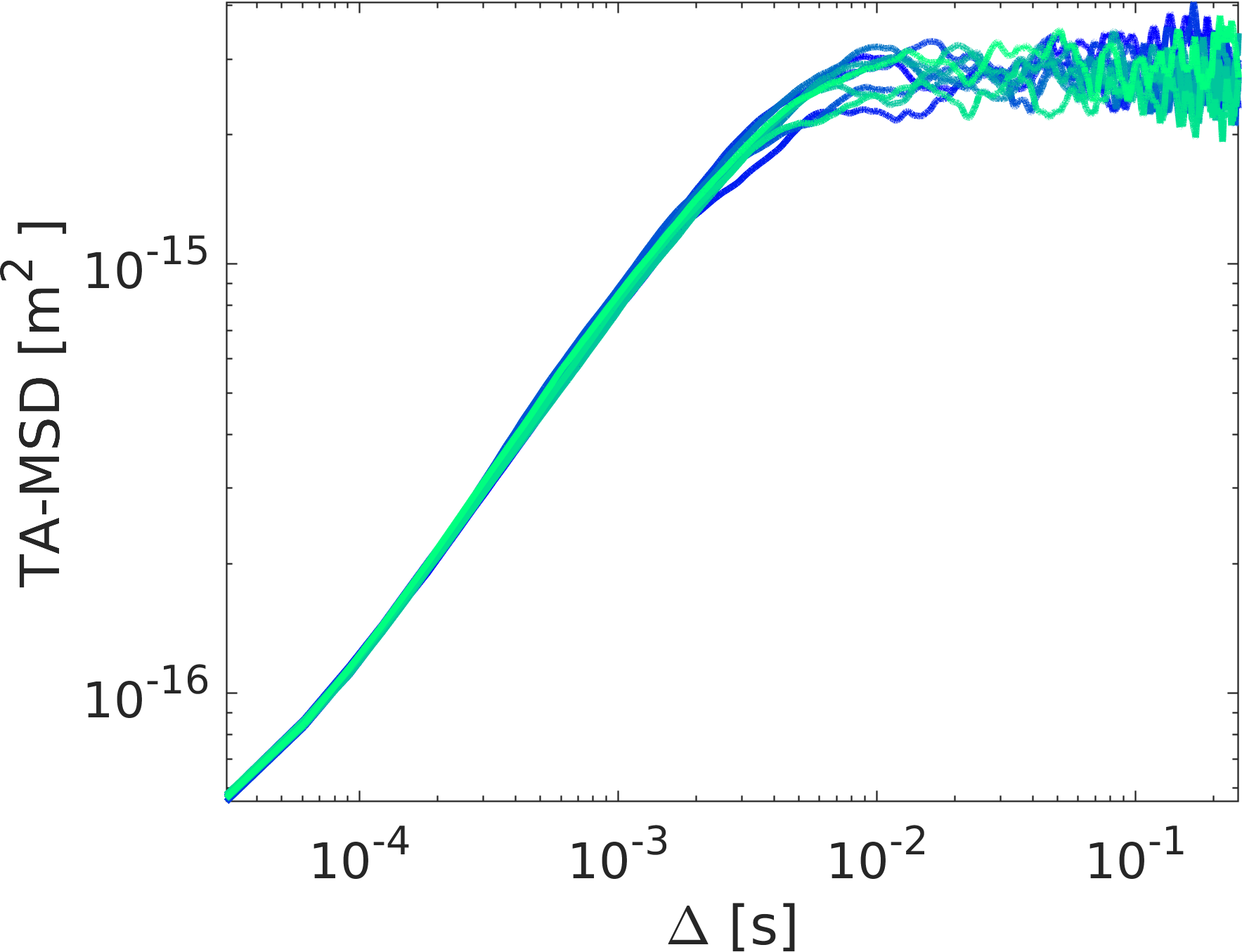}
			\label{fig:SomeTAMsd}}
		\caption{{ (Upper panel) Schematic representation of the method used to build an ensemble out of a single long time-series of position, from a single $2.5$ seconds trajectory to an ensemble of 10 individual $0.25$ seconds sub-trajectories.
		(Lower panel) Associated 10 individual TA-MSD.
		}}
	\end{center}
\end{figure}

On Fig. \ref{fig:ExplanationEnsemble} (a) we show a schematic representation of how a statistical ensemble is built out of a single time-series of position.
On each individual sub-trajectory, we can compute the time-averaged mean-square-displacement (TA-MSD)
\be
	\textrm{TA-MSD} \equiv  \overline{\delta^2_x}(\Delta) \equiv \frac{1}{\mathcal{T}-\Delta} \int_{0}^{\mathcal{T}-\Delta} \left( x^i_{t+\Delta} - x^i_t\right)^2 dt
\ee
where $\mathcal{T}$ is the total time of the measured sub-trajectory.
Because of the finite size of sub-trajectories, a dispersion is still visible on the ensemble TA-MSD as seen on Fig. \ref{fig:SomeTAMsd} (b).
Ergodicity, as explained in the main text, is probed by looking at the evolution of this dispersion as a function of $\mathcal{T}/\Delta$, a necessary and sufficient condition for the process to be ergodic being the vanishing of the dispersion for short $\Delta$.
However, if the experimental setup suffers from low frequency drift on the $300$-seconds long experiments, this dispersion will we combined with a systematic trend, that will lead to an overestimation of the ergodic parameter $\epsilon(\Delta) \sim \textrm{var}(\textrm{TA-MSD})$.
In this Appendix, we propose a novel method to decipher systematic from statistic dispersion of TA-MSD, by relying on their very short-time limit.
This allows to correct drifts and clearly unveil the different between ergodic normal Brownian motion in a potential from non-ergodic SR process.\\

We consider here, as in the section of the main text focused on Landauer's limit, a $300$ second long trajectory that can be recast into an ensemble of a thousand sub-trajectories of $0.3$ seconds each, diffusing in an optical potential of stiffness $\kappa_{\rm min} = 2.9 \pm 0.15 ~ \si{\pico\newton / \micro\meter}$.
We study both a normal Brownian motion diffusing in the aforementioned potential as well as an SR process in the same potential, but in which the optical potential is increased to $\kappa_{\rm max} \approx 83 ~ \si{\pico\newton / \micro\meter}$ at a rate $\lambda^{-1} = 20 ~ \omega_{\rm max}^{-1} \approx 6.1 ~ \si{\milli\second}$.
Each TA-MSD is computed as a time integral on individual sub-trajectories of total time $\mathcal{T} = 0.3$ seconds. 
By comparing TA-MSDs at different absolute times, we can detect low frequency drifts.
To achieve the best accuracy, we probe TA-MSD for very short time-lag $\Delta = 0.061 ~ \si{\milli\second}$ where the statistical dispersion of TA-MSD is the smallest.\\

\begin{figure}[htb]
	\begin{center}
		\centering{
			\includegraphics[width=0.35\textwidth]{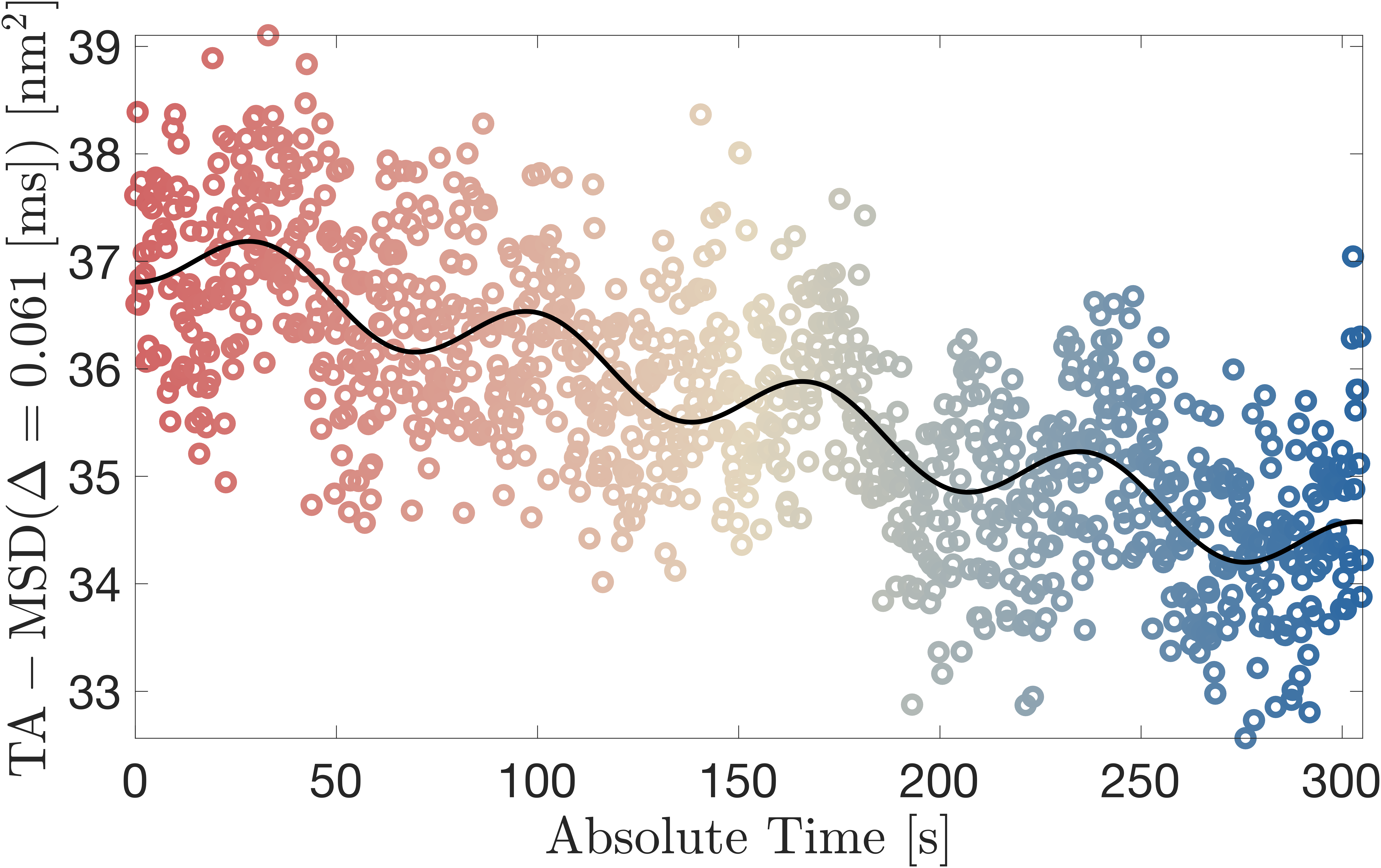}
			\label{fig:DriftCorrection}}\\
		\centering{
			\includegraphics[width=0.35\textwidth]{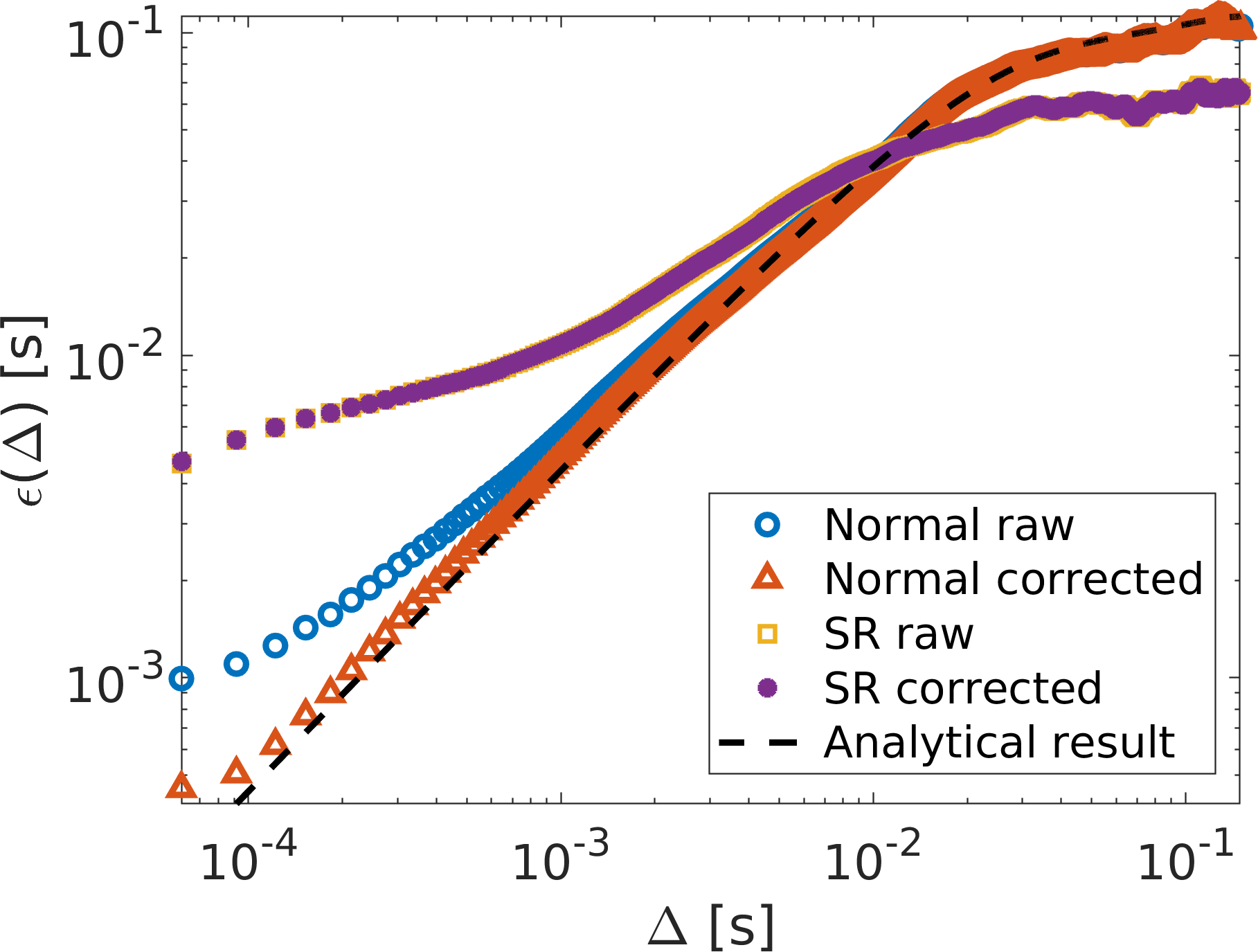}
			\label{fig:ErgodicParamCorrection}}
		\caption{{ (Upper panel) Short time-lag TA-MSD $\overline{\delta_{x_i}^2}(\Delta = 0.061 [\textrm{ms}]))$ for each individual chronologically ordered sub-trajectories as a function of the absolute time in seconds. We clearly observe a systematic trend, that is well captured by the combination of linear decrease and a $\approx 10 ~ \si{\second}$ sinusoidal evolution. The magnitude of the drift is of the order of $1.4 ~ \si{\nano\meter}$ of mean displacement.
		(Lower panel) Associated ergodic parameter, both for a normal Brownian motion in a potential of stiffness $\kappa_{\rm min}$ and for an SR process in the same potential. We show the raw data measured from the calibrated photodiode as well as the same observable on data where the $1.4 ~ \si{\nano\meter}$ mean displacement drift on the $300~\si{\second}$ has been corrected.
		}}
	\end{center}
\end{figure}

On Fig. \ref{fig:DriftCorrection} (a) we show the value of TA-MSD at $\Delta = 0.061 ~ \si{\milli\second}$ from blue to green for each chronologically ordered sub-trajectory as a function of the absolute time of the experiment.
It corresponds to a vertical cut in a TA-MSD plot such as displayed Fig. \ref{fig:SomeTAMsd} (b).
We can observe, superimposed to the expected dispersion a systematic trend.
We fit the behaviour with a guess function, combination of negative constant slope and a $10$ second sinusoidal oscillation.
The fit gives the black continuous line on Fig. \ref{fig:DriftCorrection} (a) and is used to correct as a function of time the calibration factor used to convert the measured voltages into meter.\\

On both the raw measured data and the data corrected with the aforementioned method, we compute the ergodic parameter $\epsilon(\Delta)$ probing the dispersion of TA-MSD as a function of lag-time $\Delta$.
On Fig. \ref{fig:DriftCorrection} (b) we show the effect of drift correction.
We show the ergodic parameter $\epsilon(\Delta)$ for a normal Brownian without correction (blue circles) and with calibration correction (red triangles).
Remarkably, the minute correction fitted on the short time-limit of TA-MSD very neatly recast the ergodic parameter on the expected analytical result (black dashed line), computed for a Brownian motion in a potential of stiffness $\kappa = 83 ~ \si{\pico\newton / \micro\meter}$.
This proves that the short time-lag deviation of $\epsilon$ for normal Brownian motion is solely due to drift and not to a physical ergodicity breaking.
On the other hand the ergodic parameter evaluated on the SR process is strongly departing from this trend and does not vanishes for short $\Delta$.
Furthermore, the magnitudes of $\epsilon$ probed are significantly larger and are therefore not affected by the drift (the statistical dispersion of TA-MSD is larger than the systematic trend) as seen in the equality of the ergodic parameter for the raw data (yellow squares) and corrected data (purple stars).
This assesses the validity of the test: the deviation for SR process is not due to a drift, but to a physical ergodicity breaking.

\bibliography{BiblioSR.bib}

\begin{thebibliography}{73}%
\makeatletter
\providecommand \@ifxundefined [1]{%
 \@ifx{#1\undefined}
}%
\providecommand \@ifnum [1]{%
 \ifnum #1\expandafter \@firstoftwo
 \else \expandafter \@secondoftwo
 \fi
}%
\providecommand \@ifx [1]{%
 \ifx #1\expandafter \@firstoftwo
 \else \expandafter \@secondoftwo
 \fi
}%
\providecommand \natexlab [1]{#1}%
\providecommand \enquote  [1]{``#1''}%
\providecommand \bibnamefont  [1]{#1}%
\providecommand \bibfnamefont [1]{#1}%
\providecommand \citenamefont [1]{#1}%
\providecommand \href@noop [0]{\@secondoftwo}%
\providecommand \href [0]{\begingroup \@sanitize@url \@href}%
\providecommand \@href[1]{\@@startlink{#1}\@@href}%
\providecommand \@@href[1]{\endgroup#1\@@endlink}%
\providecommand \@sanitize@url [0]{\catcode `\\12\catcode `\$12\catcode
  `\&12\catcode `\#12\catcode `\^12\catcode `\_12\catcode `\%12\relax}%
\providecommand \@@startlink[1]{}%
\providecommand \@@endlink[0]{}%
\providecommand \url  [0]{\begingroup\@sanitize@url \@url }%
\providecommand \@url [1]{\endgroup\@href {#1}{\urlprefix }}%
\providecommand \urlprefix  [0]{URL }%
\providecommand \Eprint [0]{\href }%
\providecommand \doibase [0]{https://doi.org/}%
\providecommand \selectlanguage [0]{\@gobble}%
\providecommand \bibinfo  [0]{\@secondoftwo}%
\providecommand \bibfield  [0]{\@secondoftwo}%
\providecommand \translation [1]{[#1]}%
\providecommand \BibitemOpen [0]{}%
\providecommand \bibitemStop [0]{}%
\providecommand \bibitemNoStop [0]{.\EOS\space}%
\providecommand \EOS [0]{\spacefactor3000\relax}%
\providecommand \BibitemShut  [1]{\csname bibitem#1\endcsname}%
\let\auto@bib@innerbib\@empty
\bibitem [{\citenamefont {Evans}\ and\ \citenamefont
  {Majumdar}(2011)}]{Evans2011}%
  \BibitemOpen
  \bibfield  {author} {\bibinfo {author} {\bibfnamefont {M.~R.}\ \bibnamefont
  {Evans}}\ and\ \bibinfo {author} {\bibfnamefont {S.~N.}\ \bibnamefont
  {Majumdar}},\ }\bibfield  {title} {\bibinfo {title} {Diffusion with
  stochastic resetting},\ }\href
  {https://doi.org/10.1103/PhysRevLett.106.160601} {\bibfield  {journal}
  {\bibinfo  {journal} {Phys. Rev. Lett.}\ }\textbf {\bibinfo {volume} {106}},\
  \bibinfo {pages} {160601} (\bibinfo {year} {2011})}\BibitemShut {NoStop}%
\bibitem [{\citenamefont {Fuchs}\ \emph {et~al.}(2016)\citenamefont {Fuchs},
  \citenamefont {Goldt},\ and\ \citenamefont {Seifert}}]{Fuchs2016}%
  \BibitemOpen
  \bibfield  {author} {\bibinfo {author} {\bibfnamefont {J.}~\bibnamefont
  {Fuchs}}, \bibinfo {author} {\bibfnamefont {S.}~\bibnamefont {Goldt}},\ and\
  \bibinfo {author} {\bibfnamefont {U.}~\bibnamefont {Seifert}},\ }\bibfield
  {title} {\bibinfo {title} {Stochastic thermodynamics of resetting},\ }\href
  {https://doi.org/10.1209/0295-5075/113/60009} {\bibfield  {journal} {\bibinfo
   {journal} {EPL (Europhysics Letters)}\ }\textbf {\bibinfo {volume} {113}},\
  \bibinfo {pages} {60009} (\bibinfo {year} {2016})}\BibitemShut {NoStop}%
\bibitem [{\citenamefont {Rold{\'a}n}\ \emph {et~al.}(2014)\citenamefont
  {Rold{\'a}n}, \citenamefont {Mart{\'i}nez}, \citenamefont {Parrondo},\ and\
  \citenamefont {Petrov}}]{Roldan2014}%
  \BibitemOpen
  \bibfield  {author} {\bibinfo {author} {\bibfnamefont {{\'E}.}~\bibnamefont
  {Rold{\'a}n}}, \bibinfo {author} {\bibfnamefont {I.~A.}\ \bibnamefont
  {Mart{\'i}nez}}, \bibinfo {author} {\bibfnamefont {J.~M.~R.}\ \bibnamefont
  {Parrondo}},\ and\ \bibinfo {author} {\bibfnamefont {D.}~\bibnamefont
  {Petrov}},\ }\bibfield  {title} {\bibinfo {title} {Universal features in the
  energetics of symmetry breaking},\ }\href {https://doi.org/10.1038/nphys2940}
  {\bibfield  {journal} {\bibinfo  {journal} {Nature Physics}\ }\textbf
  {\bibinfo {volume} {10}},\ \bibinfo {pages} {457} (\bibinfo {year}
  {2014})}\BibitemShut {NoStop}%
\bibitem [{\citenamefont {Ciliberto}\ and\ \citenamefont
  {Lutz}(2019)}]{Ciliberto2019}%
  \BibitemOpen
  \bibfield  {author} {\bibinfo {author} {\bibfnamefont {S.}~\bibnamefont
  {Ciliberto}}\ and\ \bibinfo {author} {\bibfnamefont {E.}~\bibnamefont
  {Lutz}},\ }\bibfield  {title} {\bibinfo {title} {The physics of information:
  From maxwell to landauer},\ }\href@noop {} {\bibfield  {journal} {\bibinfo
  {journal} {Energy Limits in Computation: A Review of Landauer’s Principle,
  Theory and Experiments}\ ,\ \bibinfo {pages} {155}} (\bibinfo {year}
  {2019})}\BibitemShut {NoStop}%
\bibitem [{\citenamefont {Lutz}\ and\ \citenamefont
  {Ciliberto}(2015)}]{Lutz2015}%
  \BibitemOpen
  \bibfield  {author} {\bibinfo {author} {\bibfnamefont {E.}~\bibnamefont
  {Lutz}}\ and\ \bibinfo {author} {\bibfnamefont {S.}~\bibnamefont
  {Ciliberto}},\ }\bibfield  {title} {\bibinfo {title} {Information: From
  maxwell’s demon to landauer’s eraser},\ }\href
  {https://doi.org/10.1063/PT.3.2912} {\bibfield  {journal} {\bibinfo
  {journal} {Physics Today}\ }\textbf {\bibinfo {volume} {68}},\ \bibinfo
  {pages} {30} (\bibinfo {year} {2015})},\ \Eprint
  {https://arxiv.org/abs/https://doi.org/10.1063/PT.3.2912}
  {https://doi.org/10.1063/PT.3.2912} \BibitemShut {NoStop}%
\bibitem [{\citenamefont {Manrubia}\ and\ \citenamefont
  {Zanette}(1999)}]{Manrubia1999}%
  \BibitemOpen
  \bibfield  {author} {\bibinfo {author} {\bibfnamefont {S.~C.}\ \bibnamefont
  {Manrubia}}\ and\ \bibinfo {author} {\bibfnamefont {D.~H.}\ \bibnamefont
  {Zanette}},\ }\bibfield  {title} {\bibinfo {title} {Stochastic multiplicative
  processes with reset events},\ }\href
  {https://doi.org/10.1103/PhysRevE.59.4945} {\bibfield  {journal} {\bibinfo
  {journal} {Phys. Rev. E}\ }\textbf {\bibinfo {volume} {59}},\ \bibinfo
  {pages} {4945} (\bibinfo {year} {1999})}\BibitemShut {NoStop}%
\bibitem [{\citenamefont {Kusmierz}\ \emph {et~al.}(2014)\citenamefont
  {Kusmierz}, \citenamefont {Majumdar}, \citenamefont {Sabhapandit},\ and\
  \citenamefont {Schehr}}]{kusmierz2014first}%
  \BibitemOpen
  \bibfield  {author} {\bibinfo {author} {\bibfnamefont {L.}~\bibnamefont
  {Kusmierz}}, \bibinfo {author} {\bibfnamefont {S.~N.}\ \bibnamefont
  {Majumdar}}, \bibinfo {author} {\bibfnamefont {S.}~\bibnamefont
  {Sabhapandit}},\ and\ \bibinfo {author} {\bibfnamefont {G.}~\bibnamefont
  {Schehr}},\ }\bibfield  {title} {\bibinfo {title} {First order transition for
  the optimal search time of l{\'e}vy flights with resetting},\ }\href@noop {}
  {\bibfield  {journal} {\bibinfo  {journal} {Physical review letters}\
  }\textbf {\bibinfo {volume} {113}},\ \bibinfo {pages} {220602} (\bibinfo
  {year} {2014})}\BibitemShut {NoStop}%
\bibitem [{\citenamefont {Montero}\ \emph {et~al.}(2017)\citenamefont
  {Montero}, \citenamefont {Mas{\'o}-Puigdellosas},\ and\ \citenamefont
  {Villarroel}}]{Montero2017}%
  \BibitemOpen
  \bibfield  {author} {\bibinfo {author} {\bibfnamefont {M.}~\bibnamefont
  {Montero}}, \bibinfo {author} {\bibfnamefont {A.}~\bibnamefont
  {Mas{\'o}-Puigdellosas}},\ and\ \bibinfo {author} {\bibfnamefont
  {J.}~\bibnamefont {Villarroel}},\ }\bibfield  {title} {\bibinfo {title}
  {Continuous-time random walks with reset events},\ }\href
  {https://doi.org/10.1140/epjb/e2017-80348-4} {\bibfield  {journal} {\bibinfo
  {journal} {The European Physical Journal B}\ }\textbf {\bibinfo {volume}
  {90}},\ \bibinfo {pages} {176} (\bibinfo {year} {2017})}\BibitemShut
  {NoStop}%
\bibitem [{\citenamefont {Chechkin}\ and\ \citenamefont
  {Sokolov}(2018)}]{Chechkin2018}%
  \BibitemOpen
  \bibfield  {author} {\bibinfo {author} {\bibfnamefont {A.}~\bibnamefont
  {Chechkin}}\ and\ \bibinfo {author} {\bibfnamefont {I.~M.}\ \bibnamefont
  {Sokolov}},\ }\bibfield  {title} {\bibinfo {title} {Random search with
  resetting: A unified renewal approach},\ }\href
  {https://doi.org/10.1103/PhysRevLett.121.050601} {\bibfield  {journal}
  {\bibinfo  {journal} {Phys. Rev. Lett.}\ }\textbf {\bibinfo {volume} {121}},\
  \bibinfo {pages} {050601} (\bibinfo {year} {2018})}\BibitemShut {NoStop}%
\bibitem [{\citenamefont {Evans}\ \emph {et~al.}(2020)\citenamefont {Evans},
  \citenamefont {Majumdar},\ and\ \citenamefont {Schehr}}]{Evans2020}%
  \BibitemOpen
  \bibfield  {author} {\bibinfo {author} {\bibfnamefont {M.~R.}\ \bibnamefont
  {Evans}}, \bibinfo {author} {\bibfnamefont {S.~N.}\ \bibnamefont
  {Majumdar}},\ and\ \bibinfo {author} {\bibfnamefont {G.}~\bibnamefont
  {Schehr}},\ }\bibfield  {title} {\bibinfo {title} {Stochastic resetting and
  applications},\ }\href {https://doi.org/10.1088/1751-8121/ab7cfe} {\bibfield
  {journal} {\bibinfo  {journal} {Journal of Physics A: Mathematical and
  Theoretical}\ }\textbf {\bibinfo {volume} {53}},\ \bibinfo {pages} {193001}
  (\bibinfo {year} {2020})}\BibitemShut {NoStop}%
\bibitem [{\citenamefont {Magoni}\ \emph {et~al.}(2020)\citenamefont {Magoni},
  \citenamefont {Majumdar},\ and\ \citenamefont {Schehr}}]{Magoni2020}%
  \BibitemOpen
  \bibfield  {author} {\bibinfo {author} {\bibfnamefont {M.}~\bibnamefont
  {Magoni}}, \bibinfo {author} {\bibfnamefont {S.~N.}\ \bibnamefont
  {Majumdar}},\ and\ \bibinfo {author} {\bibfnamefont {G.}~\bibnamefont
  {Schehr}},\ }\bibfield  {title} {\bibinfo {title} {Ising model with
  stochastic resetting},\ }\href
  {https://doi.org/10.1103/PhysRevResearch.2.033182} {\bibfield  {journal}
  {\bibinfo  {journal} {Phys. Rev. Research}\ }\textbf {\bibinfo {volume}
  {2}},\ \bibinfo {pages} {033182} (\bibinfo {year} {2020})}\BibitemShut
  {NoStop}%
\bibitem [{\citenamefont {Kumar}\ \emph {et~al.}(2020)\citenamefont {Kumar},
  \citenamefont {Sadekar},\ and\ \citenamefont {Basu}}]{Kumar2020}%
  \BibitemOpen
  \bibfield  {author} {\bibinfo {author} {\bibfnamefont {V.}~\bibnamefont
  {Kumar}}, \bibinfo {author} {\bibfnamefont {O.}~\bibnamefont {Sadekar}},\
  and\ \bibinfo {author} {\bibfnamefont {U.}~\bibnamefont {Basu}},\ }\bibfield
  {title} {\bibinfo {title} {Active brownian motion in two dimensions under
  stochastic resetting},\ }\href {https://doi.org/10.1103/PhysRevE.102.052129}
  {\bibfield  {journal} {\bibinfo  {journal} {Phys. Rev. E}\ }\textbf {\bibinfo
  {volume} {102}},\ \bibinfo {pages} {052129} (\bibinfo {year}
  {2020})}\BibitemShut {NoStop}%
\bibitem [{\citenamefont {Perfetto}\ \emph {et~al.}(2021)\citenamefont
  {Perfetto}, \citenamefont {Carollo}, \citenamefont {Magoni},\ and\
  \citenamefont {Lesanovsky}}]{Perfetto2021}%
  \BibitemOpen
  \bibfield  {author} {\bibinfo {author} {\bibfnamefont {G.}~\bibnamefont
  {Perfetto}}, \bibinfo {author} {\bibfnamefont {F.}~\bibnamefont {Carollo}},
  \bibinfo {author} {\bibfnamefont {M.}~\bibnamefont {Magoni}},\ and\ \bibinfo
  {author} {\bibfnamefont {I.}~\bibnamefont {Lesanovsky}},\ }\bibfield  {title}
  {\bibinfo {title} {Designing nonequilibrium states of quantum matter through
  stochastic resetting},\ }\href {https://doi.org/10.1103/PhysRevB.104.L180302}
  {\bibfield  {journal} {\bibinfo  {journal} {Phys. Rev. B}\ }\textbf {\bibinfo
  {volume} {104}},\ \bibinfo {pages} {L180302} (\bibinfo {year}
  {2021})}\BibitemShut {NoStop}%
\bibitem [{\citenamefont {Stanislavsky}\ and\ \citenamefont
  {Weron}(2021)}]{Stanislavsky2021}%
  \BibitemOpen
  \bibfield  {author} {\bibinfo {author} {\bibfnamefont {A.}~\bibnamefont
  {Stanislavsky}}\ and\ \bibinfo {author} {\bibfnamefont {A.}~\bibnamefont
  {Weron}},\ }\bibfield  {title} {\bibinfo {title} {Optimal non-gaussian search
  with stochastic resetting},\ }\href
  {https://doi.org/10.1103/PhysRevE.104.014125} {\bibfield  {journal} {\bibinfo
   {journal} {Phys. Rev. E}\ }\textbf {\bibinfo {volume} {104}},\ \bibinfo
  {pages} {014125} (\bibinfo {year} {2021})}\BibitemShut {NoStop}%
\bibitem [{\citenamefont {Boyer}\ and\ \citenamefont
  {Solis-Salas}(2014)}]{Boyer2014}%
  \BibitemOpen
  \bibfield  {author} {\bibinfo {author} {\bibfnamefont {D.}~\bibnamefont
  {Boyer}}\ and\ \bibinfo {author} {\bibfnamefont {C.}~\bibnamefont
  {Solis-Salas}},\ }\bibfield  {title} {\bibinfo {title} {Random walks with
  preferential relocations to places visited in the past and their application
  to biology},\ }\href {https://doi.org/10.1103/PhysRevLett.112.240601}
  {\bibfield  {journal} {\bibinfo  {journal} {Phys. Rev. Lett.}\ }\textbf
  {\bibinfo {volume} {112}},\ \bibinfo {pages} {240601} (\bibinfo {year}
  {2014})}\BibitemShut {NoStop}%
\bibitem [{\citenamefont {Rold\'an}\ \emph {et~al.}(2016)\citenamefont
  {Rold\'an}, \citenamefont {Lisica}, \citenamefont {S\'anchez-Taltavull},\
  and\ \citenamefont {Grill}}]{Roldan2016}%
  \BibitemOpen
  \bibfield  {author} {\bibinfo {author} {\bibfnamefont {E.}~\bibnamefont
  {Rold\'an}}, \bibinfo {author} {\bibfnamefont {A.}~\bibnamefont {Lisica}},
  \bibinfo {author} {\bibfnamefont {D.}~\bibnamefont {S\'anchez-Taltavull}},\
  and\ \bibinfo {author} {\bibfnamefont {S.~W.}\ \bibnamefont {Grill}},\
  }\bibfield  {title} {\bibinfo {title} {Stochastic resetting in backtrack
  recovery by rna polymerases},\ }\href
  {https://doi.org/10.1103/PhysRevE.93.062411} {\bibfield  {journal} {\bibinfo
  {journal} {Phys. Rev. E}\ }\textbf {\bibinfo {volume} {93}},\ \bibinfo
  {pages} {062411} (\bibinfo {year} {2016})}\BibitemShut {NoStop}%
\bibitem [{\citenamefont {Bressloff}(2020)}]{Bressloff2020}%
  \BibitemOpen
  \bibfield  {author} {\bibinfo {author} {\bibfnamefont {P.~C.}\ \bibnamefont
  {Bressloff}},\ }\bibfield  {title} {\bibinfo {title} {Modeling active
  cellular transport as a directed search process with stochastic resetting and
  delays},\ }\href {https://doi.org/10.1088/1751-8121/ab9fb7} {\bibfield
  {journal} {\bibinfo  {journal} {Journal of Physics A: Mathematical and
  Theoretical}\ }\textbf {\bibinfo {volume} {53}},\ \bibinfo {pages} {355001}
  (\bibinfo {year} {2020})}\BibitemShut {NoStop}%
\bibitem [{\citenamefont {Roberts}\ \emph {et~al.}(2024)\citenamefont
  {Roberts}, \citenamefont {Sezik},\ and\ \citenamefont
  {Lardet}}]{roberts2024ratchet}%
  \BibitemOpen
  \bibfield  {author} {\bibinfo {author} {\bibfnamefont {C.}~\bibnamefont
  {Roberts}}, \bibinfo {author} {\bibfnamefont {E.}~\bibnamefont {Sezik}},\
  and\ \bibinfo {author} {\bibfnamefont {E.}~\bibnamefont {Lardet}},\
  }\bibfield  {title} {\bibinfo {title} {Ratchet-mediated resetting: Current,
  efficiency, and exact solution},\ }\href@noop {} {\bibfield  {journal}
  {\bibinfo  {journal} {arXiv preprint arXiv:2405.10698}\ } (\bibinfo {year}
  {2024})}\BibitemShut {NoStop}%
\bibitem [{\citenamefont {Pal}\ \emph {et~al.}(2024)\citenamefont {Pal},
  \citenamefont {Boyer}, \citenamefont {Dagdug},\ and\ \citenamefont
  {Pal}}]{pal2024channel}%
  \BibitemOpen
  \bibfield  {author} {\bibinfo {author} {\bibfnamefont {S.}~\bibnamefont
  {Pal}}, \bibinfo {author} {\bibfnamefont {D.}~\bibnamefont {Boyer}}, \bibinfo
  {author} {\bibfnamefont {L.}~\bibnamefont {Dagdug}},\ and\ \bibinfo {author}
  {\bibfnamefont {A.}~\bibnamefont {Pal}},\ }\bibfield  {title} {\bibinfo
  {title} {Channel-facilitated transport under resetting dynamics},\
  }\href@noop {} {\bibfield  {journal} {\bibinfo  {journal} {arXiv preprint
  arXiv:2407.14157}\ } (\bibinfo {year} {2024})}\BibitemShut {NoStop}%
\bibitem [{\citenamefont {Blumer}\ \emph {et~al.}(2022)\citenamefont {Blumer},
  \citenamefont {Reuveni},\ and\ \citenamefont {Hirshberg}}]{Blumer2022}%
  \BibitemOpen
  \bibfield  {author} {\bibinfo {author} {\bibfnamefont {O.}~\bibnamefont
  {Blumer}}, \bibinfo {author} {\bibfnamefont {S.}~\bibnamefont {Reuveni}},\
  and\ \bibinfo {author} {\bibfnamefont {B.}~\bibnamefont {Hirshberg}},\
  }\bibfield  {title} {\bibinfo {title} {Stochastic resetting for enhanced
  sampling},\ }\href {https://doi.org/10.1021/acs.jpclett.2c03055} {\bibfield
  {journal} {\bibinfo  {journal} {The Journal of Physical Chemistry Letters}\
  ,\ \bibinfo {pages} {11230}} (\bibinfo {year} {2022})}\BibitemShut {NoStop}%
\bibitem [{\citenamefont {Blumer}\ \emph {et~al.}(2024)\citenamefont {Blumer},
  \citenamefont {Reuveni},\ and\ \citenamefont
  {Hirshberg}}]{blumer2024combining}%
  \BibitemOpen
  \bibfield  {author} {\bibinfo {author} {\bibfnamefont {O.}~\bibnamefont
  {Blumer}}, \bibinfo {author} {\bibfnamefont {S.}~\bibnamefont {Reuveni}},\
  and\ \bibinfo {author} {\bibfnamefont {B.}~\bibnamefont {Hirshberg}},\
  }\bibfield  {title} {\bibinfo {title} {Combining stochastic resetting with
  metadynamics to speed-up molecular dynamics simulations},\ }\href@noop {}
  {\bibfield  {journal} {\bibinfo  {journal} {Nature Communications}\ }\textbf
  {\bibinfo {volume} {15}},\ \bibinfo {pages} {240} (\bibinfo {year}
  {2024})}\BibitemShut {NoStop}%
\bibitem [{\citenamefont {Tal-Friedman}\ \emph {et~al.}(2020)\citenamefont
  {Tal-Friedman}, \citenamefont {Pal}, \citenamefont {Sekhon}, \citenamefont
  {Reuveni},\ and\ \citenamefont {Roichman}}]{TalFriedman2020}%
  \BibitemOpen
  \bibfield  {author} {\bibinfo {author} {\bibfnamefont {O.}~\bibnamefont
  {Tal-Friedman}}, \bibinfo {author} {\bibfnamefont {A.}~\bibnamefont {Pal}},
  \bibinfo {author} {\bibfnamefont {A.}~\bibnamefont {Sekhon}}, \bibinfo
  {author} {\bibfnamefont {S.}~\bibnamefont {Reuveni}},\ and\ \bibinfo {author}
  {\bibfnamefont {Y.}~\bibnamefont {Roichman}},\ }\bibfield  {title} {\bibinfo
  {title} {Experimental realization of diffusion with stochastic resetting},\
  }\href {https://doi.org/10.1021/acs.jpclett.0c02122} {\bibfield  {journal}
  {\bibinfo  {journal} {The Journal of Physical Chemistry Letters}\ }\textbf
  {\bibinfo {volume} {11}},\ \bibinfo {pages} {7350} (\bibinfo {year}
  {2020})}\BibitemShut {NoStop}%
\bibitem [{\citenamefont {Besga}\ \emph {et~al.}(2020)\citenamefont {Besga},
  \citenamefont {Bovon}, \citenamefont {Petrosyan}, \citenamefont {Majumdar},\
  and\ \citenamefont {Ciliberto}}]{Besga2020}%
  \BibitemOpen
  \bibfield  {author} {\bibinfo {author} {\bibfnamefont {B.}~\bibnamefont
  {Besga}}, \bibinfo {author} {\bibfnamefont {A.}~\bibnamefont {Bovon}},
  \bibinfo {author} {\bibfnamefont {A.}~\bibnamefont {Petrosyan}}, \bibinfo
  {author} {\bibfnamefont {S.~N.}\ \bibnamefont {Majumdar}},\ and\ \bibinfo
  {author} {\bibfnamefont {S.}~\bibnamefont {Ciliberto}},\ }\bibfield  {title}
  {\bibinfo {title} {Optimal mean first-passage time for a brownian searcher
  subjected to resetting: Experimental and theoretical results},\ }\href
  {https://doi.org/10.1103/PhysRevResearch.2.032029} {\bibfield  {journal}
  {\bibinfo  {journal} {Phys. Rev. Research}\ }\textbf {\bibinfo {volume}
  {2}},\ \bibinfo {pages} {032029} (\bibinfo {year} {2020})}\BibitemShut
  {NoStop}%
\bibitem [{\citenamefont {Besga}\ \emph {et~al.}(2021)\citenamefont {Besga},
  \citenamefont {Faisant}, \citenamefont {Petrosyan}, \citenamefont
  {Ciliberto},\ and\ \citenamefont {Majumdar}}]{Besga2021}%
  \BibitemOpen
  \bibfield  {author} {\bibinfo {author} {\bibfnamefont {B.}~\bibnamefont
  {Besga}}, \bibinfo {author} {\bibfnamefont {F.}~\bibnamefont {Faisant}},
  \bibinfo {author} {\bibfnamefont {A.}~\bibnamefont {Petrosyan}}, \bibinfo
  {author} {\bibfnamefont {S.}~\bibnamefont {Ciliberto}},\ and\ \bibinfo
  {author} {\bibfnamefont {S.~N.}\ \bibnamefont {Majumdar}},\ }\bibfield
  {title} {\bibinfo {title} {Dynamical phase transition in the first-passage
  probability of a brownian motion},\ }\href
  {https://doi.org/10.1103/PhysRevE.104.L012102} {\bibfield  {journal}
  {\bibinfo  {journal} {Phys. Rev. E}\ }\textbf {\bibinfo {volume} {104}},\
  \bibinfo {pages} {L012102} (\bibinfo {year} {2021})}\BibitemShut {NoStop}%
\bibitem [{\citenamefont {Altshuler}\ \emph {et~al.}(2024)\citenamefont
  {Altshuler}, \citenamefont {Bonomo}, \citenamefont {Gorohovsky},
  \citenamefont {Marchini}, \citenamefont {Rosen}, \citenamefont
  {Tal-Friedman}, \citenamefont {Reuveni},\ and\ \citenamefont
  {Roichman}}]{Altshuler2023}%
  \BibitemOpen
  \bibfield  {author} {\bibinfo {author} {\bibfnamefont {A.}~\bibnamefont
  {Altshuler}}, \bibinfo {author} {\bibfnamefont {O.~L.}\ \bibnamefont
  {Bonomo}}, \bibinfo {author} {\bibfnamefont {N.}~\bibnamefont {Gorohovsky}},
  \bibinfo {author} {\bibfnamefont {S.}~\bibnamefont {Marchini}}, \bibinfo
  {author} {\bibfnamefont {E.}~\bibnamefont {Rosen}}, \bibinfo {author}
  {\bibfnamefont {O.}~\bibnamefont {Tal-Friedman}}, \bibinfo {author}
  {\bibfnamefont {S.}~\bibnamefont {Reuveni}},\ and\ \bibinfo {author}
  {\bibfnamefont {Y.}~\bibnamefont {Roichman}},\ }\bibfield  {title} {\bibinfo
  {title} {Environmental memory facilitates search with home returns},\
  }\href@noop {} {\bibfield  {journal} {\bibinfo  {journal} {Physical Review
  Research}\ }\textbf {\bibinfo {volume} {6}},\ \bibinfo {pages} {023255}
  (\bibinfo {year} {2024})}\BibitemShut {NoStop}%
\bibitem [{\citenamefont {Sokolov}(2023)}]{Sokolov2023}%
  \BibitemOpen
  \bibfield  {author} {\bibinfo {author} {\bibfnamefont {I.~M.}\ \bibnamefont
  {Sokolov}},\ }\bibfield  {title} {\bibinfo {title} {Linear response and
  fluctuation-dissipation relations for brownian motion under resetting},\
  }\href {https://doi.org/10.1103/PhysRevLett.130.067101} {\bibfield  {journal}
  {\bibinfo  {journal} {Phys. Rev. Lett.}\ }\textbf {\bibinfo {volume} {130}},\
  \bibinfo {pages} {067101} (\bibinfo {year} {2023})}\BibitemShut {NoStop}%
\bibitem [{\citenamefont {Evans}\ \emph {et~al.}(2013)\citenamefont {Evans},
  \citenamefont {Majumdar},\ and\ \citenamefont {Mallick}}]{evans2013optimal}%
  \BibitemOpen
  \bibfield  {author} {\bibinfo {author} {\bibfnamefont {M.~R.}\ \bibnamefont
  {Evans}}, \bibinfo {author} {\bibfnamefont {S.~N.}\ \bibnamefont
  {Majumdar}},\ and\ \bibinfo {author} {\bibfnamefont {K.}~\bibnamefont
  {Mallick}},\ }\bibfield  {title} {\bibinfo {title} {Optimal diffusive search:
  nonequilibrium resetting versus equilibrium dynamics},\ }\href@noop {}
  {\bibfield  {journal} {\bibinfo  {journal} {Journal of Physics A:
  Mathematical and Theoretical}\ }\textbf {\bibinfo {volume} {46}},\ \bibinfo
  {pages} {185001} (\bibinfo {year} {2013})}\BibitemShut {NoStop}%
\bibitem [{\citenamefont {Sunil}\ \emph {et~al.}(2023)\citenamefont {Sunil},
  \citenamefont {Blythe}, \citenamefont {Evans},\ and\ \citenamefont
  {Majumdar}}]{Sunil2023}%
  \BibitemOpen
  \bibfield  {author} {\bibinfo {author} {\bibfnamefont {J.~C.}\ \bibnamefont
  {Sunil}}, \bibinfo {author} {\bibfnamefont {R.~A.}\ \bibnamefont {Blythe}},
  \bibinfo {author} {\bibfnamefont {M.~R.}\ \bibnamefont {Evans}},\ and\
  \bibinfo {author} {\bibfnamefont {S.~N.}\ \bibnamefont {Majumdar}},\
  }\bibfield  {title} {\bibinfo {title} {The cost of stochastic resetting},\
  }\bibfield  {journal} {\bibinfo  {journal} {Arxiv}\ }\href
  {https://doi.org/arXiv:2304.09348} {arXiv:2304.09348} (\bibinfo {year}
  {2023})\BibitemShut {NoStop}%
\bibitem [{\citenamefont {Pal}\ and\ \citenamefont {Rahav}(2017)}]{Pal2017}%
  \BibitemOpen
  \bibfield  {author} {\bibinfo {author} {\bibfnamefont {A.}~\bibnamefont
  {Pal}}\ and\ \bibinfo {author} {\bibfnamefont {S.}~\bibnamefont {Rahav}},\
  }\bibfield  {title} {\bibinfo {title} {Integral fluctuation theorems for
  stochastic resetting systems},\ }\href
  {https://doi.org/10.1103/PhysRevE.96.062135} {\bibfield  {journal} {\bibinfo
  {journal} {Phys. Rev. E}\ }\textbf {\bibinfo {volume} {96}},\ \bibinfo
  {pages} {062135} (\bibinfo {year} {2017})}\BibitemShut {NoStop}%
\bibitem [{\citenamefont {Gupta}\ \emph
  {et~al.}(2020{\natexlab{a}})\citenamefont {Gupta}, \citenamefont {Plata},\
  and\ \citenamefont {Pal}}]{Gupta2020_thermo}%
  \BibitemOpen
  \bibfield  {author} {\bibinfo {author} {\bibfnamefont {D.}~\bibnamefont
  {Gupta}}, \bibinfo {author} {\bibfnamefont {C.~A.}\ \bibnamefont {Plata}},\
  and\ \bibinfo {author} {\bibfnamefont {A.}~\bibnamefont {Pal}},\ }\bibfield
  {title} {\bibinfo {title} {Work fluctuations and jarzynski equality in
  stochastic resetting},\ }\href
  {https://doi.org/10.1103/PhysRevLett.124.110608} {\bibfield  {journal}
  {\bibinfo  {journal} {Phys. Rev. Lett.}\ }\textbf {\bibinfo {volume} {124}},\
  \bibinfo {pages} {110608} (\bibinfo {year} {2020}{\natexlab{a}})}\BibitemShut
  {NoStop}%
\bibitem [{\citenamefont {Pal}\ \emph {et~al.}(2021)\citenamefont {Pal},
  \citenamefont {Reuveni},\ and\ \citenamefont {Rahav}}]{Pal2021}%
  \BibitemOpen
  \bibfield  {author} {\bibinfo {author} {\bibfnamefont {A.}~\bibnamefont
  {Pal}}, \bibinfo {author} {\bibfnamefont {S.}~\bibnamefont {Reuveni}},\ and\
  \bibinfo {author} {\bibfnamefont {S.}~\bibnamefont {Rahav}},\ }\bibfield
  {title} {\bibinfo {title} {Thermodynamic uncertainty relation for systems
  with unidirectional transitions},\ }\href
  {https://doi.org/10.1103/PhysRevResearch.3.013273} {\bibfield  {journal}
  {\bibinfo  {journal} {Phys. Rev. Research}\ }\textbf {\bibinfo {volume}
  {3}},\ \bibinfo {pages} {013273} (\bibinfo {year} {2021})}\BibitemShut
  {NoStop}%
\bibitem [{\citenamefont {Gupta}\ and\ \citenamefont
  {Jayannavar}(2022)}]{Gupta2022}%
  \BibitemOpen
  \bibfield  {author} {\bibinfo {author} {\bibfnamefont {S.}~\bibnamefont
  {Gupta}}\ and\ \bibinfo {author} {\bibfnamefont {A.~M.}\ \bibnamefont
  {Jayannavar}},\ }\bibfield  {title} {\bibinfo {title} {Stochastic resetting:
  A (very) brief review},\ }\bibfield  {journal} {\bibinfo  {journal}
  {Frontiers in Physics}\ }\textbf {\bibinfo {volume} {10}},\ \href
  {https://doi.org/10.3389/fphy.2022.789097} {10.3389/fphy.2022.789097}
  (\bibinfo {year} {2022})\BibitemShut {NoStop}%
\bibitem [{\citenamefont {Gupta}\ and\ \citenamefont
  {Plata}(2022)}]{Gupta2022Thermo}%
  \BibitemOpen
  \bibfield  {author} {\bibinfo {author} {\bibfnamefont {D.}~\bibnamefont
  {Gupta}}\ and\ \bibinfo {author} {\bibfnamefont {C.~A.}\ \bibnamefont
  {Plata}},\ }\bibfield  {title} {\bibinfo {title} {Work fluctuations for
  diffusion dynamics submitted to stochastic return},\ }\href
  {https://doi.org/10.1088/1367-2630/aca25e} {\bibfield  {journal} {\bibinfo
  {journal} {New Journal of Physics}\ }\textbf {\bibinfo {volume} {24}},\
  \bibinfo {pages} {113034} (\bibinfo {year} {2022})}\BibitemShut {NoStop}%
\bibitem [{\citenamefont {Mori}\ \emph {et~al.}(2023)\citenamefont {Mori},
  \citenamefont {Olsen},\ and\ \citenamefont {Krishnamurthy}}]{Mori2022}%
  \BibitemOpen
  \bibfield  {author} {\bibinfo {author} {\bibfnamefont {F.}~\bibnamefont
  {Mori}}, \bibinfo {author} {\bibfnamefont {K.~S.}\ \bibnamefont {Olsen}},\
  and\ \bibinfo {author} {\bibfnamefont {S.}~\bibnamefont {Krishnamurthy}},\
  }\bibfield  {title} {\bibinfo {title} {Entropy production of resetting
  processes},\ }\href {https://doi.org/10.1103/PhysRevResearch.5.023103}
  {\bibfield  {journal} {\bibinfo  {journal} {Phys. Rev. Res.}\ }\textbf
  {\bibinfo {volume} {5}},\ \bibinfo {pages} {023103} (\bibinfo {year}
  {2023})}\BibitemShut {NoStop}%
\bibitem [{\citenamefont {Toyabe}\ \emph {et~al.}(2010)\citenamefont {Toyabe},
  \citenamefont {Sagawa}, \citenamefont {Ueda}, \citenamefont {Muneyuki},\ and\
  \citenamefont {Sano}}]{toyabe_experimental_2010}%
  \BibitemOpen
  \bibfield  {author} {\bibinfo {author} {\bibfnamefont {S.}~\bibnamefont
  {Toyabe}}, \bibinfo {author} {\bibfnamefont {T.}~\bibnamefont {Sagawa}},
  \bibinfo {author} {\bibfnamefont {M.}~\bibnamefont {Ueda}}, \bibinfo {author}
  {\bibfnamefont {E.}~\bibnamefont {Muneyuki}},\ and\ \bibinfo {author}
  {\bibfnamefont {M.}~\bibnamefont {Sano}},\ }\bibfield  {title} {\bibinfo
  {title} {Experimental demonstration of information to energy conversion and
  validation of the generalized jarzynski equality},\ }\href
  {https://doi.org/10.1038/nphys1821} {\bibfield  {journal} {\bibinfo
  {journal} {Nature Physics}\ }\textbf {\bibinfo {volume} {6}},\ \bibinfo
  {pages} {988} (\bibinfo {year} {2010})}\BibitemShut {NoStop}%
\bibitem [{\citenamefont {Admon}\ \emph {et~al.}(2018)\citenamefont {Admon},
  \citenamefont {Rahav},\ and\ \citenamefont
  {Roichman}}]{admon_experimental_2018}%
  \BibitemOpen
  \bibfield  {author} {\bibinfo {author} {\bibfnamefont {T.}~\bibnamefont
  {Admon}}, \bibinfo {author} {\bibfnamefont {S.}~\bibnamefont {Rahav}},\ and\
  \bibinfo {author} {\bibfnamefont {Y.}~\bibnamefont {Roichman}},\ }\bibfield
  {title} {\bibinfo {title} {Experimental realization of an information machine
  with tunable temporal correlations},\ }\href
  {https://doi.org/10.1103/PhysRevLett.121.180601} {\bibfield  {journal}
  {\bibinfo  {journal} {Physical Review Letters}\ }\textbf {\bibinfo {volume}
  {121}},\ \bibinfo {pages} {180601} (\bibinfo {year} {2018})}\BibitemShut
  {NoStop}%
\bibitem [{\citenamefont {Gupta}\ \emph
  {et~al.}(2020{\natexlab{b}})\citenamefont {Gupta}, \citenamefont {Plata},
  \citenamefont {Kundu},\ and\ \citenamefont {Pal}}]{Gupta2020_NonInst}%
  \BibitemOpen
  \bibfield  {author} {\bibinfo {author} {\bibfnamefont {D.}~\bibnamefont
  {Gupta}}, \bibinfo {author} {\bibfnamefont {C.~A.}\ \bibnamefont {Plata}},
  \bibinfo {author} {\bibfnamefont {A.}~\bibnamefont {Kundu}},\ and\ \bibinfo
  {author} {\bibfnamefont {A.}~\bibnamefont {Pal}},\ }\bibfield  {title}
  {\bibinfo {title} {Stochastic resetting with stochastic returns using
  external trap},\ }\href {https://doi.org/10.1088/1751-8121/abcf0b} {\bibfield
   {journal} {\bibinfo  {journal} {Journal of Physics A: Mathematical and
  Theoretical}\ }\textbf {\bibinfo {volume} {54}},\ \bibinfo {pages} {025003}
  (\bibinfo {year} {2020}{\natexlab{b}})}\BibitemShut {NoStop}%
\bibitem [{\citenamefont {Bodrova}\ and\ \citenamefont
  {Sokolov}(2020)}]{Bodrova2020}%
  \BibitemOpen
  \bibfield  {author} {\bibinfo {author} {\bibfnamefont {A.~S.}\ \bibnamefont
  {Bodrova}}\ and\ \bibinfo {author} {\bibfnamefont {I.~M.}\ \bibnamefont
  {Sokolov}},\ }\bibfield  {title} {\bibinfo {title} {Resetting processes with
  noninstantaneous return},\ }\href
  {https://doi.org/10.1103/PhysRevE.101.052130} {\bibfield  {journal} {\bibinfo
   {journal} {Phys. Rev. E}\ }\textbf {\bibinfo {volume} {101}},\ \bibinfo
  {pages} {052130} (\bibinfo {year} {2020})}\BibitemShut {NoStop}%
\bibitem [{\citenamefont {Mercado-V{\'{a}}squez}\ \emph
  {et~al.}(2020)\citenamefont {Mercado-V{\'{a}}squez}, \citenamefont {Boyer},
  \citenamefont {Majumdar},\ and\ \citenamefont {Schehr}}]{MercadoVasquez2020}%
  \BibitemOpen
  \bibfield  {author} {\bibinfo {author} {\bibfnamefont {G.}~\bibnamefont
  {Mercado-V{\'{a}}squez}}, \bibinfo {author} {\bibfnamefont {D.}~\bibnamefont
  {Boyer}}, \bibinfo {author} {\bibfnamefont {S.~N.}\ \bibnamefont
  {Majumdar}},\ and\ \bibinfo {author} {\bibfnamefont {G.}~\bibnamefont
  {Schehr}},\ }\bibfield  {title} {\bibinfo {title} {Intermittent resetting
  potentials},\ }\href {https://doi.org/10.1088/1742-5468/abc1d9} {\bibfield
  {journal} {\bibinfo  {journal} {Journal of Statistical Mechanics: Theory and
  Experiment}\ }\textbf {\bibinfo {volume} {2020}},\ \bibinfo {pages} {113203}
  (\bibinfo {year} {2020})}\BibitemShut {NoStop}%
\bibitem [{\citenamefont {Santra}\ \emph {et~al.}(2021)\citenamefont {Santra},
  \citenamefont {Das},\ and\ \citenamefont {Nath}}]{Santra2021}%
  \BibitemOpen
  \bibfield  {author} {\bibinfo {author} {\bibfnamefont {I.}~\bibnamefont
  {Santra}}, \bibinfo {author} {\bibfnamefont {S.}~\bibnamefont {Das}},\ and\
  \bibinfo {author} {\bibfnamefont {S.~K.}\ \bibnamefont {Nath}},\ }\bibfield
  {title} {\bibinfo {title} {Brownian motion under intermittent harmonic
  potentials},\ }\href {https://doi.org/10.1088/1751-8121/ac12a0} {\bibfield
  {journal} {\bibinfo  {journal} {Journal of Physics A: Mathematical and
  Theoretical}\ }\textbf {\bibinfo {volume} {54}},\ \bibinfo {pages} {334001}
  (\bibinfo {year} {2021})}\BibitemShut {NoStop}%
\bibitem [{\citenamefont {Esposito}\ and\ \citenamefont {den
  Broeck}(2011)}]{Esposito2011}%
  \BibitemOpen
  \bibfield  {author} {\bibinfo {author} {\bibfnamefont {M.}~\bibnamefont
  {Esposito}}\ and\ \bibinfo {author} {\bibfnamefont {C.~V.}\ \bibnamefont {den
  Broeck}},\ }\bibfield  {title} {\bibinfo {title} {Second law and landauer
  principle far from equilibrium},\ }\href
  {https://doi.org/10.1209/0295-5075/95/40004} {\bibfield  {journal} {\bibinfo
  {journal} {{EPL} (Europhysics Letters)}\ }\textbf {\bibinfo {volume} {95}},\
  \bibinfo {pages} {40004} (\bibinfo {year} {2011})}\BibitemShut {NoStop}%
\bibitem [{\citenamefont {Szilard}(1929)}]{szilard1929}%
  \BibitemOpen
  \bibfield  {author} {\bibinfo {author} {\bibfnamefont {L.}~\bibnamefont
  {Szilard}},\ }\bibfield  {title} {\bibinfo {title} {\"uber die
  entropieverminderung in einem thermodynamischen system bei eingriffen
  intelligenter wesen},\ }\href@noop {} {\bibfield  {journal} {\bibinfo
  {journal} {Zeitschrift f{\"u}r Physik}\ }\textbf {\bibinfo {volume} {53}},\
  \bibinfo {pages} {840} (\bibinfo {year} {1929})}\BibitemShut {NoStop}%
\bibitem [{\citenamefont {Parrondo}\ \emph {et~al.}(2015)\citenamefont
  {Parrondo}, \citenamefont {Horowitz},\ and\ \citenamefont
  {Sagawa}}]{Parrondo2015}%
  \BibitemOpen
  \bibfield  {author} {\bibinfo {author} {\bibfnamefont {J.~M.~R.}\
  \bibnamefont {Parrondo}}, \bibinfo {author} {\bibfnamefont {J.~M.}\
  \bibnamefont {Horowitz}},\ and\ \bibinfo {author} {\bibfnamefont
  {T.}~\bibnamefont {Sagawa}},\ }\bibfield  {title} {\bibinfo {title}
  {Thermodynamics of information},\ }\href {https://doi.org/10.1038/nphys3230}
  {\bibfield  {journal} {\bibinfo  {journal} {Nature Physics}\ }\textbf
  {\bibinfo {volume} {11}},\ \bibinfo {pages} {131} (\bibinfo {year}
  {2015})}\BibitemShut {NoStop}%
\bibitem [{Sup()}]{Supplemental}%
  \BibitemOpen
  \bibfield  {title} {\bibinfo {title} {See {S}upplemental {M}aterial at [url
  will be inserted by publisher] for a detailed account of the experimental and
  numerical methods used in this work as well as theoretical derivations; the
  derivation of the stationary distribution of a particle undergoing poissonian
  sr in a harmonic trap; the derivation of the system, medium and resetting
  entropies used in the main manuscript; the derivation of the quasistatic
  limit of the external applied work; the derivation of the minimal reversible
  work, from an explicit measure of the information content at each resetting
  event and finally the experimental construction of the ergodicity criterium
  used, as well as the method of drift correction used},\ }\href@noop {} {\
  }\BibitemShut {NoStop}%
\bibitem [{\citenamefont {Pal}(2015)}]{Pal2015}%
  \BibitemOpen
  \bibfield  {author} {\bibinfo {author} {\bibfnamefont {A.}~\bibnamefont
  {Pal}},\ }\bibfield  {title} {\bibinfo {title} {Diffusion in a potential
  landscape with stochastic resetting},\ }\href
  {https://doi.org/10.1103/PhysRevE.91.012113} {\bibfield  {journal} {\bibinfo
  {journal} {Phys. Rev. E}\ }\textbf {\bibinfo {volume} {91}},\ \bibinfo
  {pages} {012113} (\bibinfo {year} {2015})}\BibitemShut {NoStop}%
\bibitem [{\citenamefont {Gradshteyn}\ and\ \citenamefont
  {Ryzhik}(2007)}]{gradshteyn2007}%
  \BibitemOpen
  \bibfield  {author} {\bibinfo {author} {\bibfnamefont {I.~S.}\ \bibnamefont
  {Gradshteyn}}\ and\ \bibinfo {author} {\bibfnamefont {I.~M.}\ \bibnamefont
  {Ryzhik}},\ }\href@noop {} {\emph {\bibinfo {title} {Table of integrals,
  series, and products}}},\ \bibinfo {edition} {seventh}\ ed.\ (\bibinfo
  {publisher} {Elsevier/Academic Press, Amsterdam},\ \bibinfo {year} {2007})\
  pp.\ \bibinfo {pages} {1170+1171}\BibitemShut {NoStop}%
\bibitem [{\citenamefont {Seifert}(2005)}]{Seifert2005}%
  \BibitemOpen
  \bibfield  {author} {\bibinfo {author} {\bibfnamefont {U.}~\bibnamefont
  {Seifert}},\ }\bibfield  {title} {\bibinfo {title} {Entropy production along
  a stochastic trajectory and an integral fluctuation theorem},\ }\href
  {https://doi.org/10.1103/PhysRevLett.95.040602} {\bibfield  {journal}
  {\bibinfo  {journal} {Phys. Rev. Lett.}\ }\textbf {\bibinfo {volume} {95}},\
  \bibinfo {pages} {040602} (\bibinfo {year} {2005})}\BibitemShut {NoStop}%
\bibitem [{\citenamefont {Sekimoto}(1998)}]{Sekimoto1998}%
  \BibitemOpen
  \bibfield  {author} {\bibinfo {author} {\bibfnamefont {K.}~\bibnamefont
  {Sekimoto}},\ }\bibfield  {title} {\bibinfo {title} {{Langevin Equation and
  Thermodynamics}},\ }\href {https://doi.org/10.1143/PTPS.130.17} {\bibfield
  {journal} {\bibinfo  {journal} {Progress of Theoretical Physics Supplement}\
  }\textbf {\bibinfo {volume} {130}},\ \bibinfo {pages} {17} (\bibinfo {year}
  {1998})},\ \Eprint
  {https://arxiv.org/abs/https://academic.oup.com/ptps/article-pdf/doi/10.1143/PTPS.130.17/5213518/130-17.pdf}
  {https://academic.oup.com/ptps/article-pdf/doi/10.1143/PTPS.130.17/5213518/130-17.pdf}
  \BibitemShut {NoStop}%
\bibitem [{\citenamefont {Sekimoto}(2010)}]{SekimotoBook}%
  \BibitemOpen
  \bibfield  {author} {\bibinfo {author} {\bibfnamefont {K.}~\bibnamefont
  {Sekimoto}},\ }\href {https://link.springer.com/} {\emph {\bibinfo {title}
  {Stochastic energetics}}}\ (\bibinfo  {publisher} {Springer-Verlag, Berlin
  Heidelberg},\ \bibinfo {year} {2010})\BibitemShut {NoStop}%
\bibitem [{\citenamefont {Ciliberto}(2017)}]{CilibertoPRX2017}%
  \BibitemOpen
  \bibfield  {author} {\bibinfo {author} {\bibfnamefont {S.}~\bibnamefont
  {Ciliberto}},\ }\bibfield  {title} {\bibinfo {title} {Experiments in
  stochastic thermodynamics: Short history and perspectives},\ }\href
  {https://doi.org/10.1103/PhysRevX.7.021051} {\bibfield  {journal} {\bibinfo
  {journal} {Phys. Rev. X}\ }\textbf {\bibinfo {volume} {7}},\ \bibinfo {pages}
  {021051} (\bibinfo {year} {2017})}\BibitemShut {NoStop}%
\bibitem [{\citenamefont {Bechhoefer}\ \emph {et~al.}(2020)\citenamefont
  {Bechhoefer}, \citenamefont {Ciliberto}, \citenamefont {Pigolotti},\ and\
  \citenamefont {Rold{\'{a}}n}}]{Bechhoefer2020}%
  \BibitemOpen
  \bibfield  {author} {\bibinfo {author} {\bibfnamefont {J.}~\bibnamefont
  {Bechhoefer}}, \bibinfo {author} {\bibfnamefont {S.}~\bibnamefont
  {Ciliberto}}, \bibinfo {author} {\bibfnamefont {S.}~\bibnamefont
  {Pigolotti}},\ and\ \bibinfo {author} {\bibfnamefont {E.}~\bibnamefont
  {Rold{\'{a}}n}},\ }\bibfield  {title} {\bibinfo {title} {Stochastic
  thermodynamics: experiment and theory},\ }\href
  {https://doi.org/10.1088/1742-5468/ab7f35} {\bibfield  {journal} {\bibinfo
  {journal} {J. Stat. Mech.}\ }\textbf {\bibinfo {volume} {2020}},\ \bibinfo
  {pages} {064001} (\bibinfo {year} {2020})}\BibitemShut {NoStop}%
\bibitem [{\citenamefont {Rosales-Cabara}\ \emph {et~al.}(2020)\citenamefont
  {Rosales-Cabara}, \citenamefont {Manfredi}, \citenamefont {Schnoering},
  \citenamefont {Hervieux}, \citenamefont {Mertz},\ and\ \citenamefont
  {Genet}}]{Cabara2020}%
  \BibitemOpen
  \bibfield  {author} {\bibinfo {author} {\bibfnamefont {Y.}~\bibnamefont
  {Rosales-Cabara}}, \bibinfo {author} {\bibfnamefont {G.}~\bibnamefont
  {Manfredi}}, \bibinfo {author} {\bibfnamefont {G.}~\bibnamefont
  {Schnoering}}, \bibinfo {author} {\bibfnamefont {P.-A.}\ \bibnamefont
  {Hervieux}}, \bibinfo {author} {\bibfnamefont {L.}~\bibnamefont {Mertz}},\
  and\ \bibinfo {author} {\bibfnamefont {C.}~\bibnamefont {Genet}},\ }\bibfield
   {title} {\bibinfo {title} {Optimal protocols and universal time-energy bound
  in brownian thermodynamics},\ }\href
  {https://doi.org/10.1103/PhysRevResearch.2.012012} {\bibfield  {journal}
  {\bibinfo  {journal} {Phys. Rev. Research}\ }\textbf {\bibinfo {volume}
  {2}},\ \bibinfo {pages} {012012} (\bibinfo {year} {2020})}\BibitemShut
  {NoStop}%
\bibitem [{\citenamefont {B{\'e}rut}\ \emph {et~al.}(2012)\citenamefont
  {B{\'e}rut}, \citenamefont {Arakelyan}, \citenamefont {Petrosyan},
  \citenamefont {Ciliberto}, \citenamefont {Dillenschneider},\ and\
  \citenamefont {Lutz}}]{Berut2012}%
  \BibitemOpen
  \bibfield  {author} {\bibinfo {author} {\bibfnamefont {A.}~\bibnamefont
  {B{\'e}rut}}, \bibinfo {author} {\bibfnamefont {A.}~\bibnamefont
  {Arakelyan}}, \bibinfo {author} {\bibfnamefont {A.}~\bibnamefont
  {Petrosyan}}, \bibinfo {author} {\bibfnamefont {S.}~\bibnamefont
  {Ciliberto}}, \bibinfo {author} {\bibfnamefont {R.}~\bibnamefont
  {Dillenschneider}},\ and\ \bibinfo {author} {\bibfnamefont {E.}~\bibnamefont
  {Lutz}},\ }\bibfield  {title} {\bibinfo {title} {Experimental verification of
  landauer's principle linking information and thermodynamics},\ }\href
  {https://doi.org/10.1038/nature10872} {\bibfield  {journal} {\bibinfo
  {journal} {Nature}\ }\textbf {\bibinfo {volume} {483}},\ \bibinfo {pages}
  {187} (\bibinfo {year} {2012})}\BibitemShut {NoStop}%
\bibitem [{\citenamefont {Jun}\ \emph {et~al.}(2014)\citenamefont {Jun},
  \citenamefont {Gavrilov},\ and\ \citenamefont {Bechhoefer}}]{Jun2014}%
  \BibitemOpen
  \bibfield  {author} {\bibinfo {author} {\bibfnamefont {Y.}~\bibnamefont
  {Jun}}, \bibinfo {author} {\bibfnamefont {M.~c.~v.}\ \bibnamefont
  {Gavrilov}},\ and\ \bibinfo {author} {\bibfnamefont {J.}~\bibnamefont
  {Bechhoefer}},\ }\bibfield  {title} {\bibinfo {title} {High-precision test of
  landauer's principle in a feedback trap},\ }\href
  {https://doi.org/10.1103/PhysRevLett.113.190601} {\bibfield  {journal}
  {\bibinfo  {journal} {Phys. Rev. Lett.}\ }\textbf {\bibinfo {volume} {113}},\
  \bibinfo {pages} {190601} (\bibinfo {year} {2014})}\BibitemShut {NoStop}%
\bibitem [{\citenamefont {Touchette}(2009)}]{touchette2009large}%
  \BibitemOpen
  \bibfield  {author} {\bibinfo {author} {\bibfnamefont {H.}~\bibnamefont
  {Touchette}},\ }\bibfield  {title} {\bibinfo {title} {The large deviation
  approach to statistical mechanics},\ }\href@noop {} {\bibfield  {journal}
  {\bibinfo  {journal} {Physics Reports}\ }\textbf {\bibinfo {volume} {478}},\
  \bibinfo {pages} {1} (\bibinfo {year} {2009})}\BibitemShut {NoStop}%
\bibitem [{\citenamefont {Stojkoski}\ \emph {et~al.}(2022)\citenamefont
  {Stojkoski}, \citenamefont {Sandev}, \citenamefont {Kocarev},\ and\
  \citenamefont {Pal}}]{Stojkoski2022}%
  \BibitemOpen
  \bibfield  {author} {\bibinfo {author} {\bibfnamefont {V.}~\bibnamefont
  {Stojkoski}}, \bibinfo {author} {\bibfnamefont {T.}~\bibnamefont {Sandev}},
  \bibinfo {author} {\bibfnamefont {L.}~\bibnamefont {Kocarev}},\ and\ \bibinfo
  {author} {\bibfnamefont {A.}~\bibnamefont {Pal}},\ }\bibfield  {title}
  {\bibinfo {title} {Autocorrelation functions and ergodicity in diffusion with
  stochastic resetting},\ }\href {https://doi.org/10.1088/1751-8121/ac4ce9}
  {\bibfield  {journal} {\bibinfo  {journal} {Journal of Physics A:
  Mathematical and Theoretical}\ }\textbf {\bibinfo {volume} {55}},\ \bibinfo
  {pages} {104003} (\bibinfo {year} {2022})}\BibitemShut {NoStop}%
\bibitem [{\citenamefont {Barkai}\ \emph {et~al.}(2023)\citenamefont {Barkai},
  \citenamefont {Flaquer-Galm\'~es},\ and\ \citenamefont
  {M\'endez}}]{Barkai2023}%
  \BibitemOpen
  \bibfield  {author} {\bibinfo {author} {\bibfnamefont {E.}~\bibnamefont
  {Barkai}}, \bibinfo {author} {\bibfnamefont {R.}~\bibnamefont
  {Flaquer-Galm\'~es}},\ and\ \bibinfo {author} {\bibfnamefont
  {V.}~\bibnamefont {M\'endez}},\ }\bibfield  {title} {\bibinfo {title}
  {Ergodic properties of brownian motion under stochastic resetting},\ }\href
  {https://doi.org/10.1103/PhysRevE.108.064102} {\bibfield  {journal} {\bibinfo
   {journal} {Phys. Rev. E}\ }\textbf {\bibinfo {volume} {108}},\ \bibinfo
  {pages} {064102} (\bibinfo {year} {2023})}\BibitemShut {NoStop}%
\bibitem [{\citenamefont {Metzler}\ and\ \citenamefont
  {Klafter}(2000)}]{METZLER2000}%
  \BibitemOpen
  \bibfield  {author} {\bibinfo {author} {\bibfnamefont {R.}~\bibnamefont
  {Metzler}}\ and\ \bibinfo {author} {\bibfnamefont {J.}~\bibnamefont
  {Klafter}},\ }\bibfield  {title} {\bibinfo {title} {The random walk's guide
  to anomalous diffusion: a fractional dynamics approach},\ }\href
  {https://doi.org/https://doi.org/10.1016/S0370-1573(00)00070-3} {\bibfield
  {journal} {\bibinfo  {journal} {Physics Reports}\ }\textbf {\bibinfo {volume}
  {339}},\ \bibinfo {pages} {1} (\bibinfo {year} {2000})}\BibitemShut {NoStop}%
\bibitem [{\citenamefont {Deng}\ and\ \citenamefont
  {Barkai}(2009)}]{Barkai2009}%
  \BibitemOpen
  \bibfield  {author} {\bibinfo {author} {\bibfnamefont {W.}~\bibnamefont
  {Deng}}\ and\ \bibinfo {author} {\bibfnamefont {E.}~\bibnamefont {Barkai}},\
  }\bibfield  {title} {\bibinfo {title} {Ergodic properties of fractional
  brownian-langevin motion},\ }\href
  {https://doi.org/10.1103/PhysRevE.79.011112} {\bibfield  {journal} {\bibinfo
  {journal} {Phys. Rev. E}\ }\textbf {\bibinfo {volume} {79}},\ \bibinfo
  {pages} {011112} (\bibinfo {year} {2009})}\BibitemShut {NoStop}%
\bibitem [{\citenamefont {Cherstvy}\ and\ \citenamefont
  {Metzler}(2015)}]{Metzler2015}%
  \BibitemOpen
  \bibfield  {author} {\bibinfo {author} {\bibfnamefont {A.~G.}\ \bibnamefont
  {Cherstvy}}\ and\ \bibinfo {author} {\bibfnamefont {R.}~\bibnamefont
  {Metzler}},\ }\bibfield  {title} {\bibinfo {title} {Ergodicity breaking,
  ageing, and confinement in generalized diffusion processes with position and
  time dependent diffusivity},\ }\href
  {https://doi.org/10.1088/1742-5468/2015/05/p05010} {\bibfield  {journal}
  {\bibinfo  {journal} {J. Stat. Mech.}\ }\textbf {\bibinfo {volume} {2015}},\
  \bibinfo {pages} {P05010} (\bibinfo {year} {2015})}\BibitemShut {NoStop}%
\bibitem [{\citenamefont {Metzler}\ \emph
  {et~al.}(2014{\natexlab{a}})\citenamefont {Metzler}, \citenamefont {Jeon},
  \citenamefont {Cherstvy},\ and\ \citenamefont
  {Barkai}}]{metzler2014anomalous}%
  \BibitemOpen
  \bibfield  {author} {\bibinfo {author} {\bibfnamefont {R.}~\bibnamefont
  {Metzler}}, \bibinfo {author} {\bibfnamefont {J.-H.}\ \bibnamefont {Jeon}},
  \bibinfo {author} {\bibfnamefont {A.~G.}\ \bibnamefont {Cherstvy}},\ and\
  \bibinfo {author} {\bibfnamefont {E.}~\bibnamefont {Barkai}},\ }\bibfield
  {title} {\bibinfo {title} {Anomalous diffusion models and their properties:
  non-stationarity, non-ergodicity, and ageing at the centenary of single
  particle tracking},\ }\href@noop {} {\bibfield  {journal} {\bibinfo
  {journal} {Physical Chemistry Chemical Physics}\ }\textbf {\bibinfo {volume}
  {16}},\ \bibinfo {pages} {24128} (\bibinfo {year}
  {2014}{\natexlab{a}})}\BibitemShut {NoStop}%
\bibitem [{\citenamefont {Li}\ \emph {et~al.}(2019)\citenamefont {Li},
  \citenamefont {Sentissi}, \citenamefont {Azzini}, \citenamefont {Schnoering},
  \citenamefont {Canaguier-Durand},\ and\ \citenamefont {Genet}}]{Li2019}%
  \BibitemOpen
  \bibfield  {author} {\bibinfo {author} {\bibfnamefont {M.}~\bibnamefont
  {Li}}, \bibinfo {author} {\bibfnamefont {O.}~\bibnamefont {Sentissi}},
  \bibinfo {author} {\bibfnamefont {S.}~\bibnamefont {Azzini}}, \bibinfo
  {author} {\bibfnamefont {G.}~\bibnamefont {Schnoering}}, \bibinfo {author}
  {\bibfnamefont {A.}~\bibnamefont {Canaguier-Durand}},\ and\ \bibinfo {author}
  {\bibfnamefont {C.}~\bibnamefont {Genet}},\ }\bibfield  {title} {\bibinfo
  {title} {Subfemtonewton force fields measured with ergodic brownian
  ensembles},\ }\href {https://doi.org/10.1103/PhysRevA.100.063816} {\bibfield
  {journal} {\bibinfo  {journal} {Phys. Rev. A}\ }\textbf {\bibinfo {volume}
  {100}},\ \bibinfo {pages} {063816} (\bibinfo {year} {2019})}\BibitemShut
  {NoStop}%
\bibitem [{\citenamefont {Goerlich}\ \emph {et~al.}(2021)\citenamefont
  {Goerlich}, \citenamefont {Li}, \citenamefont {Albert}, \citenamefont
  {Manfredi}, \citenamefont {Hervieux},\ and\ \citenamefont
  {Genet}}]{Ergodicity2021}%
  \BibitemOpen
  \bibfield  {author} {\bibinfo {author} {\bibfnamefont {R.}~\bibnamefont
  {Goerlich}}, \bibinfo {author} {\bibfnamefont {M.}~\bibnamefont {Li}},
  \bibinfo {author} {\bibfnamefont {S.}~\bibnamefont {Albert}}, \bibinfo
  {author} {\bibfnamefont {G.}~\bibnamefont {Manfredi}}, \bibinfo {author}
  {\bibfnamefont {P.-A.}\ \bibnamefont {Hervieux}},\ and\ \bibinfo {author}
  {\bibfnamefont {C.}~\bibnamefont {Genet}},\ }\bibfield  {title} {\bibinfo
  {title} {Noise and ergodic properties of brownian motion in an optical
  tweezer: Looking at regime crossovers in an ornstein-uhlenbeck process},\
  }\href {https://doi.org/10.1103/PhysRevE.103.032132} {\bibfield  {journal}
  {\bibinfo  {journal} {Phys. Rev. E}\ }\textbf {\bibinfo {volume} {103}},\
  \bibinfo {pages} {032132} (\bibinfo {year} {2021})}\BibitemShut {NoStop}%
\bibitem [{\citenamefont {Metzler}\ \emph
  {et~al.}(2014{\natexlab{b}})\citenamefont {Metzler}, \citenamefont {Jeon},
  \citenamefont {Cherstvy},\ and\ \citenamefont {Barkai}}]{Metzler2014}%
  \BibitemOpen
  \bibfield  {author} {\bibinfo {author} {\bibfnamefont {R.}~\bibnamefont
  {Metzler}}, \bibinfo {author} {\bibfnamefont {J.-H.}\ \bibnamefont {Jeon}},
  \bibinfo {author} {\bibfnamefont {A.~G.}\ \bibnamefont {Cherstvy}},\ and\
  \bibinfo {author} {\bibfnamefont {E.}~\bibnamefont {Barkai}},\ }\bibfield
  {title} {\bibinfo {title} {Anomalous diffusion models and their properties:
  non-stationarity{,} non- ergodicity{,} and ageing at the centenary of single
  particle tracking},\ }\href {https://doi.org/10.1039/C4CP03465A} {\bibfield
  {journal} {\bibinfo  {journal} {Phys. Chem. Chem. Phys.}\ }\textbf {\bibinfo
  {volume} {16}},\ \bibinfo {pages} {24128} (\bibinfo {year}
  {2014}{\natexlab{b}})}\BibitemShut {NoStop}%
\bibitem [{\citenamefont {Stojkoski}\ \emph {et~al.}(2021)\citenamefont
  {Stojkoski}, \citenamefont {Sandev}, \citenamefont {Kocarev},\ and\
  \citenamefont {Pal}}]{Stojkoski2021}%
  \BibitemOpen
  \bibfield  {author} {\bibinfo {author} {\bibfnamefont {V.}~\bibnamefont
  {Stojkoski}}, \bibinfo {author} {\bibfnamefont {T.}~\bibnamefont {Sandev}},
  \bibinfo {author} {\bibfnamefont {L.}~\bibnamefont {Kocarev}},\ and\ \bibinfo
  {author} {\bibfnamefont {A.}~\bibnamefont {Pal}},\ }\bibfield  {title}
  {\bibinfo {title} {Geometric brownian motion under stochastic resetting: A
  stationary yet nonergodic process},\ }\href
  {https://doi.org/10.1103/PhysRevE.104.014121} {\bibfield  {journal} {\bibinfo
   {journal} {Phys. Rev. E}\ }\textbf {\bibinfo {volume} {104}},\ \bibinfo
  {pages} {014121} (\bibinfo {year} {2021})}\BibitemShut {NoStop}%
\bibitem [{\citenamefont {Wang}\ \emph {et~al.}(2021)\citenamefont {Wang},
  \citenamefont {Cherstvy}, \citenamefont {Kantz}, \citenamefont {Metzler},\
  and\ \citenamefont {Sokolov}}]{Wang2021}%
  \BibitemOpen
  \bibfield  {author} {\bibinfo {author} {\bibfnamefont {W.}~\bibnamefont
  {Wang}}, \bibinfo {author} {\bibfnamefont {A.~G.}\ \bibnamefont {Cherstvy}},
  \bibinfo {author} {\bibfnamefont {H.}~\bibnamefont {Kantz}}, \bibinfo
  {author} {\bibfnamefont {R.}~\bibnamefont {Metzler}},\ and\ \bibinfo {author}
  {\bibfnamefont {I.~M.}\ \bibnamefont {Sokolov}},\ }\bibfield  {title}
  {\bibinfo {title} {Time averaging and emerging nonergodicity upon resetting
  of fractional brownian motion and heterogeneous diffusion processes},\ }\href
  {https://doi.org/10.1103/PhysRevE.104.024105} {\bibfield  {journal} {\bibinfo
   {journal} {Phys. Rev. E}\ }\textbf {\bibinfo {volume} {104}},\ \bibinfo
  {pages} {024105} (\bibinfo {year} {2021})}\BibitemShut {NoStop}%
\bibitem [{\citenamefont {Cherstvy}\ \emph {et~al.}(2008)\citenamefont
  {Cherstvy}, \citenamefont {Kolomeisky},\ and\ \citenamefont
  {Kornyshev}}]{Cherstvy2008}%
  \BibitemOpen
  \bibfield  {author} {\bibinfo {author} {\bibfnamefont {A.~G.}\ \bibnamefont
  {Cherstvy}}, \bibinfo {author} {\bibfnamefont {A.~B.}\ \bibnamefont
  {Kolomeisky}},\ and\ \bibinfo {author} {\bibfnamefont {A.~A.}\ \bibnamefont
  {Kornyshev}},\ }\bibfield  {title} {\bibinfo {title} {Protein dna
  interactions: Reaching and recognizing the targets},\ }\href
  {https://doi.org/10.1021/jp076432e} {\bibfield  {journal} {\bibinfo
  {journal} {The Journal of Physical Chemistry B}\ }\textbf {\bibinfo {volume}
  {112}},\ \bibinfo {pages} {4741} (\bibinfo {year} {2008})}\BibitemShut
  {NoStop}%
\bibitem [{\citenamefont {Olsen}\ \emph {et~al.}(2023)\citenamefont {Olsen},
  \citenamefont {Gupta}, \citenamefont {Mori},\ and\ \citenamefont
  {Krishnamurthy}}]{olsen2023thermodynamic}%
  \BibitemOpen
  \bibfield  {author} {\bibinfo {author} {\bibfnamefont {K.~S.}\ \bibnamefont
  {Olsen}}, \bibinfo {author} {\bibfnamefont {D.}~\bibnamefont {Gupta}},
  \bibinfo {author} {\bibfnamefont {F.}~\bibnamefont {Mori}},\ and\ \bibinfo
  {author} {\bibfnamefont {S.}~\bibnamefont {Krishnamurthy}},\ }\bibfield
  {title} {\bibinfo {title} {Thermodynamic cost of finite-time stochastic
  resetting},\ }\href@noop {} {\bibfield  {journal} {\bibinfo  {journal} {arXiv
  preprint arXiv:2310.11267}\ } (\bibinfo {year} {2023})}\BibitemShut {NoStop}%
\bibitem [{\citenamefont {Olsen}\ and\ \citenamefont
  {Gupta}(2024)}]{olsen2024thermodynamic}%
  \BibitemOpen
  \bibfield  {author} {\bibinfo {author} {\bibfnamefont {K.~S.}\ \bibnamefont
  {Olsen}}\ and\ \bibinfo {author} {\bibfnamefont {D.}~\bibnamefont {Gupta}},\
  }\bibfield  {title} {\bibinfo {title} {Thermodynamic work of partial
  resetting},\ }\href@noop {} {\bibfield  {journal} {\bibinfo  {journal}
  {Journal of Physics A: Mathematical and Theoretical}\ }\textbf {\bibinfo
  {volume} {57}},\ \bibinfo {pages} {245001} (\bibinfo {year}
  {2024})}\BibitemShut {NoStop}%
\bibitem [{\citenamefont {Santra}\ \emph {et~al.}(2020)\citenamefont {Santra},
  \citenamefont {Basu},\ and\ \citenamefont {Sabhapandit}}]{Santra2020}%
  \BibitemOpen
  \bibfield  {author} {\bibinfo {author} {\bibfnamefont {I.}~\bibnamefont
  {Santra}}, \bibinfo {author} {\bibfnamefont {U.}~\bibnamefont {Basu}},\ and\
  \bibinfo {author} {\bibfnamefont {S.}~\bibnamefont {Sabhapandit}},\
  }\bibfield  {title} {\bibinfo {title} {Run-and-tumble particles in two
  dimensions under stochastic resetting conditions},\ }\href
  {https://doi.org/10.1088/1742-5468/abc7b7} {\bibfield  {journal} {\bibinfo
  {journal} {Journal of Statistical Mechanics: Theory and Experiment}\ }\textbf
  {\bibinfo {volume} {2020}},\ \bibinfo {pages} {113206} (\bibinfo {year}
  {2020})}\BibitemShut {NoStop}%
\bibitem [{\citenamefont {Abdoli}\ and\ \citenamefont
  {Sharma}(2021)}]{Abdoli2021}%
  \BibitemOpen
  \bibfield  {author} {\bibinfo {author} {\bibfnamefont {I.}~\bibnamefont
  {Abdoli}}\ and\ \bibinfo {author} {\bibfnamefont {A.}~\bibnamefont
  {Sharma}},\ }\bibfield  {title} {\bibinfo {title} {Stochastic resetting of
  active brownian particles with lorentz force},\ }\href
  {https://doi.org/10.1039/D0SM01773F} {\bibfield  {journal} {\bibinfo
  {journal} {Soft Matter}\ }\textbf {\bibinfo {volume} {17}},\ \bibinfo {pages}
  {1307} (\bibinfo {year} {2021})}\BibitemShut {NoStop}%
\bibitem [{\citenamefont {Volpe}\ and\ \citenamefont
  {Volpe}(2013)}]{Volpe2013}%
  \BibitemOpen
  \bibfield  {author} {\bibinfo {author} {\bibfnamefont {G.}~\bibnamefont
  {Volpe}}\ and\ \bibinfo {author} {\bibfnamefont {G.}~\bibnamefont {Volpe}},\
  }\bibfield  {title} {\bibinfo {title} {Simulation of a brownian particle in
  an optical trap},\ }\href {https://doi.org/10.1119/1.4772632} {\bibfield
  {journal} {\bibinfo  {journal} {Am. J. Phys.}\ }\textbf {\bibinfo {volume}
  {81}},\ \bibinfo {pages} {224} (\bibinfo {year} {2013})},\ \Eprint
  {https://arxiv.org/abs/https://doi.org/10.1119/1.4772632}
  {https://doi.org/10.1119/1.4772632} \BibitemShut {NoStop}%
\bibitem [{\citenamefont {Busiello}\ \emph {et~al.}(2020)\citenamefont
  {Busiello}, \citenamefont {Gupta},\ and\ \citenamefont
  {Maritan}}]{Busiello2020}%
  \BibitemOpen
  \bibfield  {author} {\bibinfo {author} {\bibfnamefont {D.~M.}\ \bibnamefont
  {Busiello}}, \bibinfo {author} {\bibfnamefont {D.}~\bibnamefont {Gupta}},\
  and\ \bibinfo {author} {\bibfnamefont {A.}~\bibnamefont {Maritan}},\
  }\bibfield  {title} {\bibinfo {title} {Entropy production in systems with
  unidirectional transitions},\ }\href
  {https://doi.org/10.1103/PhysRevResearch.2.023011} {\bibfield  {journal}
  {\bibinfo  {journal} {Phys. Rev. Research}\ }\textbf {\bibinfo {volume}
  {2}},\ \bibinfo {pages} {023011} (\bibinfo {year} {2020})}\BibitemShut
  {NoStop}%
\end{thebibliography}%

\end{document}